%% file: main.tex
\documentclass[sigplan,screen]{acmart}

\usepackage{pervasives}
\include{revision}
\include{format-hacks}

\renewcommand{\paragraph}[1]{\noindentparagraph{#1}}


\setcopyright{acmlicensed}
\copyrightyear{2020}
\acmYear{2020}
\acmConference[PLDI '20]{Proceedings of the 41st ACM SIGPLAN International Conference on Programming Language Design and Implementation}{June 15--20, 2020}{London, UK}
\acmBooktitle{Proceedings of the 41st ACM SIGPLAN International Conference on Programming Language Design and Implementation (PLDI '20), June 15--20, 2020, London, UK}
\acmPrice{15.00}
\acmDOI{10.1145/3385412.3385974}
\acmISBN{978-1-4503-7613-6/20/06}
\renewcommand{\shortauthors}{Nigam, Atapattu, Thomas,
Li, Bauer, Ye, Koti, Sampson, and Zhang}

\begin{document}

\title[Predictable Accelerator Design with Time-Sensitive Affine Types]{Predictable Accelerator Design \\
with Time-Sensitive Affine Types}
\ifdraft
\subtitle{Rev \Revision}
\fi

\begin{abstract}
Field-programmable gate arrays (FPGAs) provide an opportunity to co-design applications with hardware accelerators,
yet they remain difficult to program.
\emph{High-level synthesis} (HLS) tools promise to raise the level of abstraction by compiling C or \cxx to accelerator designs.
Repurposing legacy software languages, however, requires complex heuristics to map imperative code onto hardware structures.
We find that the black-box heuristics in HLS can be \emph{unpredictable:}
changing parameters in the program that should improve performance can counterintuitively yield slower and larger designs.
This paper proposes a type system that restricts HLS to programs that can predictably compile to hardware accelerators.
The key idea is to model consumable hardware resources with a \emph{time-sensitive affine type system} that prevents simultaneous uses of the same hardware structure.
We implement the type system in \sys, a language that compiles to HLS \cxx,
and show that it can reduce the size of HLS parameter spaces while accepting Pareto-optimal designs.
\end{abstract}

\maketitle

\section{Introduction}

While Moore's law may not be dead yet, its stalled returns for traditional CPUs have sparked renewed interest in specialized hardware accelerators~\cite{goldenage}, for domains from machine learning~\cite{tpu} to genomics~\cite{darwin}.
Reconfigurable hardware---namely, field-programmable gate arrays (FPGAs)---offer some of the benefits of specialization without the cost of custom silicon.
FPGAs can accelerate code in domains from databases~\cite{linqits} to networking~\cite{tonic}
and have driven vast efficiency improvements in Microsoft's data centers~\cite{catapult, brainwave}.

However, FPGAs are hard to program.
The gold-standard programming model for FPGAs is register transfer level (RTL) design in hardware description languages such as Verilog, VHDL, Bluespec, and Chisel~\cite{nikhil:bluespec, bachrach:chisel}.
RTL requires digital design expertise:
akin to assembly languages for CPUs, RTL is irreplaceable for
manual performance tuning, but it is too explicit and verbose for rapid iteration~\cite{gotcha-again}.

FPGA vendors offer
\emph{high-level synthesis} (HLS) or ``C-to-gates'' tools~\cite{vivadohls,
cong2011hls, pilato:bambu, canis:legup} that translate annotated
subsets of C and \cxx
to RTL.
Repurposing a legacy software languages, however, has drawbacks:
the resulting language subset is small and
difficult to specify,
and minor code edits can cause large
swings in hardware efficiency.
We find empirically that smoothly changing source-level hints can cause wild variations in accelerator performance.
Semantically, \emph{there is no HLS programming
language:} there is only the subset of \cxx that a particular version of a
particular compiler supports.
%

This paper describes a type system that restricts HLS to programs whose
hardware implementation is clear.
The goal is \emph{predictable} architecture generation: the hardware implications are observable in the source code,
and costly implementation decisions require explicit permission from the programmer.
Instead of silently generating bad hardware for difficult input programs, the type system yields errors that help guide the programmer toward a better design.
The result is a language that can express a subset of the architectures that HLS can---but it does so predictably.
%
%
%

The central insight is that an affine type system~\cite{alms}
can model the restrictions of hardware implementation.
Components in a hardware design are finite and expendable: a
subcircuit or a memory can only do one thing at a time, so a
program needs to avoid conflicting uses.
Previous research has shown how to apply substructural
type systems to model classic computational resources such as memory allocations and file handles~\cite{cyclone-region, linear-haskell, rust,
alms} and to enforce exclusion for safe shared-memory
parallelism~\cite{gordon:refim, baker:useonce, pony-deny}.
Unlike those classic resources, however, the availability of hardware components changes with time.
%
We extend affine types with \emph{time sensitivity} to express that
repeated uses of the same hardware is safe as long as they are
temporally separated.

We describe \sys, a programming language for predictable accelerator design. \sys differs from traditional HLS in two ways: (1) \sys makes the hardware implementation for each language construct manifest in the source code instead of leaving this decision up to the HLS middle-end, and (2) \sys uses its \emph{time-sensitive affine types} to reason about the hardware constraints and reject programs that would require complex transformation to implement in hardware. We implement a compiler for \sys that emits annotated \cxx for a commercial HLS toolchain.
We show that predictability pitfalls exist in both industrial and recent academic tools and that \sys's reasoning can help alleviate these issues.

%
%
%
%

The contributions of this paper are:
\vspace{-0.1em}
\begin{itemize}

\item We identify predictability pitfalls in HLS and measure their effects in an industrial tool in \Cref{sec:principles}.

\item We design \sys (\Cref{sec:language}), a language that restricts HLS to predictable design spaces by modeling hardware constraints using \emph{time-sensitive affine types}.

\item We formalize a time-sensitive affine type system and prove syntactic type soundness in \Cref{sec:formalism}.

\item We empirically demonstrate \sys's effectiveness in rejecting
  unpredictable design points and its ability to make
  area--performance trade-offs in common accelerator designs in \Cref{sec:evaluation}.

\end{itemize}

\section{Predictability Pitfalls in Traditional HLS}
\label{sec:principles}

\begin{figure}
    \includegraphics[width=\linewidth]{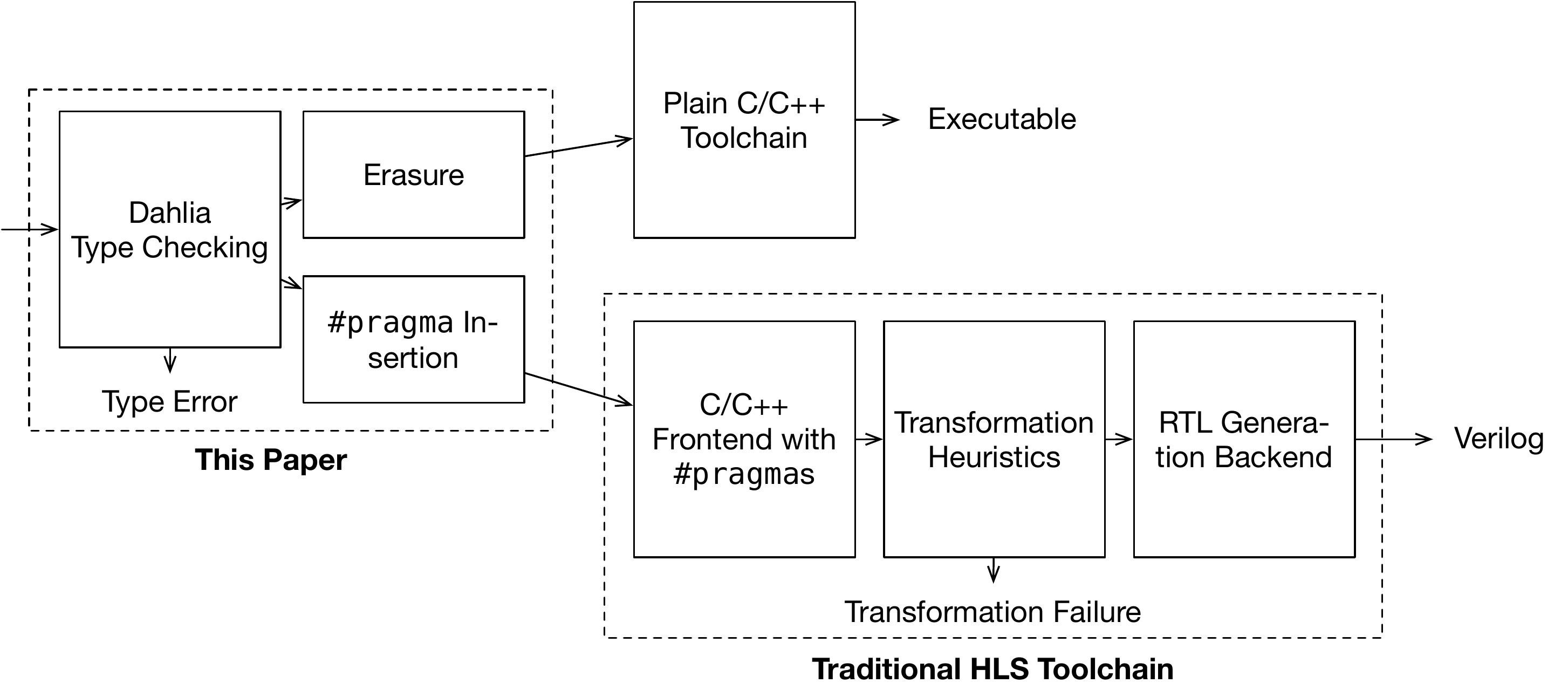}
    \caption{Overview of a traditional high-level synthesis toolchain and how \sys{} layers type safety on top.
    }
    \label{fig:overview}
\end{figure}

%
\cref{fig:overview} depicts the design of a traditional high-level synthesis (HLS) compiler.
A typical HLS tool adopts an existing open-source C/\cxx frontend and adds a
set of \emph{transformation heuristics} that attempt to map software constructs
onto hardware elements along with a backend that generates RTL
code~\cite{xpilot,canis:legup}.
The transformation step typically relies on a constraint solver, such as an LP or SAT solver, to satisfy resource, layout, and timing requirements~\cite{gupta:spark, sdc-sched}.
Programmers can add \code{#pragma} hints to guide the transformation---for
example, to duplicate loop bodies or to share functional units.
%

HLS tools are best-effort compilers: they make a heuristic effort to translate
\emph{any} valid C/\cxx program to RTL, regardless of the consequences for the
generated accelerator architecture.
%
Sometimes, the mapping constraints are unsatisfiable, so the compiler selectively ignores some \code{#pragma} hints or issues an error.
The generated accelerator's efficiency depends on the interaction between the code, the hints, and the transformation heuristics that use them.

The standard approach prioritizes automation over predictability.
%
%
Small code changes
can yield large shifts in the generated architecture.
When performance is poor, the compiler provides little guidance about how to improve it.
%
Pruning such \emph{unpredictable} points from the design space would let programmers
explore smaller, smoother parameter spaces.




\subsection{An Example in HLS}
\label{sec:principles:ex}

\begin{figure}
\begin{lstlisting}[language=C,xrightmargin=0.4cm,aboveskip=0.2cm]
int m1[512][512], m2[512][512], prod[512][512];
int sum;
for (int i = 0; i < 512; i++) {
  for (int j = 0; j < 512; j++) {
    sum = 0;
    for (int k = 0; k < 512; k++) {
      sum += m1[i][k] * m2[k][j];
    }
    prod[i][j] = sum; } }
\end{lstlisting}
\caption{Dense matrix multiplication in HLS-friendly C.}
\label{fig:gemm-ncubed}
\end{figure}

\begin{figure*}
    \centering
    \hspace{2em}
    \begin{subfigure}[b]{0.20\linewidth}
        \centering
        \includegraphics[width=\linewidth]{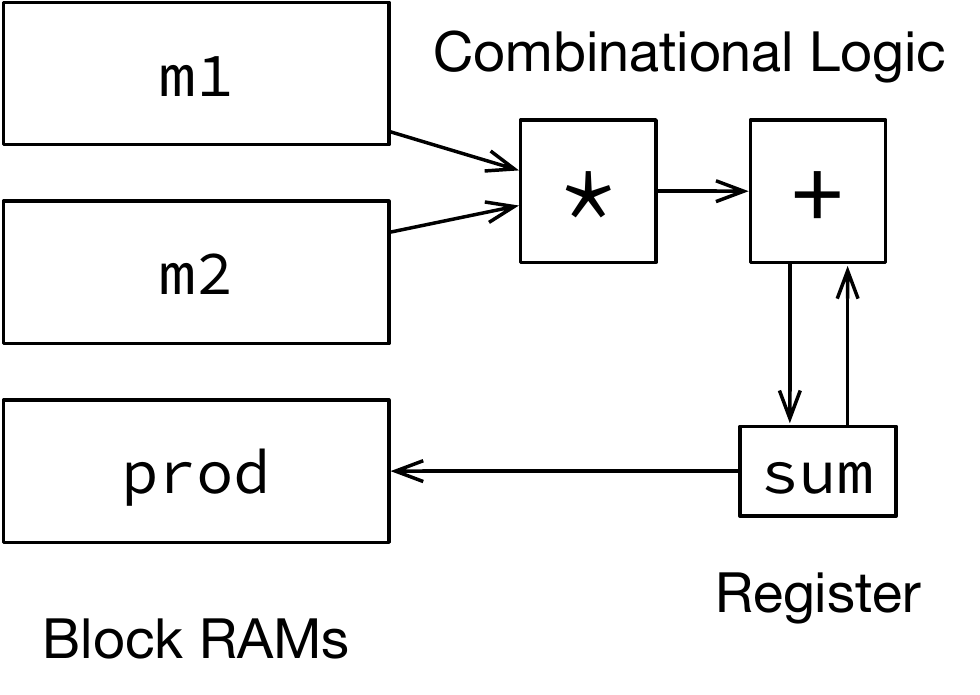}
        \caption{The original code.}
        \label{fig:hls-example:orig}
    \end{subfigure}
    \hfill
    \begin{subfigure}[b]{0.33\linewidth}
        \centering
        \includegraphics[width=\linewidth]{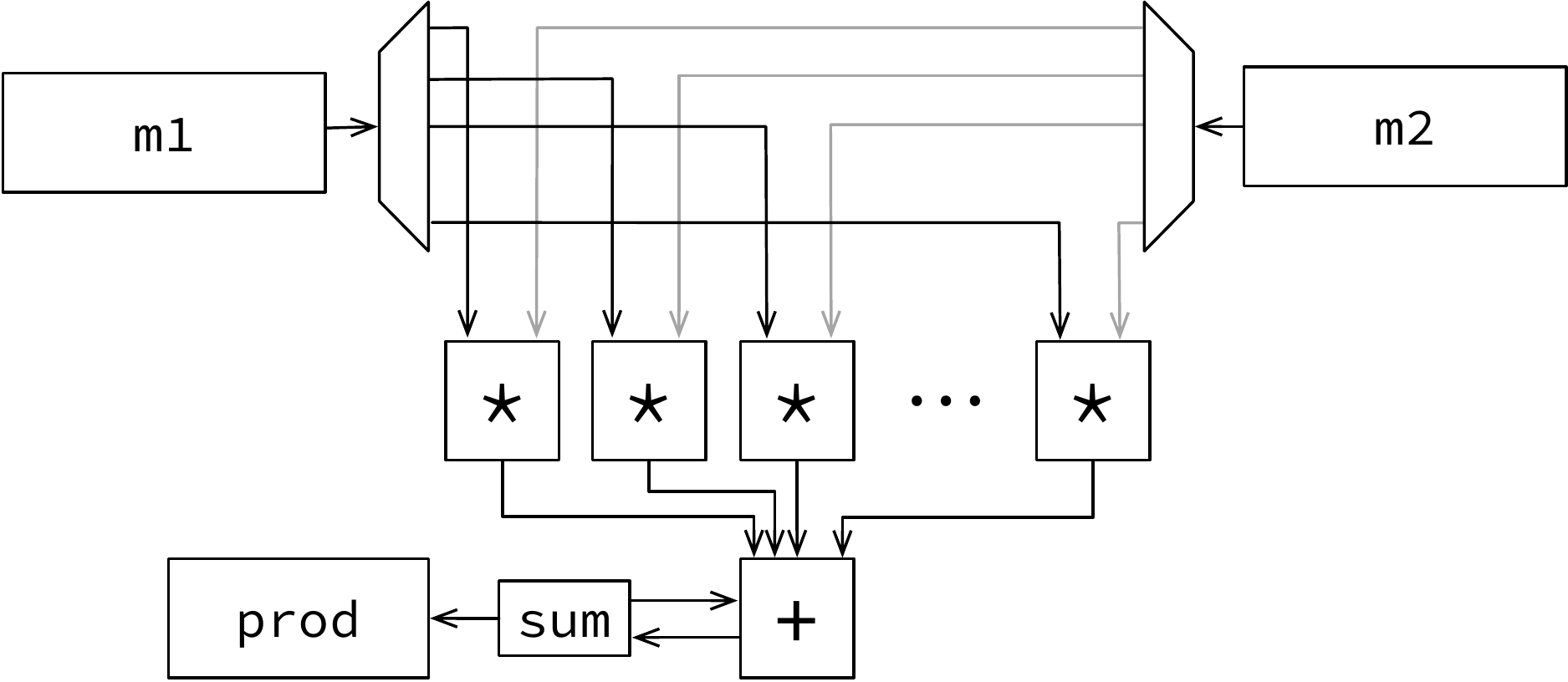}
        \caption{With unrolling.}
        \label{fig:hls-example:mux}
    \end{subfigure}
    \hfill
    \begin{subfigure}[b]{0.27\linewidth}
        \centering
        \includegraphics[width=\linewidth]{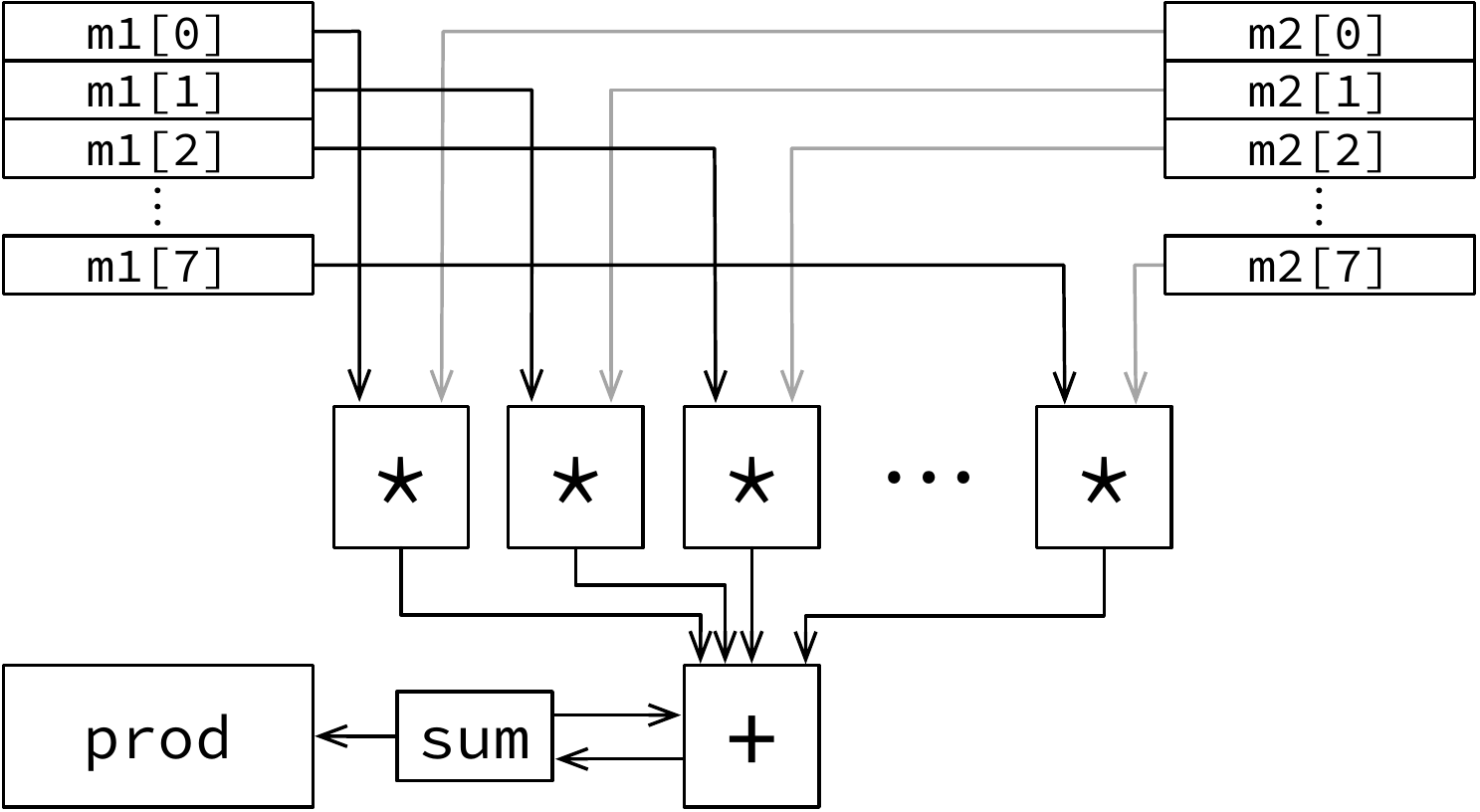}
        \caption{With unrolling and banking.}
        \label{fig:hls-example:unrolled}
    \end{subfigure}
    \hspace{2em}
    \caption{Three accelerator implementations of the matrix multiplication in \cref{fig:gemm-ncubed}.}
    \label{fig:hls-example}
\end{figure*}

\begin{figure*}
  \centering
  \begin{subfigure}[b]{0.27\linewidth}
    \centering
    \includegraphics[width=\linewidth]{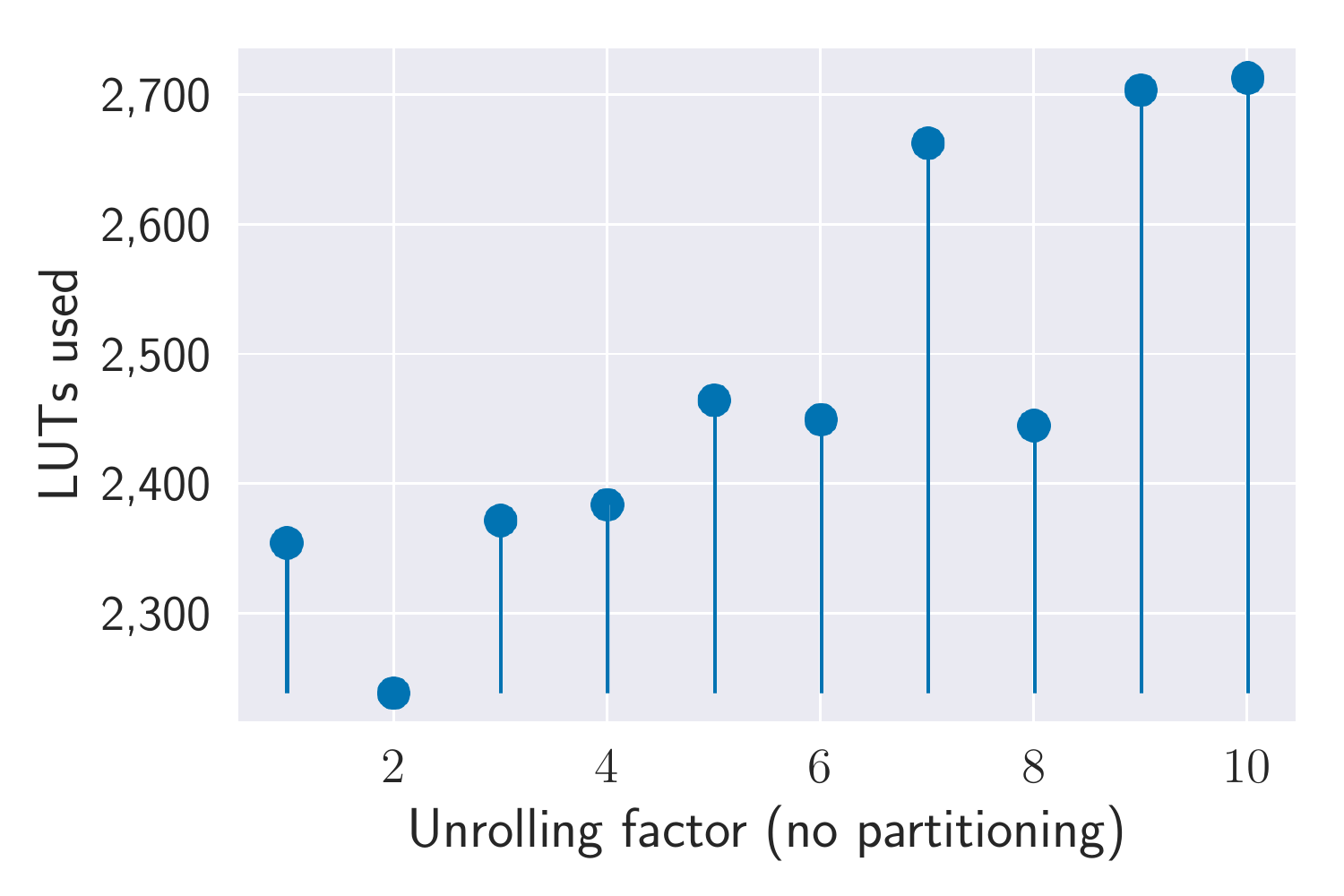}
    \includegraphics[width=\linewidth]{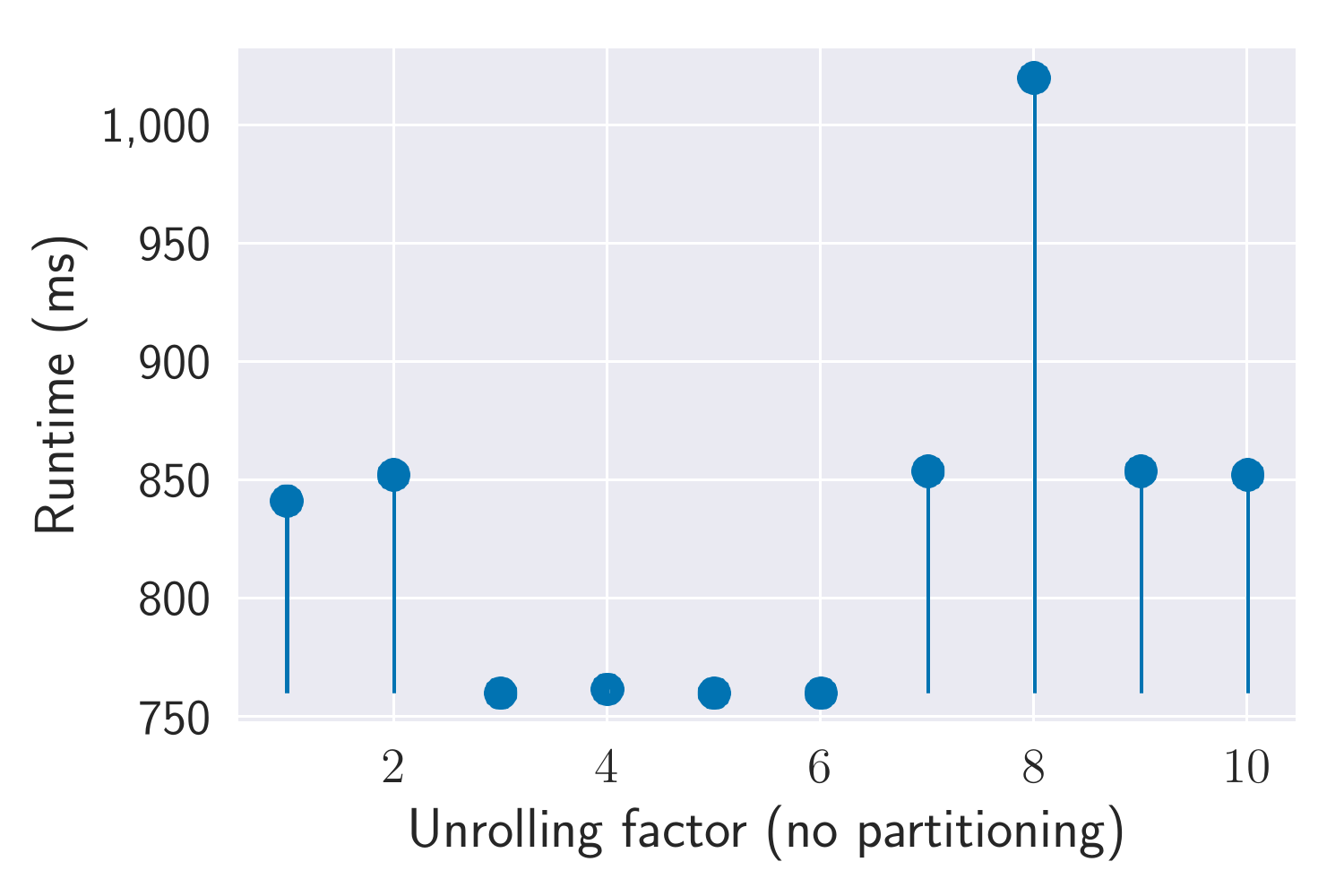}
    \caption{Unrolling without partitioning.}
    \label{fig:gemm-sensitivity:no-bank-unroll}
  \end{subfigure}
  \hspace{0.03\linewidth}
  \begin{subfigure}[b]{0.27\linewidth}
    \centering
    \includegraphics[width=\linewidth]{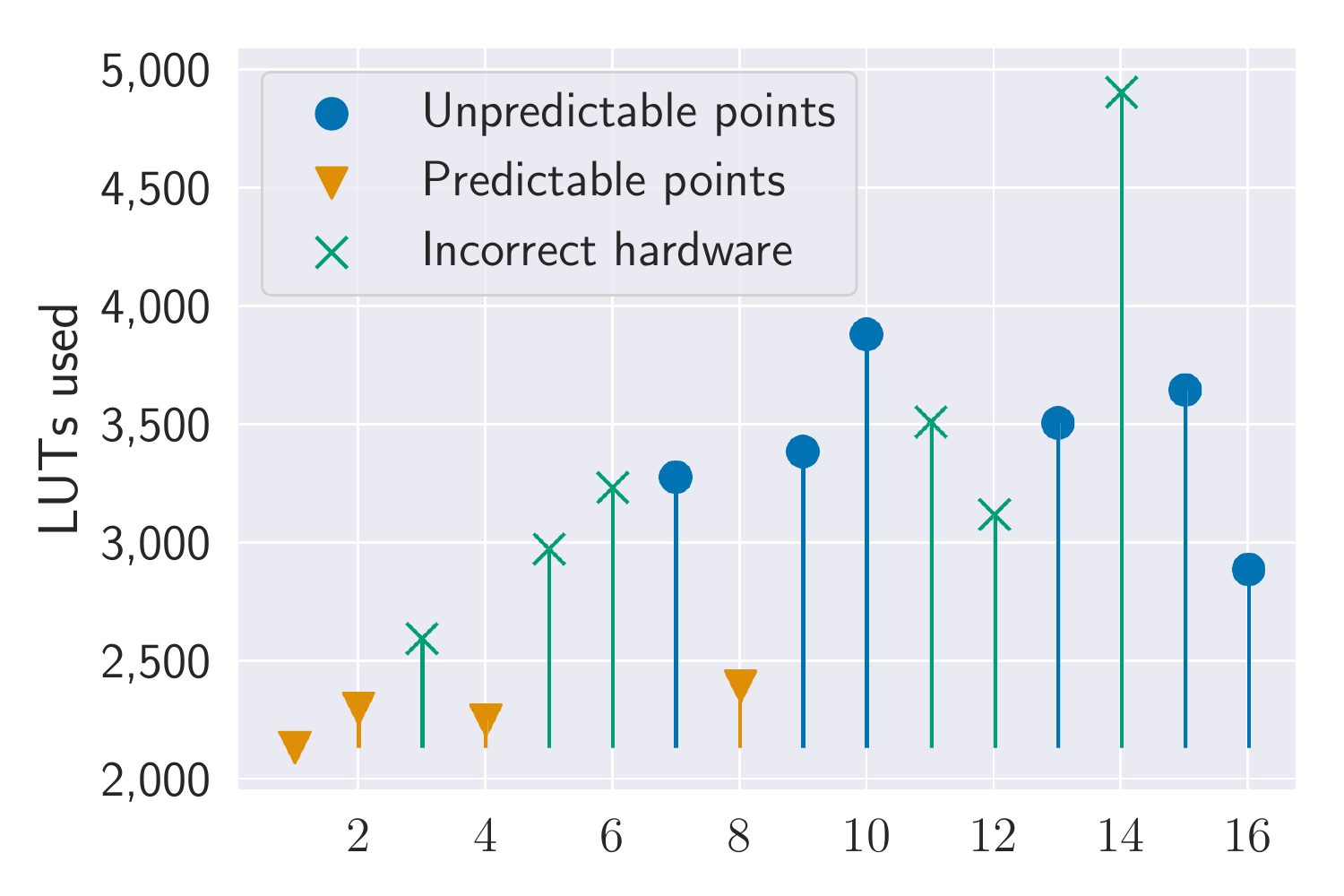}
    \includegraphics[width=\linewidth]{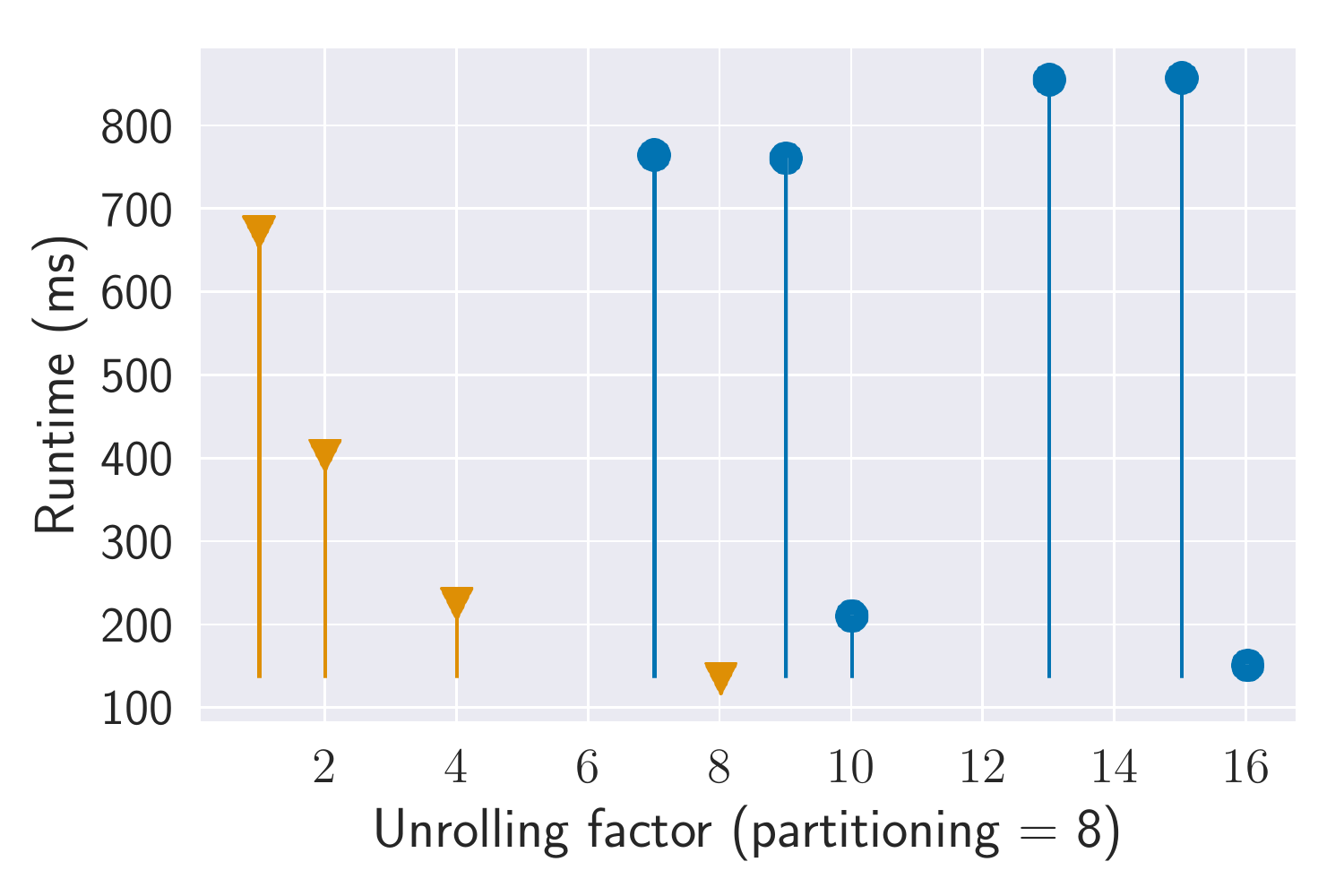}
    \caption{Unrolling with 8-way partitioning.}
    \label{fig:gemm-sensitivity:vary-unroll}
  \end{subfigure}
  \hspace{0.03\linewidth}
  \begin{subfigure}[b]{0.27\linewidth}
    \centering
    \includegraphics[width=\linewidth]{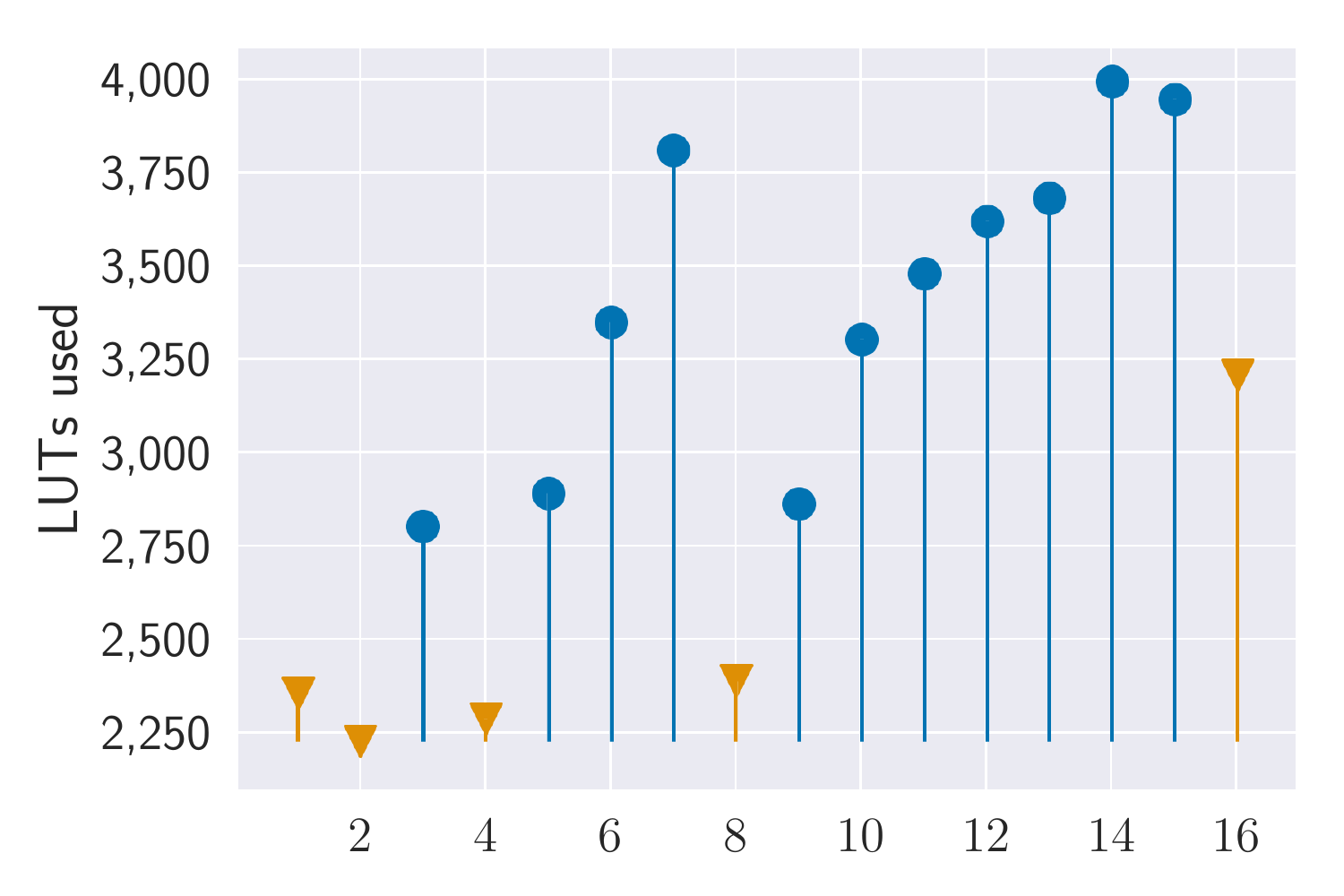}
    \includegraphics[width=\linewidth]{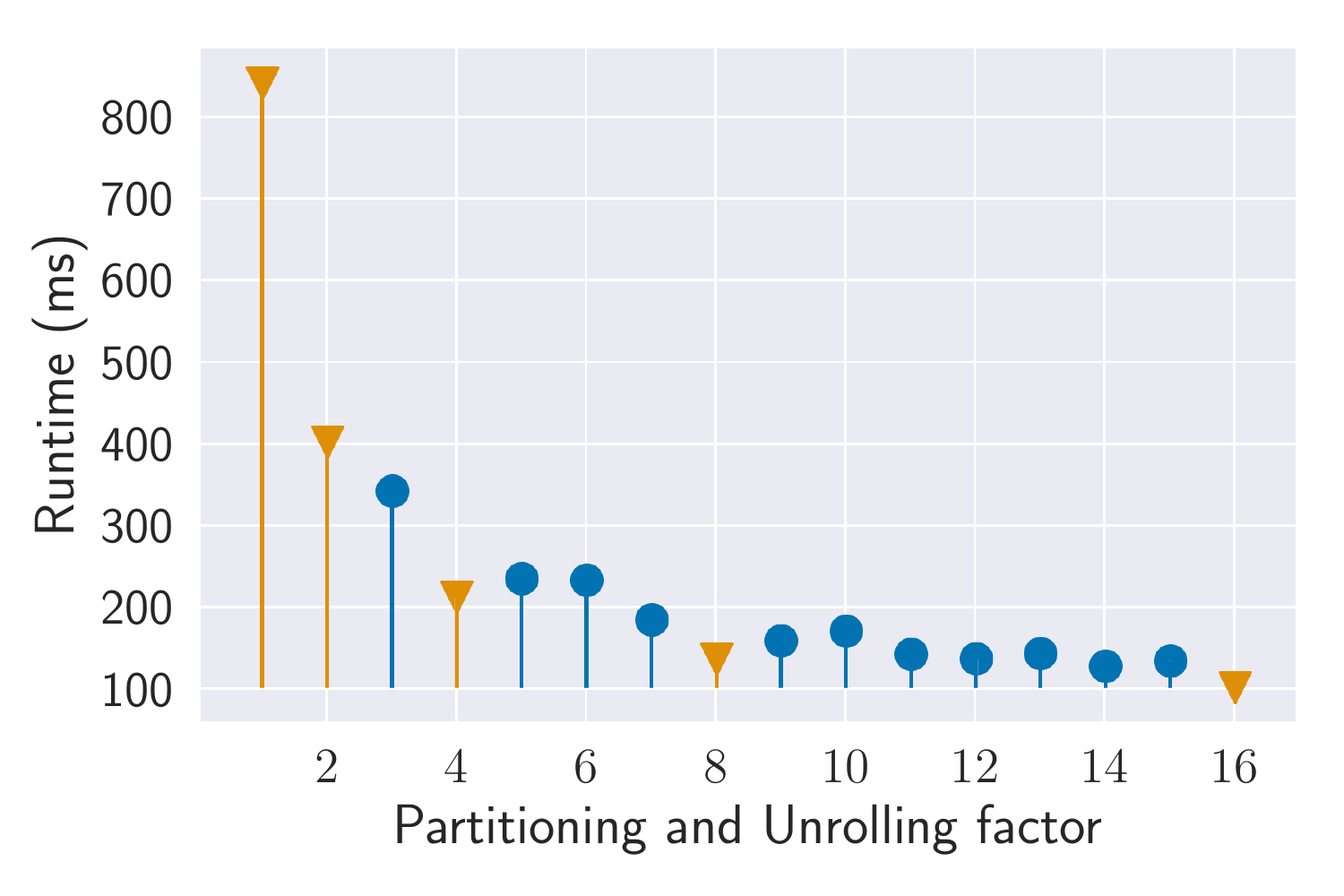}
    \caption{Unrolling and banking in lockstep.}
    \label{fig:gemm-sensitivity:vary-both}
  \end{subfigure}
  \caption{Look-up table count (top) and execution latency (bottom) for
  the kernel in \cref{fig:gemm-ncubed} with varying parameters.
   }
  \label{fig:gemm-sensitivity}
\end{figure*}
%

Programming with HLS centers on arrays and loops, which correspond to memory banks and logic blocks.
\cref{fig:gemm-ncubed} shows the C code for a matrix multiplication kernel.
This section imagines the journey of a programmer attempting to use HLS to
generate a fast FPGA-based accelerator from this code.
We use Xilinx's SDAccel~\cite{sdaccel} compiler (v2018.3.op) and target an UltraScale+ VU9P FGPA on an
AWS F1 instance~\cite{awsf1} to perform the experiments in this section.

\paragraph{Initial accelerator.}

Our imaginary programmer might first try compiling the code verbatim.
The HLS tool maps the arrays \code{m1}, \code{m2}, and
\code{prod} onto on-chip memories.
FPGAs have SRAM arrays, called \emph{block RAMs} (BRAMs), that
the compiler allocates for this purpose.
The loop body becomes combinational logic consisting of a multiplier, an adder, and an accumulator register.
\cref{fig:hls-example:orig}
depicts this configuration.

This design, while functional, does not harness any parallelism that an FPGA can offer.
The two key metrics for evaluating an accelerator design are performance and area, i.e., the amount of physical chip resources that the accelerator occupies.
This initial configuration computes the matrix product in 841.1 ms and occupies 2,355 of the device's lookup tables (LUTs).
However, the target FPGA device has over 1 million LUTs, so
the programmer's next job is to expend more of the FPGA area to improve performance.

\paragraph{Loop unrolling.}

The standard tool that HLS offers for expressing parallelism is an \code{UNROLL} annotation, which duplicates the logic for a loop body.
A programmer might attempt to obtain a better accelerator design by adding this annotation to the innermost loop on lines 6--8 in \cref{fig:gemm-ncubed}:
\begin{lstlisting}[language=C]
#pragma HLS UNROLL FACTOR=8
\end{lstlisting}
This unrolling directive instructs the HLS tool to create 8 copies of the multiplier and adder, called \emph{processing elements} (PEs), and attempt to run them in parallel.
Loop unrolling represents an area--performance trade-off: programmers can reasonably expect greater unrolling factors to consume more of the FPGA chip but yield lower-latency execution.

The \code{UNROLL} directive alone, however, fails to achieve this objective.
\Cref{fig:gemm-sensitivity:no-bank-unroll} shows the effect of various unrolling factors on this code in area (LUT count) and performance (latency).
There is no clear trend: greater unrolling yields unpredictably better and worse designs.
The problem is that the accelerator's memories now bottleneck the parallelism provided by the PEs.
The BRAMs in an FPGA have a fixed, small number of \emph{ports}, so they can only service one or two reads or writes at a time.
So while the HLS tool obeys the programmer's \code{UNROLL} request to duplicate PEs, its scheduling must serialize their execution.
\cref{fig:hls-example:mux} shows how the HLS tool must insert additional \emph{multiplexing} hardware to connect the multipliers to the single-ported memories.
The additional hardware and the lack of parallelism yields the unpredictable performance and area for different PE counts.

\paragraph{Memory banking to match parallelism.}
To achieve expected speedups from parallelism, accelerators need to use multiple memories.
HLS tools provide annotations to \emph{partition} arrays, allocating multiple BRAMs and increasing the access throughput.
The programmer can insert these partitioning annotations to allocate 8 BRAMs per input memory:
\begin{lstlisting}[language=C]
#pragma HLS ARRAY_PARTITION VARIABLE=m1 FACTOR=8
#pragma HLS ARRAY_PARTITION VARIABLE=m2 FACTOR=8
\end{lstlisting}
Banking uses several physical memories, each of
which stores a subset of the array's data. The compiler partitions the
array using a ``round-robin'' policy to enable parallel access. In this example, elements 0 and 8 go in bank 0, elements 1 and 9 go in bank 1, etc.:
\begin{center}
  \begin{minipage}{0.80\columnwidth}
    \includegraphics[%
    width=\linewidth]{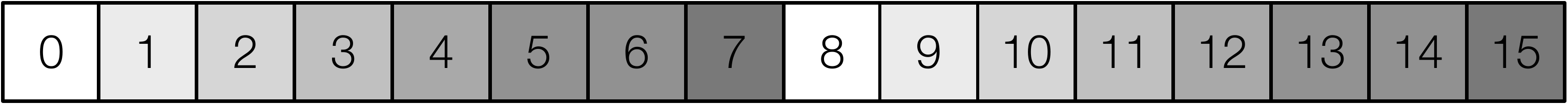}
  \end{minipage}
\end{center}
(Each shade represents a different memory bank.)
\cref{fig:hls-example:unrolled} shows the resulting architecture, which requires
no multiplexing and allows memory parallel access.

Combining banking and unrolling, however, unearths another source of unpredictable performance.
While the HLS tool produces a good result when both the banking factors and the loop unrolling factor are 8, other design choices perform worse.
\Cref{fig:gemm-sensitivity:vary-unroll} shows the effect of varying the
unrolling factor while keeping the arrays partitioned with factor 8.
Again, the area and performance varies unpredictably with the unrolling factor.
Reducing the unrolling factor from 9 to 8 can counter-intuitively \emph{improve} both performance and area.
In our experiments, some unrolling factors yield hardware that produces incorrect results. (We show the area but omit the running time for these configurations.)

The problem is that some partitioning/unrolling combinations yield much simpler hardware than others.
When both the unrolling and the banking factors are 8, each parallel PE need only access a single bank, as in \cref{fig:hls-example:unrolled}.
The first PE needs to access elements 0, 8, 16, and so on---and because the array elements are ``striped'' across the banks, all of these values live in the first bank.
With unrolling factor 9, however, the first PE needs to access values from \emph{every} bank, which requires complicated memory indirection hardware.
With unrolling factor 4, the indirection cost is smaller---the first PE needs to access only bank 0 and bank 4.

From the programmer's perspective, the HLS compiler silently enforces an unwritten rule:
\emph{When the unrolling factor divides the banking factor, the area is good and parallelism predictably improves performance. Otherwise, all bets are off.}
\Cref{fig:gemm-sensitivity:vary-unroll} labels the points where the unrolling factor divides the banking factor as \emph{predictable points}.
The HLS compiler emits no errors or warnings for any parameter setting.

\paragraph{Banking vs. array size.}

Even if we imagine that a programmer carefully ensures that banking factors exactly match unrolling factors, another pitfall awaits them when choosing the amount of parallelism.
\cref{fig:gemm-sensitivity:vary-both} shows the effects of
varying the banking and unrolling factor in our kernel \emph{together}.
The LUT count again varies wildly.

The problem is that, when the banking and unrolling factors do not evenly divide the sizes of the arrays involved, the accelerator needs extra hardware to cope with the ``leftover'' elements.
The memory banks are unevenly sized,
and the PEs need extra hardware to selectively disable themselves on the final iteration to avoid out-of-bounds memory accesses.

Again, there is a predictable subset of design points when the programmer obeys
the unwritten rule:
\emph{An array's banking factor should divide the array size.}
\cref{fig:gemm-sensitivity:vary-both} highlights the predictable points that follow this rule.
Among this subset, the performance reliably improves with increasing parallelism and the area cost scales proportionally.

\subsection{Enforcing the Unwritten Rules}

The underlying problem in each of these sources of unpredictability is that the traditional HLS tool prioritizes automation over programmer control.
While automation can seem convenient,
mapping heuristics give rise to implicit rules that, when violated, silently produce bad hardware instead of reporting a useful error.

This paper instead prioritizes the predictability of hardware generation and making architectural decisions obvious in the source code. HLS tools \emph{already} contain such a predictable subset hidden within their unrestricted input language. By modeling resource constraints, we can separate out this well-behaved fragment.
\cref{fig:overview} shows how our checker augments a traditional HLS toolchain by lifting hidden compiler reasoning into the source code and rejecting potentially unpredictable programs.

The challenge, however, is that the ``unwritten rules'' of HLS are never explicitly encoded anywhere---they arise implicitly from non-local interactions between program structure, hints, and heuristics.
%
A na\"ive syntactic enforcement strategy
would be too conservative---it would struggle to allow flexible, fine-grained sharing of hardware resources.

We design a type system that models the constraints of hardware implementation to enforce these constraints in a composable, formal way.
Our type system addresses \emph{target-independent} issues---it prevents problems that would occur even on an arbitrarily large FPGA.
We do not attempt to rule out resource exhaustion problems because they would tie programs to specific target devices.
We see that kind of quantitative resource reasoning as important future work.

%

\section{The \sys{} Language}
\label{sec:language}
\sys's type system enforces a safety property:
that the number of simultaneous reads and writes to a given memory bank may not exceed the number of ports.
While traditional HLS tools enforce this requirement with scheduling heuristics,
\sys enforces it at the source level using types.


The key ideas in \sys{} are
(1) using substructural typing to reason about consumable hardware resources
and (2) expressing time ordering in the language to reason about when resources are available.
This section describes these two core features (\cref{sec:memtypes,sec:spcomp})
and then shows how \sys{} builds on them to yield a language that is flexible enough to express real programs (\cref{sec:bank,sec:loops,sec:combine,sec:views}).

\subsection{Affine Memory Types}\label{sec:memtypes}

The foundation of \sys{}'s type system is its reasoning about memories.
The problem in \Cref{sec:principles:ex}'s example is conflicting simultaneous
accesses to the design's memories.
The number of reads and writes supported by a memory per cycle is limited by the
number of ports in the memory.
HLS tools automatically detect potential read/write conflicts and schedule
accesses across clock cycles to avoid errors.
\sys{} instead makes this reasoning about conflicts explicit by enforcing an
affine restriction on memories.

Memories are defined by giving their type and size:
\begin{lstlisting}
let A: float[10];
\end{lstlisting}
The type of \code{A} is \code|mem float[10]|, denoting a single-ported memory that holds 10
floating-point values.
Each \sys memory corresponds to an on-chip BRAM in the FPGA.
Memories resemble C or Java arrays: programs read and mutate the contents via subscripting, as in \code{A[5] := 4.2}.
Because they represent static physical resources in the generated hardware, memory types differ from plain value types like \code{float} by preventing duplication and aliasing:
\begin{lstlisting}
let x = A[0];  // OK: x is a float.
let B = A;     // Error: cannot copy memories.
\end{lstlisting}
The affine restriction on memories disallows reads and writes to a memory that might occur at the same time:
\begin{lstlisting}
let x = A[0];  // OK
A[1] := 1;     // Error: Previous read consumed A.
\end{lstlisting}
While type-checking \code{A}, the \sys compiler removes \code{A} from the typing context.
Subsequent uses of \code{A} are errors, with one exception:
identical reads to the same memory location are allowed.
This program is valid, for example:
\begin{lstlisting}
let x = A[0];
let y = A[0];  // OK: Reading the same address.
\end{lstlisting}
The type system uses access capabilities to check reads and writes~\cite{fluet:linear,gordon:rely}.
A read expression such as \code{A[0]} acquires a non-affine \emph{read capability} for index $0$ in the current scope, which permits unlimited reads to the same location but prevents the acquisition of other capabilities for \code{A}.
The generated hardware reads once from \code{A} and distributes the result to both variables \code{x} and \code{y}, as in this equivalent code:
\begin{lstlisting}
let tmp = A[0];  let x = tmp;  let y = tmp;
\end{lstlisting}
However, memory writes use affine \emph{write capabilities}, which are use-once resources: multiple simultaneous writes to the same memory location remain illegal.

\subsection{Ordered and Unordered Composition}
\label{sec:spcomp}

A key HLS optimization is parallelizing execution of independent code.
This optimization lets HLS compilers parallelize and reorder dependency-free statements connected by \code{;}
when the hardware constraints allow it---critically, when they do not need to access the same memory banks.

\sys makes these parallelism opportunities explicit by distinguishing between \emph{ordered} and \emph{unordered} composition.
The C-style \code{;} connector is unordered:
the compiler is free to reorder and parallelize the statements on either side while respecting their data dependencies.
A second connector,
\code{---}, is ordered: in
\code{A --- B}, statement \code{A} must execute before \code{B}.

\sys prevents resource conflicts in unordered composition but allows two statements in ordered composition to use the same resources.
For example, \sys accepts this program that would be illegal when joined by the \code{;} connector:
\begin{lstlisting}
let x = A[0]
---
A[1] := 1
\end{lstlisting}
In the type checker, ordered composition \emph{restores} the affine resources that
were consumed in the first command before checking the second command.
The capabilities for all memories are discarded, and the program can acquire fresh capabilities to read and write any memory.
%

Together, ordered and unordered composition can express complex concurrent designs:
\begin{lstlisting}
let A: float[10];  let B: float[10];
{
  let x = A[0] + 1
  ---
  B[1] := A[1] + x  // OK
};
let y = B[0];  // Error: B already consumed.
\end{lstlisting}
The statements composed with \code{---} are ordered with each other
but \emph{unordered} with the last line.
The read therefore must not conflict with either of the first two statements.

\paragraph{Logical time.}
From the programmer's perspective, a chain of ordered computations executes over a series of \emph{logical time steps}.
Logical time in \sys does not directly reflect physical time (i.e., clock cycles).
Instead, the HLS backend is responsible for allocating cycles to logical time steps in a way that preserves the ordering of memory accesses.
For example, a long logical time step containing an integer division might require multiple clock cycles to complete,
and the compiler may optimize away unneeded time steps that do not separate memory accesses.
Regardless of optimizations, however, a well-typed \sys program requires at least enough ordered composition to ensure that memory accesses do not conflict.

\paragraph{Local variables as wires \& registers.}
\label{sec:locals}

Local variables, defined using the \code{let} construct, do not share the affine restrictions of memories.
Programs can freely read and write to local variables without restriction, and
unordered composition respects the dependencies induced by local
variables:
\begin{lstlisting}
let x = 0;  x := x + 1;  let y = x;  // All OK
\end{lstlisting}
In hardware, local variables manifest as wires or registers.
The choice depends on the allocation of physical clock cycles:
values that persist across clock cycles require registers.
Consider this example consisting of two logical time steps:
\begin{lstlisting}
let x = A[0] + 1  ---  B[0] := A[1] + x
\end{lstlisting}
The compiler must implement the two logical time steps in different clock cycles,
so it must use a register to hold \code{x}.
In the absence of optimizations, registers appear whenever a variable's live range crosses a logical time step boundary.
Therefore, programmers can minimize the use of registers by reducing the live ranges of variables or by reducing the amount of sequential composition.

\subsection{Memory Banking}
\label{sec:bank}

As \Cref{sec:principles:ex} details, HLS tools can \emph{bank}
memories into disjoint components to allow parallel access.
\sys{} memory declarations support bank annotations:
\begin{lstlisting}
let A: float[8 bank 4];
\end{lstlisting}
%
In a memory type \code|mem $t$[$n$ bank $m$]|, the banking factor $m$ must
evenly divide the size $n$ to yield equally-sized banks.
HLS tools, in contrast, allow uneven banking and silently insert
additional hardware to account for it (see \Cref{sec:principles:ex}).

\paragraph{Affine restrictions for banks.}

\sys tracks an affine resource for each memory bank.
To physically address a bank, the syntax
\code|$M${$b$}[$i$]| denotes the $i$th element of $M$'s $b$th bank.
This program is legal, for example:
\begin{lstlisting}
let A: float[10 bank 2];
A{0}[0] := 1;
A{1}[0] := 2;  // OK: Accessing a different bank.
\end{lstlisting}
%
\sys also supports logical indexing into banked arrays using
the syntax \code|$M$[$n$]| for literals $n$.
For example, \code{A[1]} is equivalent to \code|A{1}[0]| above.
Because the index is static, the type checker can automatically deduce the bank and offset.

\paragraph{Multi-ported memories.} \sys also supports reasoning about
multi-ported memories.
This syntax declares a memory where each bank has two read/write ports:
\begin{lstlisting}
let A: float{2}[10];
\end{lstlisting}
A memory provides $k$ affine resources per bank
where $k$ is the number of ports in a memory.
This rule lets multi-ported memories provide multiple read/write capabilities
in each logical time step.
For example, \sys accepts this program:
\begin{lstlisting}
let A: float{2}[10];
let x = A[0];
A[1] := x + 1;
\end{lstlisting}
\sys does not guarantee data-race freedom in the presence of multi-ported memories.
Programs are free to write to and read from the same memory location in the
same logical time step and should expect the semantics of the
underlying memory technology.
Extensions to rule out data races would resemble race detection for parallel software~\cite{locksmith, naik:chord}.

\paragraph{Multi-dimensional banking.}

Banking generalizes to multi-dimensional arrays.
Every
dimension can have an independent banking factor.
This two-dimensional memory has two banks in each dimension, a total of $2\times2=4$ banks:

\vspace{0.5ex}
\noindent
\hspace{1em}
\begin{minipage}{0.63\columnwidth}
\begin{lstlisting}[xrightmargin=0cm, xleftmargin=0cm]
let M: float[4 bank 2][4 bank 2];
\end{lstlisting}
\end{minipage}%
\hfill
\begin{minipage}{0.2\columnwidth}
  \includegraphics[
    width=\linewidth]{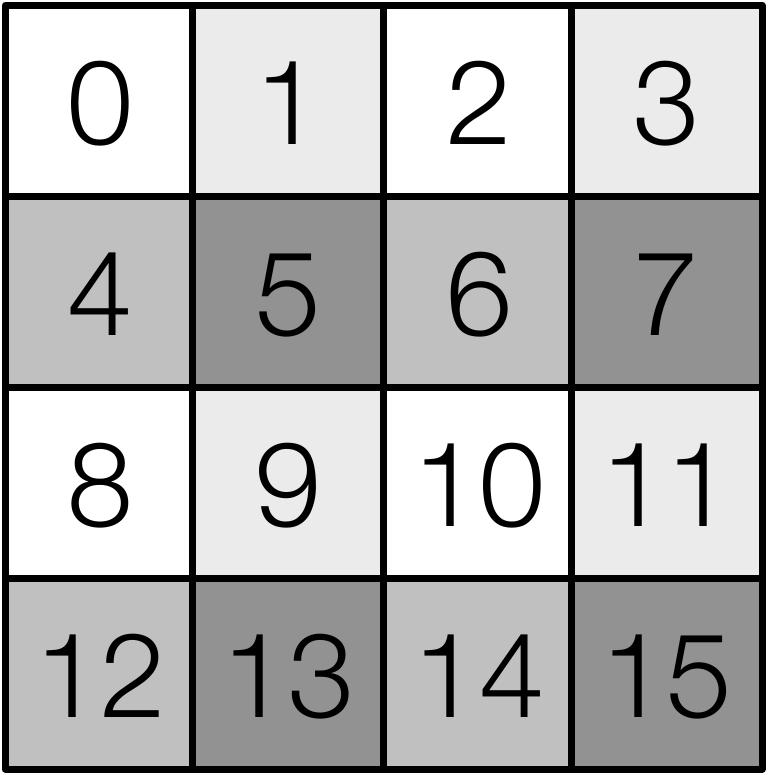}
\end{minipage}
\hspace{1em}
\vspace{1ex}

\noindent
The physical and logical memory access syntax similarly generalizes to multiple dimensions.
For example, \code|M{3}[0]| represents the element logically located at \code|M[1][1]|.

\subsection{Loops and Unrolling}\label{sec:loops}

Fine-grained parallelism is an essential optimization in hardware accelerator design.
Accelerator designers duplicate a block of logic to trade off area for performance: $n$ copies of the same logic consume $n$ times as much area while offering a theoretical $n$-way speedup.
\sys syntactically separates parallelizable \emph{doall} \code|for| loops, which must not have any cross-iteration dependencies, from sequential \code|while| loops, which may have dependencies but are not parallelizable.
Programmers can mark \code{for} loops with an \code{unroll} factor to duplicate the loop body logic and run it in parallel:
\begin{lstlisting}
for (let i = 0..10) unroll 2 { f(i) }
\end{lstlisting}
This loop is equivalent to a sequential one that iterates
half as many times and composes two copies of the body in parallel:
\begin{lstlisting}
for (let i = 0..5) {  f(2*i + 0);  f(2*i + 1)  }
\end{lstlisting}
The doall restriction is important because it allows the compiler
to run the two copies of the loop body in parallel using unordered
composition.
In traditional HLS tools, a loop unrolling annotation such as
\code{#pragma HLS unroll} is always allowed---even when the loop body makes
parallelization difficult or impossible. The toolchain will replicate the
loop body and rely on complex analysis and resource scheduling to optimize
the unrolled loop body as well as it can.

Resource conflicts in unrolled loops are errors.
For example, this loop accesses an unbanked array in parallel:
\begin{lstlisting}
let A: float[10];
for (let i = 0..10) unroll 2 {
  A[i] := compute(i)  // Error: Insufficient banks.
}
\end{lstlisting}
%

\paragraph{Unrolled memory accesses.}\label{par:unrolled-mems}
\sys uses special \emph{index types} for loop iterators to type-check memory accesses within unrolled loops.
Index types generalize integers to encode information about loop unrolling.
In this example:
\begin{lstlisting}
for (let i = 0..8) unroll 4 { A[i] }
\end{lstlisting}
The iterator \code{i} gets the type \code|idx{0..4}|, indicating that accessing an array at \code{i} will consume banks 0, 1, 2, and 3.
Type-checking a memory access with \code{i} consumes all banks indicated by its
index type.

\paragraph{Unrolling and ordered composition.}
Loop unrolling has a subtle interaction with ordered composition.
In a loop body containing \code{---}, like this:
\begin{lstlisting}
let A: float[10 bank 2];
for (let i = 0..10) unroll 2 {
  let x = A[i]
  ---
  f(x, A[0]) }
\end{lstlisting}
A naive interpretation would use parallel composition to join the loop bodies at the top level:
\begin{lstlisting}
for (let i = 0..5) {
  { let x0 = A[2*i]      ---  f(x0, A[0]) };
  { let x1 = A[2*i + 1]  ---  f(x1, A[0]) } }
\end{lstlisting}
However, this interpretation is too restrictive.
It requires \emph{all} time steps in each loop body to avoid conflicts with all other time steps.
This example would be illegal because the access to \code{A[i]} in the first time step may conflict with the access to \code{A[0]} in the second time step.
Instead, \sys reasons about unrolled loops
\emph{in lockstep} by parallelizing \emph{within} each logical time step.
The loop above is equivalent to:
\begin{lstlisting}
for (let i = 0..5) {
  { let x0 = A[2*i]; let x1 = A[2*i + 1] }
  ---
  { f(x0, A[0]);     f(x1, A[0]) } }
\end{lstlisting}
The lockstep semantics permits this unrolling because conflicts need only be avoided between unrolled copies of the same logical time step.
HLS tools must enforce a similar restriction but leave the choice to black-box heuristics.

\paragraph{Nested unrolling.}
In nested loops,
unrolled iterators
can separately access dimensions of a multi-dimensional array.
Nested loops also interact with \sys's read and write capabilities.
In this program:
\begin{lstlisting}
let A: float[8 bank 4][10 bank 5];
for (let i = 0..8) {
  for (let j = 0..10) unroll 5 {
    let x = A[i][0]
    ---
    A[i][0] := j;  // Error: Insufficient write
  } }              //        capabilities.
\end{lstlisting}
The read to array $A[i][0]$ can be proved to be safe because after desugaring, the reads turn into:
\begin{lstlisting}
let x0 = A[i][0];  let x1 = A[i][0] ...
\end{lstlisting}
The access is safe because the first access acquires a read capability for
indices \code{i} and \code{0}, so the subsequent copies are safe. Architecturally, the code entails a single read \emph{fanned out} to each parallel PE.
However, the write desugars to:
\begin{lstlisting}
A[i][0] := j;  A[i][0] := j + 1 ...
\end{lstlisting}
which causes a write conflict in the hardware.

\subsection{Combine Blocks for Reduction}
\label{sec:combine}

In traditional HLS, loops can freely include dependent operations,
as in this dot product:
\begin{lstlisting}
for (let i = 0..10) unroll 2 { dot += A[i] * B[i] }
\end{lstlisting}
However, the \code{+=} update silently introduces a dependency between every iteration which is disallowed by \sys's doall \code{for}-loops.
HLS tools heuristically analyze loops to extract and serialize dependent portions.
In \sys, programmers explicitly distinguish the non-parallelizable reduction components of \code{for} loops.
Each \code{for} can have an optional \code{combine} block that contains
sequential code to run after each unrolled iteration group of the main loop body.
For example, this loop is legal: 

\vspace{1.2ex}
\noindent
\begin{minipage}{0.45\columnwidth}
\begin{lstlisting}
for (let i = 0..10)
unroll 2 {
  let v = A[i] * B[i];
} combine {
  dot += v;
}
\end{lstlisting}
\end{minipage}%
\hfill
\begin{minipage}{0.45\columnwidth}
\vspace{-8px}
\includegraphics[width=1.5in]{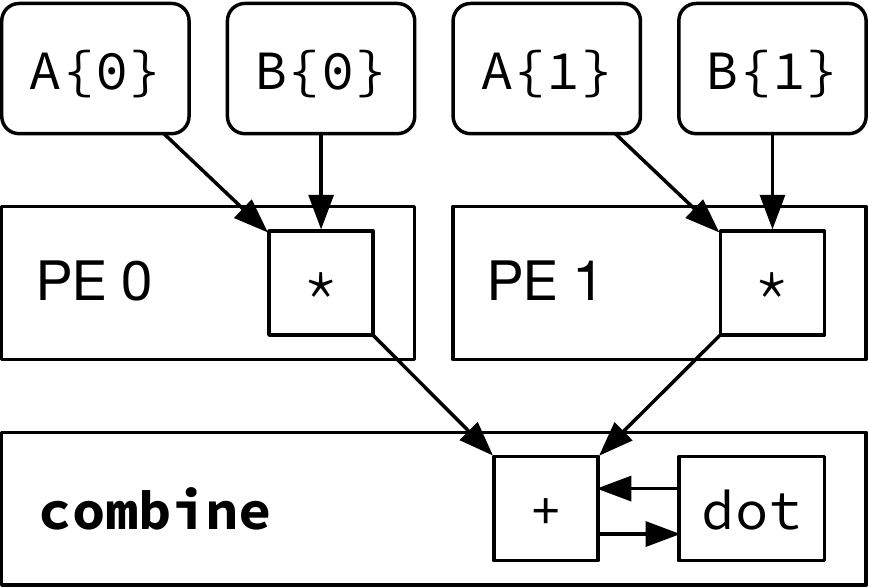}
\end{minipage}
\vspace{0.5ex}

\noindent
There are two copies of the loop body that run in parallel and feed into a single reduction tree for the \code{combine} block.

The type checker gives
special treatment to variables like \code{v} that are defined in \code{for} bodies and used in \code{combine} blocks.
In the context of the \code{combine} block, \code{v} is a \emph{combine register}, which is a tuple containing all values produced for \code{v} in the unrolled loop bodies.
\sys defines a class of functions called \emph{reducers} that
take a combine register and
return a single value (similar to a functional fold). \sys{} defines
\code{+=}, \code{-=}, \code{*=}, \code{/=} as built-in reducers with infix
syntax.


\subsection{Memory Views for Flexible Iteration}
\label{sec:views}

\begin{figure*}
  \centering
  \hspace{1em}
  \begin{subfigure}[b]{0.14\linewidth}
    \includegraphics[scale=0.374]{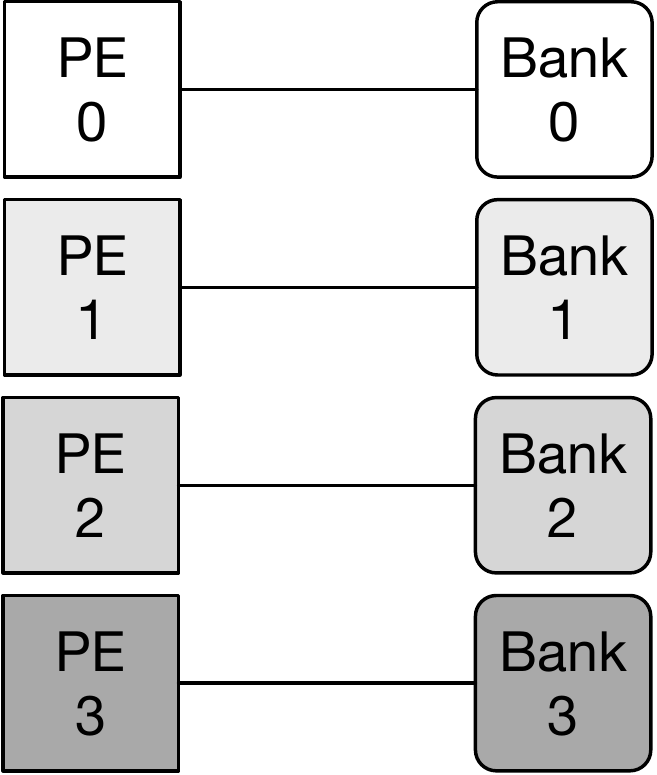}
    \caption{No view.}
    \label{fig:views:none}
  \end{subfigure}
  \hfill
  \begin{subfigure}[b]{0.14\linewidth}
    \includegraphics[scale=0.374]{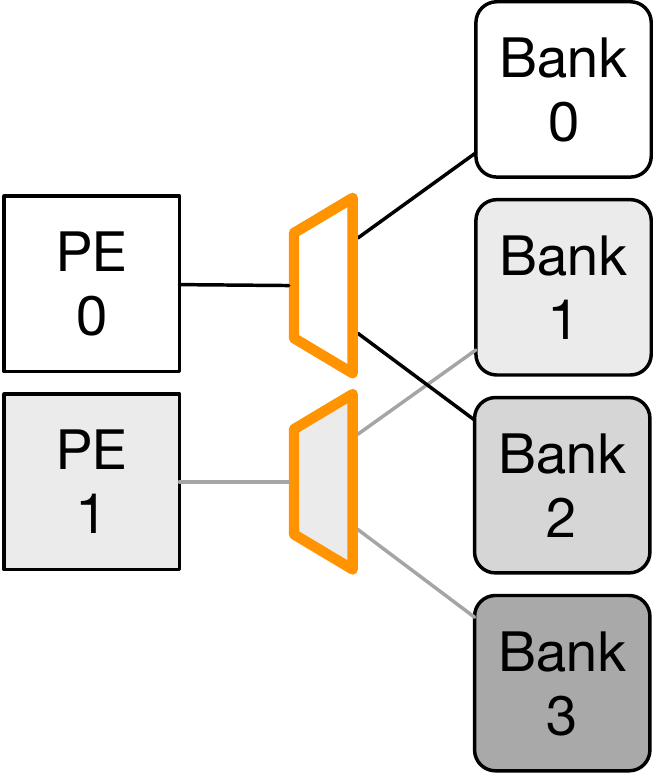}
    \caption{Shrink view.}
    \label{fig:views:shrink}
  \end{subfigure}
  \hfill
  \begin{subfigure}[b]{0.14\linewidth}
    \includegraphics[scale=0.374]{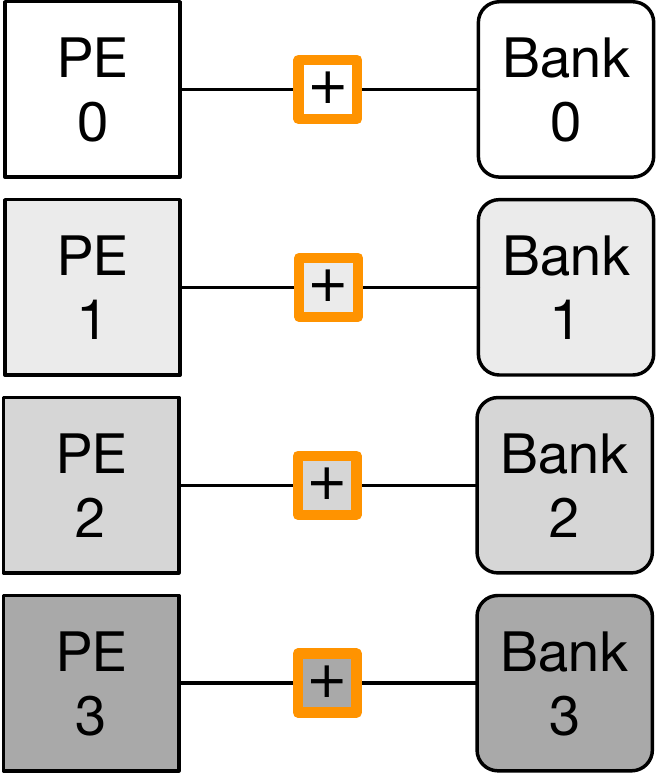}
    \caption{Suffix view.}
    \label{fig:views:suffix}
  \end{subfigure}
  \hfill
  \begin{subfigure}[b]{0.20\linewidth}
    \includegraphics[scale=0.374]{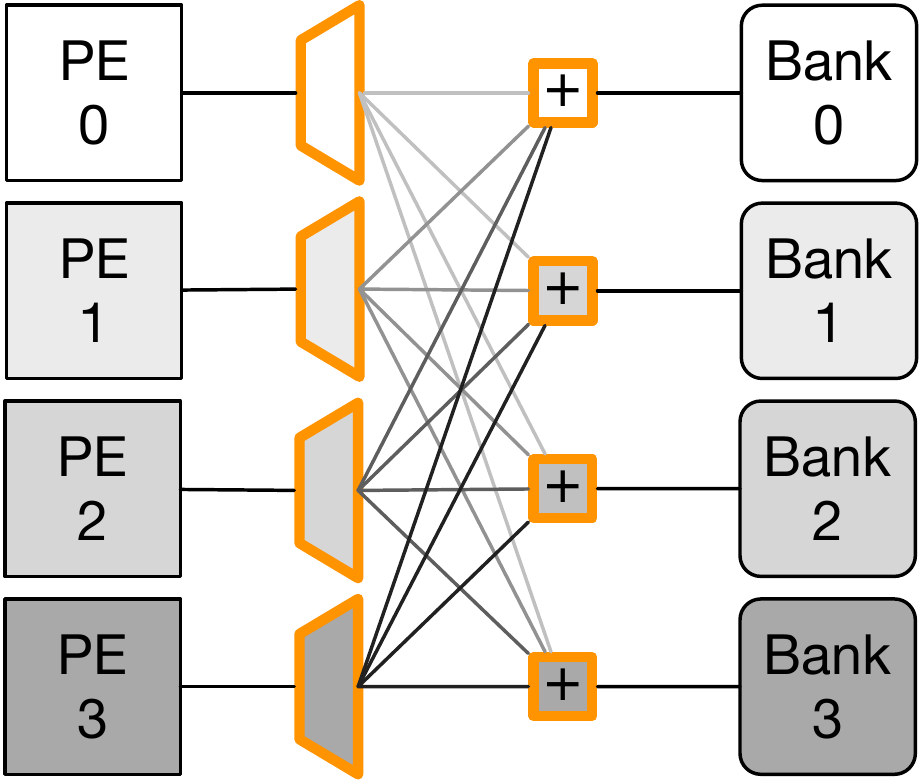}
    \caption{Shift view.}
    \label{fig:views:shift}
  \end{subfigure}
  \hfill
  \begin{subfigure}[b]{0.14\linewidth}
    \includegraphics[scale=0.374]{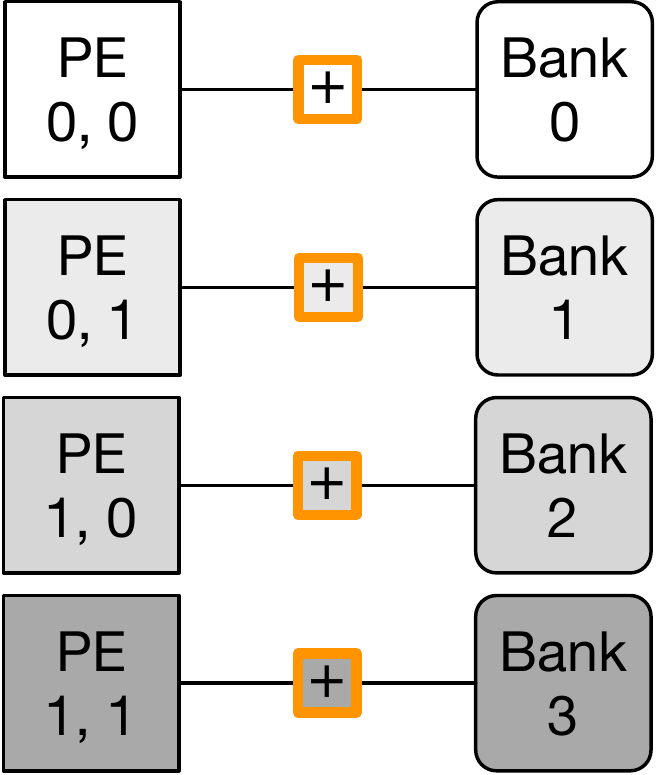}
    \caption{Split view.}
    \label{fig:views:split}
  \end{subfigure}
  \hspace{1em}
  \caption{Hardware schematics for each kind of memory view.
  Highlighted outlines indicate added hardware cost.}
  \label{fig:views}
\end{figure*}

In order to predictably generate hardware for parallel accesses, \sys statically
calculates banks accessed by each PE and guarantees that they are distinct. \cref{fig:views:none} shows the kind of hardware generated by this restriction---each PE is directly connected to a bank.

To enforce this hardware generation, \sys only allows simple indexing expressions like \code{A[i]} and \code{A[4]} and rejects arbitrary index calculations like \code{A[2*i]}.
General indexing expressions can require complex indirection hardware to allow any PE to access any memory bank. An access like \code{A[i*i]}, for example, makes it difficult to deduce which bank it would read on which iteration. For simple expressions like \code{A[j+8]}, however, the bank stride pattern is clear. Traditional HLS tools make a best-effort attempt to deduce access patterns, but subtle changes in the code can unpredictable prevent the analysis and generate bad hardware.

\sys uses \emph{memory views} to define access patterns that HLS compilers can
compile efficiently and to convince the \sys type checker that a parallel
access will be predictable.
The key idea is to offer different \emph{logical} arrangements of the same
underlying physical memory. By logically re-organizing the memory, views can
simply reuse \sys's type-checking to ensure that complex access patterns are
predictable. Furthermore, this allows views to capture the hardware cost of
an access pattern in the source code instead of relying on black-box analysis
in HLS tools. For \sys's HLS \cxx backend, views are compiled to direct memory
accesses.

The rest of this section
describes \sys's memory views and their cost in terms of hardware
required to transform bank and index values to support the iteration pattern.

\paragraph{Shrink.}
To directly connect PEs to memory banks, \sys requires the unrolling factor to match the banking factor. To allow lower unrolling factors, \sys provides \emph{shrink views}, which reduce the banking factors of an underlying memory by an integer factor.
For example:
\begin{lstlisting}
let A: float[8 bank 4];
view sh = shrink A[by 2]; // sh: float[8 bank 2]
for (let i = 0..8) unroll 2
  sh[i]; // OK: sh has 2 banks. Compiled to: A[i].
\end{lstlisting}
The example first defines a view \code|sh| with the underlying memory \code{A}
and divides its banking factor by 2.
\sys allows \code{sh[i]} here because each PE will access a distinct set of banks. The first PE accesses banks 0 and 2; the second accesses banks 1 and 3. The hardware cost of a shrink view, as \cref{fig:views:shrink} illustrates, consists of multiplexing to select the right bank on every iteration.
The access \code{sh[i]} compiles to \code{A[i]}.

\paragraph{Suffix.}

A second kind of view lets programs create small slices of a larger memory.
\sys distinguishes between suffixes that it can implement efficiently and costlier ones.
An efficient \emph{aligned suffix} view uses this syntax:
\begin{lstlisting}
view v = suffix M[by $k$ * $e$];
\end{lstlisting}
where view \code{v} starts at element $k \times e$ of the memory \code{M}.
Critically, $k$ must be the banking factor of \code{M}. This restriction
allows \sys to prove that each logical bank in the view maps to the same
physical bank while the indices are offset by the indexing expression. The
hardware cost of a suffix view is the \emph{address adapter} for each bank. A view access \code|v{$b$}[$i$]| is compiled to \code|M{$b$}[$e + i$]|.

For example, generating suffixes in a loop results in this pattern, where the
digits in each cell are the indices, the shades represent the banks, and the
highlighted outline indicates the view:
\vspace{1.5ex}

\noindent
\begin{minipage}{0.58\columnwidth}
\begin{lstlisting}
let A: float[8 bank 2];
for (let i = 0..4) {
  view s = suffix A[by 2*i];
  s[1]; // reads A[2*i + 1]
}
\end{lstlisting}
\end{minipage} %
\hfill
\begin{minipage}{0.35\columnwidth}
\vspace{-7pt}
\includegraphics[width=\linewidth]{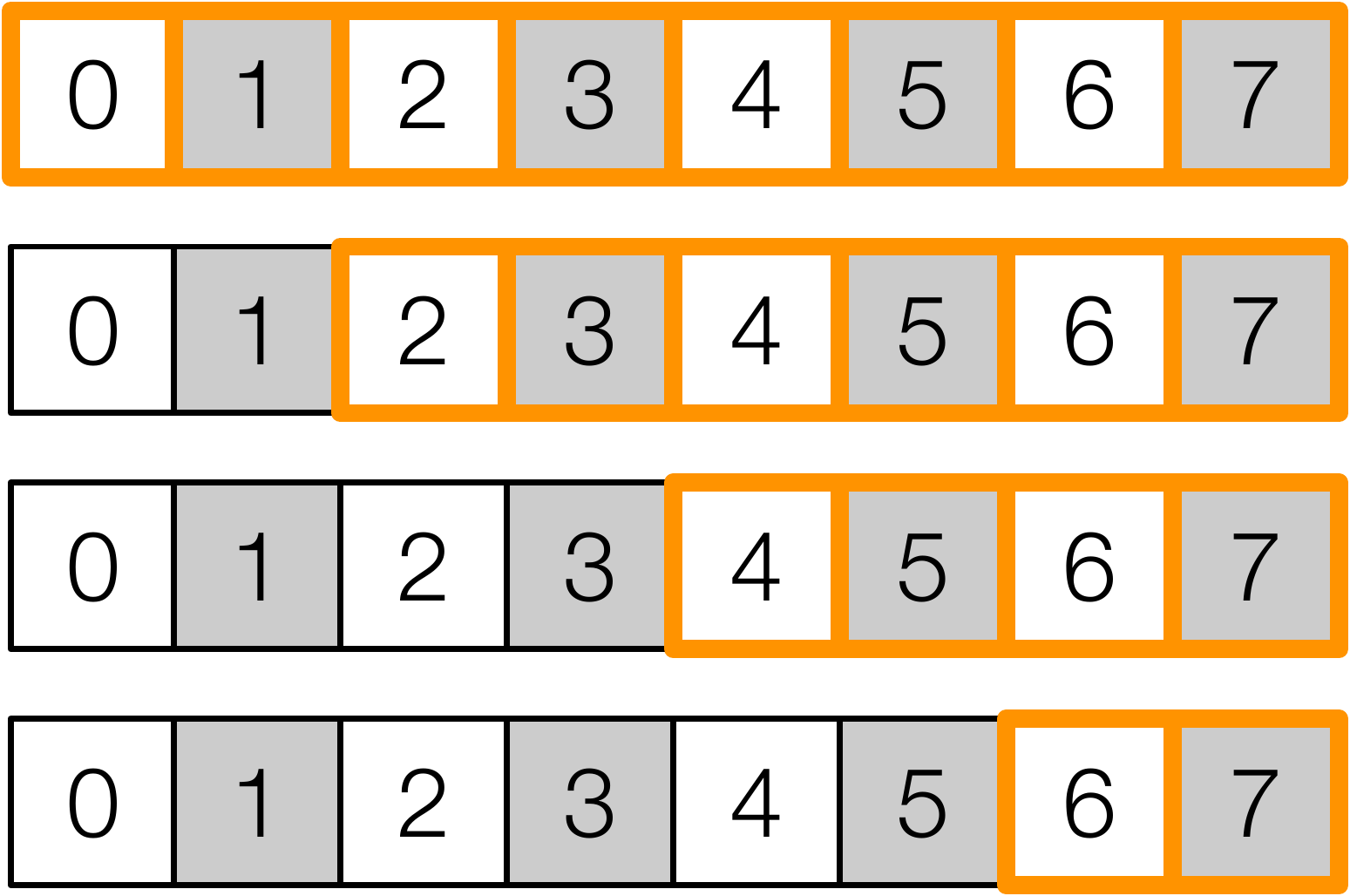}
\end{minipage}
\vspace{0.5ex}

\noindent
A suffix
view defined using \code|view v = suffix M[by k*e]| and accessed using
\code{v[i]} is compiled to \code|M[k*e + i]|.

\paragraph{Shift.}
Shifted suffixes are like standard suffixes but allow unrestricted offset
expressions:
\begin{lstlisting}
view v = shift M[by e];
\end{lstlisting}
Since \code{e} is unrestricted, \sys assumes that \emph{both} the bank and
the indices need to be adapted and that each PE accesses \emph{every} bank.
\Cref{fig:views:shift} shows the hardware cost of a shift view: each PE is
connected to every bank and the index expression is transformed using an
address adapter. The distinction between suffix and shift views allows
\sys to capture the cost of different accessing schemes.

Even in this worst-case scenario, \sys can reason about the disjointness of
bank accesses. This loop is legal:
\begin{lstlisting}
let A: float[12 bank 4];
for (let i = 0..3) {
  view r = shift A[by i*i];  // r: float[12 bank 4]
  for (let j = 0..4) unroll 4
    let x = r[j];  // accesses A[i*i + j]
}
\end{lstlisting}
The view \code|r| has a memory type, so \sys can guarantee that the inner access \code|r[j]| uses disjoint banks and is therefore safe to parallelize.
An access \code|r[$i$]| to a view declared with \code|shift M[by $e$]| compiles to \code|M[$e + i$]|.

\paragraph{Split.}
Some nested iteration patterns can be parallelized at two levels: globally,
over an entire array, and locally, over a smaller window.
This pattern arises in blocked computations, such as this dot product loop in \cxx:
\begin{lstlisting}
float A[12], B[12], sum = 0.0;
for (int i = 0; i < 6; i++)
  for (int j = 0; j < 2; j++)
    sum += A[2*i + j] * B[2*i + j];
\end{lstlisting}
Both the inner loop and the outer loop represent opportunities for
parallelization. However, \sys{} cannot prove this parallelization to be
safe:
\begin{lstlisting}
let A, B: float[12 bank 4];
view shA, shB = shrink A[by 2], B[by 2];
for (let i = 0..6) unroll 2 {
  view vA, vB = suffix shA[by 2*i], shB[by 2*i];
  for (let j = 0..2) unroll 2 {
    let v = vA[j] + vB[j];
  } combine {
    sum += v; }}
\end{lstlisting}
While \sys{} can prove that the inner accesses into the views can be
predictably parallelized, it cannot establish the disjointness of the parallel
copies of the views \code{va} and \code{vb} created by the outer unrolled loop.

\emph{Split views} allow for this reasoning. The key idea is to create logically \emph{more} dimensions than the physical memory and reusing \sys's reasoning for multidimensional memories to prove safety for such parallel accesses.
A \emph{split view} transforms a one-dimensional memory (left) into a two-dimensional memory (right):
\begin{center}
  \includegraphics[width=\columnwidth]{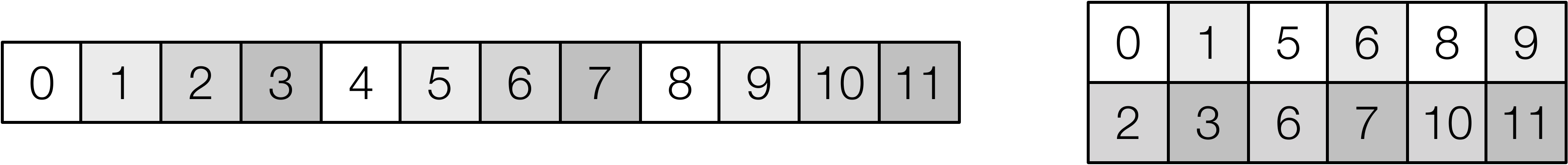}
\end{center}
Using these split-view declarations:
\begin{lstlisting}
view split_A = split A[by 2];
view split_B = split B[by 2];
\end{lstlisting}
Each view has type \code{mem float[2 bank 2][6 bank 2]}.
A row in the logical view represents a ``window'' for computation.
The above example can now unroll both loops, by changing the inner access to:
\begin{lstlisting}
let v = split_A[j][i] * split_B[j][i];
\end{lstlisting}
As \cref{fig:views:split} illustrates, split views have similar cost to aligned suffix views:
they require no bank indirection hardware because the bank index is always known statically.
They require an address adapter to compute the address within the bank from the separate coordinates. A split view declared \code|view sp = split M[by $k$]| on a memory
\code{M} with $k$ banks translates the access
\code{sp[$i$][$j$]} to
\code|M{$\text{\rm bank}$}[$\text{\rm idx}$]| where:
\[
  \text{bank} = i * k + (j \bmod b) \quad
  \text{idx} = \left\lfloor\dfrac{j}{b}\right\rfloor
\]
%






\section{Formalism}
\label{sec:formalism}

  \input{semantics}

  \renewottcommands[ott]

\begin{figure}
  \small
  \begin{align*}
    x &\in \text{variables} \enspace\enspace
    a \in \text{memories} \enspace\enspace
    n \in \text{numbers} \\
    b &::= \ottkw{true} \mid \ottkw{false} \qquad
    v ::= n \mid b \\
    e &::= v \mid
        \ottkw{bop} \; e_1 \; e_2 \mid
        x \mid
        a[e] \\
    c &::= e \mid
        \ottkw{let} \; x = e \mid
        c_1 \seqcomp c_2 \mid
        c_1 ~\textbf{;}~ c_2 \mid
        \ottkw{if} \; x \; c_1 \; c_2 \mid \\
    &\qquad \ottkw{while} \; x \; c \mid
        x := e \mid
        a[e_1] := e_2 \mid
        \ottkw{skip} \\
    \tau &::= \ottkw{bit}\langle n \rangle \mid
        \ottkw{float} \mid
        \ottkw{bool} \mid
        \ottkw{mem} \; \tau[n_1]
    \end{align*}
\caption{Abstract syntax for the \corelang core language.}
\label{fig:syntax}
\end{figure}

This section formalizes the time-sensitive affine type system that underlies \sys in a core language, \corelang.
We give both a large-step semantics, which is more intelligible, and a small-step semantics, which enables a soundness proof.

\subsection{Syntax}

\Cref{fig:syntax} lists the grammar for \corelang.
\corelang statements $c$ resemble a typical imperative language: there are
expressions, variable declarations, conditions, and simple sequential
iteration via \code{while}.
\corelang has
ordered composition $c_1 ~\rule[0.5ex]{1.5em}{0.55pt}~ c_2$
and unordered composition $c_1 ~\textbf{;}~ c_2$.
It separates memories $a$
and variables $x$ into separate syntactic categories.
\corelang programs can only declare the latter: a program runs with a
fixed set of available memories.

\subsection{Large-Step Semantics}

\corelang's large-step operational semantics is a \emph{checked semantics} that
enforces \sys's safety condition by explicitly tracking and getting stuck when it would otherwise require two conflicting accesses. Our type system (\Cref{sec:types}) aims to rule out these conflicts.

The semantics uses an environment $\sto$ mapping variable and memory names to values, which may be primitive values or memories, which in turn map indices to primitive values.
A second context, $\loc$, is the set of the
memories that the program has accessed.
$\loc$ starts empty and
accumulates memories as the program reads and writes them.

The operational semantics
consists of an expression judgment $\sigma_1, \rho_1, e \Downarrow \sigma_2, \rho_2, v$
and a command judgment $\sigma_1, \rho_1, c \Downarrow \sigma_2, \rho_2$.
We describe some relevant rules here, and the supplementary material lists the full semantics and proof~\cite{techreport}.

\paragraph{Memory accesses.}

Memories in \corelang are mutable stores of values. Banked memories in \sys can
be built up using these simpler memories.
The rule for a memory read expression $a[n]$ requires that $a$ not already be present in $\loc$,
which would indicate that the memory was previously consumed:
\begin{mathpar}
  \small
  \ottdrulelargeXXread{}
\end{mathpar}

\paragraph{Composition.}

Unordered composition accumulates the resource demands of two commands by
threading $\loc$ through:
\begin{mathpar}
  \small
  \ottdrulelargeXXpar{}
\end{mathpar}
If both commands read or write the same memory, they will conflict in $\loc$.
%
%
Ordered composition runs each command in the same initial $\loc$ environment
and merges the resulting $\loc$:
%
\begin{mathpar}
  \small
  \ottdrulelargeXXseq{}
\end{mathpar}

\subsection{Type System}
\label{sec:types}

The typing judgments have the form \defnccheckshape{}
and \defnecheckshape{}. $\gam$ is a standard typing context for variables and $\tloc$ is the affine context for memories.

\paragraph{Affine memory accesses.}

Memories are affine resources.
The rules for reads and writes check the type of the index in $\gam$ and remove the
memory from $\tloc$:
\begin{mathpar}
  \small
  \ottdrulecheckXXread{}
\end{mathpar}
\paragraph{Composition.}
%
The unordered composition rule checks
the first statement in the initial contexts and uses the resulting contexts
to check the second statement:
\begin{mathpar}
  \small
  \ottdrulecheckXXparXXcomp{}
\end{mathpar}
%
%
Ordered composition checks both commands under the same
resource set, $\Delta_1$, but threads the non-affine context through:
%
\begin{mathpar}
  \small
  \ottdrulecheckXXseqXXcomp{}
\end{mathpar}
The rule merges the resulting $\tloc$ contexts with set intersection
to yield the resources not consumed by either statement.




\subsection{Small-Step Semantics}
\label{sec:smallstep}

We also define a small-step operational semantics for \corelang upon which we
build a proof of soundness.
We claim that the small-step semantics, when iterated to a value, is equivalent to the big-step semantics.
The semantics consists of judgments \defnereduceshapesmall{} and \defncreduceshapesmall{} where
$\sigma$ and $\rho$ are the environment and the memory context respectively.
The main challenge is sequential composition,
which uses an intermediate
command form $\smash{c_1 \overset{\rho}{\sim} c_2}$ to thread $\rho$ to $c_1$ and $c_2$.
The supplementary material has full details.

\subsection{Desugaring Surface Constructs}

\corelang desugars surface language features present in \sys{}.

\paragraph{Memory banking.}

A banked memory declaration like this:
\begin{lstlisting}
let A: float[$m$ bank $n$];
\end{lstlisting}
desugars into several unbanked memories:
\begin{lstlisting}
let A_0: float[$\frac{m}{n}$];  let A_1: float[$\frac{m}{n}$]; ...
\end{lstlisting}
Desugaring transforms reads and writes of banked memories to conditional
statements that use the indexing expression to decide
which bank to access.

\paragraph{Loop unrolling.}

Desugaring of \code{for} loops uses the technique described in
\Cref{sec:loops}, translating from:
\begin{lstlisting}
for (let i = 0 .. $m$) unroll $k$ {  $c_1$ --- $c_2$ ...  }
\end{lstlisting}
into a \code{while} loop that duplicates the body:
\begin{lstlisting}
let i = 0;
while (i < $\frac{m}{k}$) {
  { $c_1[$i$\;\mapsto\;$k*i+0$]$; $c_1[$i$\;\mapsto\;$k*i+1$]$ ... }
  ---
  { $c_2[$i$\;\mapsto\;$k*i+0$]$; $c_2[$i$\;\mapsto\;$k*i+1$]$ ... }
  ...
  i++;  }
\end{lstlisting}
where $c[x \mapsto e]$ denotes substitution.

\paragraph{Memory views.}

For views' operational semantics, a desugaring based on the mathematical descriptions in \cref{sec:views} suffices.
To type-check them, however, would require tracking the underlying memory for each view (transitively, to cope with views of views) and type-level reasoning about the bank requirements of an access pattern.
Formal treatment of these types would require an extension to \corelang.

\paragraph{Multi-ported memories.}

Reasoning about memory ports requires quantitative resource tracking, as in bounded linear logic~\cite{girard:bll}.
We leave such an extension of \corelang's affine type system as future work.

\subsection{Soundness Theorem}
\label{sec:soundness}

We state a soundness theorem for \corelang's type system with respect to its checked small-step operational semantics.
\begin{theorem*}
If
$\emptyset, \tloc^* \vdash c \dashv \gam_2, \tloc_2$ and
$\emptyset, \emptyset, c \stackrel{*}{\rightarrow} \sto, \loc, c'$ and $\sigma, \rho, c' \centernot\rightarrow$, then $c' = \text{\bf{skip}}$.
\end{theorem*}
\noindent
where $\tloc^*$ is the initial affine context of memories available to a program.
The theorem states that a well-typed program never gets stuck
due to memory conflicts in $\loc$.
We prove this theorem using progress and preservation lemmas: 
\begin{lemma}[Progress]
If
$\Gamma, \Delta \vdash c \dashv \Gamma_2, \Delta_2$ and $\Gamma, \Delta \sim \sigma, \rho$, then
  $\sigma, \rho, c \rightarrow \sigma', \rho', c'$ or $c = \text{\bf{skip}}$.
\end{lemma}

\begin{lemma}[Preservation]
If
$\Gamma, \Delta \vdash c \dashv \Gamma_2, \Delta_2$ and $\Gamma, \Delta \sim \sigma, \rho$, and
$\sigma, \rho, c \rightarrow \sigma', \rho', c'$, then $\Gamma', \Delta' \vdash c' \dashv \Gamma_2', \Delta_2'$
and $\Gamma', \Delta' \sim \sigma', \rho'$.
\end{lemma}

\noindent
In these lemmas, $\Gamma, \Delta \sim \sigma, \rho$ is a well-formedness judgment stating that all variables in $\Gamma$ are in $\sigma$ and
all memories in $\Delta$ are not in $\rho$.
Using an extension of the syntax in \Cref{fig:syntax}, we prove
the lemmas by induction on the small-step relation~\cite{techreport}.

\section{Evaluation}
\label{sec:evaluation}

Our evaluation measures whether \sys's restrictions can improve predictability without sacrificing too much sheer performance.
We conduct two experiments:
(1) We perform an exhaustive design space
exploration for one kernel to determine how well the restricted design points compare to the much larger unrestricted parameter space.
(2) We port the MachSuite benchmarks~\cite{machsuite} and, where \sys yields a meaningful design space, perform a parameter sweep.





\subsection{Implementation and Experimental Setup}

We implemented a \sys compiler in \sysloc{} of Scala.
The compiler checks \sys{} programs and generates \cxx code using Xilinx Vivado HLS's
\code{#pragma} directives~\cite{vivadohls}.
We execute benchmarks on AWS F1 instances~\cite{awsf1} with 8~vCPUs, 122~GB of main
memory, and a Xilinx UltraScale+ VU9P.
We use the SDAccel development environment~\cite{sdaccel} and synthesize
the benchmarks with a target clock period of 250~MHz.
%

\subsection{Case Study: Unrestricted DSE vs.\ \sys}\label{sec:eval:dse}

\begin{figure*}
  \centering
  \begin{subfigure}[b]{0.32\linewidth}
    \centering
    \includegraphics[width=\linewidth]{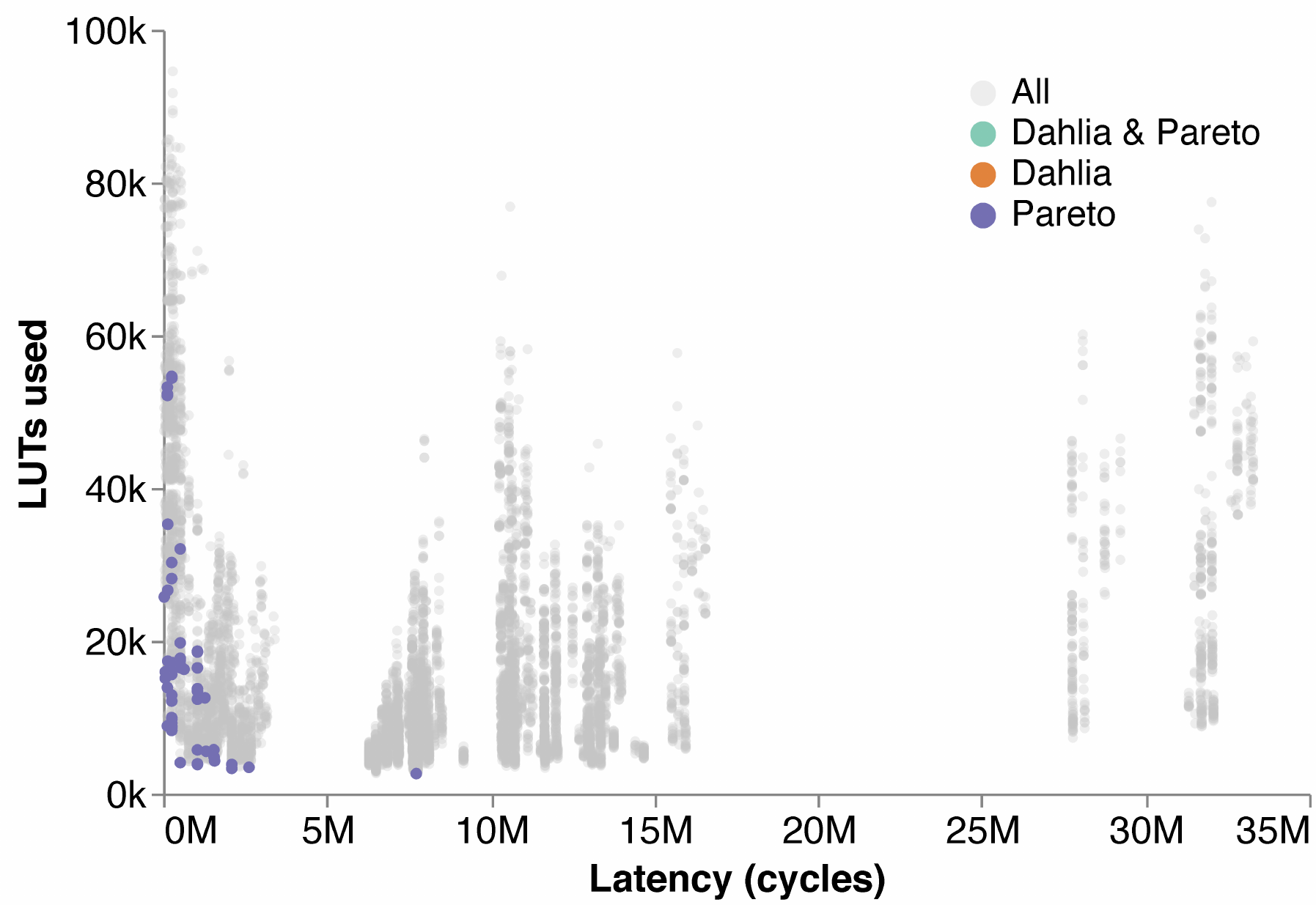}
    \caption{Pareto-optimal points.}
    \label{fig:gemm-dse:only-pareto}
  \end{subfigure}
  \hfill
  \begin{subfigure}[b]{0.32\linewidth}
    \centering
    \includegraphics[width=\linewidth]{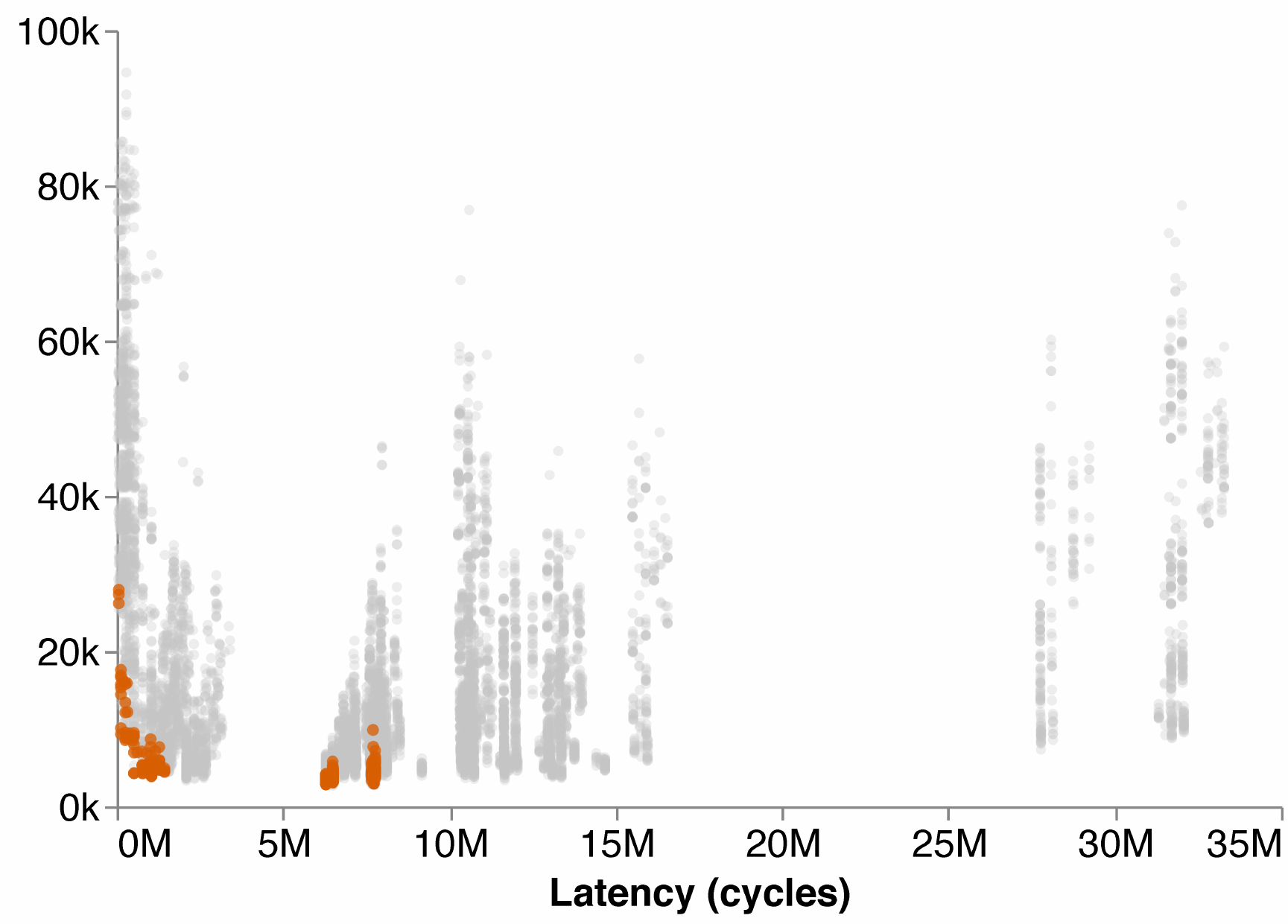}
    \caption{Points accepted by \sys.}
    \label{fig:gemm-dse:only-dahlia}
  \end{subfigure}
  \hfill
  \begin{subfigure}[b]{0.32\linewidth}
    \centering
    \includegraphics[width=\linewidth]{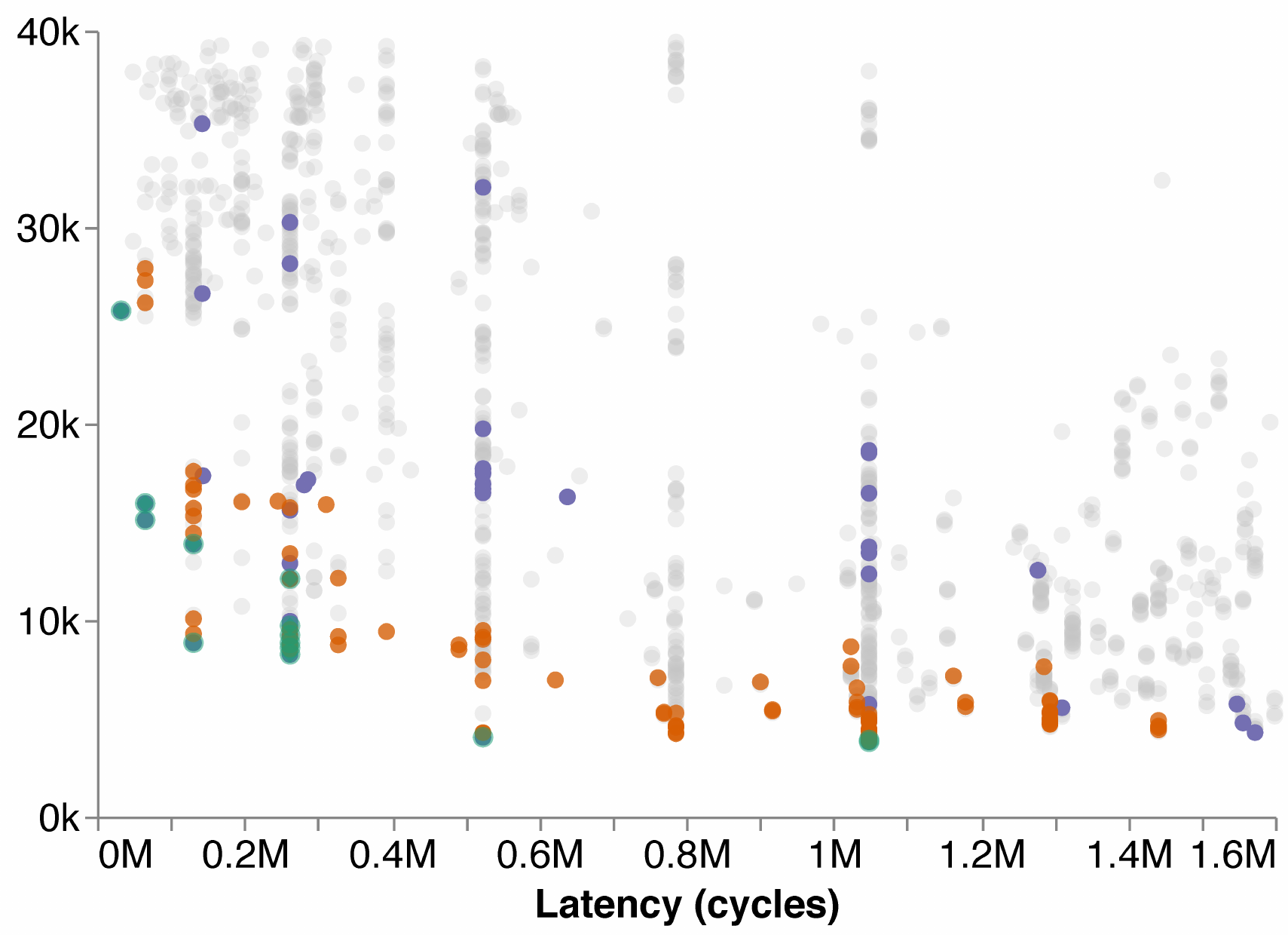}
    \caption{Cluster of Pareto points.}
    \label{fig:gemm-dse:pareto-front}
  \end{subfigure}
  \caption{Results from exhaustive design space exploration for \dseKernel.}
  \label{fig:gemm-dse}
\end{figure*}

In this section, we conduct an exhaustive design-space exploration (DSE) of a single benchmark as a case study.
Without \sys, the HLS design space is extremely large---we study how the smaller \sys-restricted design space compares.
We select a blocked matrix multiplication kernel (\dseKernel from MachSuite) for its large but tractable design space.
The kernel has 3 two-dimensional arrays (two operands and the output product) and 5 nested loops, of which the inner 3 are parallelizable.
We define parameters for the 6 banking factors (two dimensions for each memory) and 3 unrolling factors.
(A full code listing appears in the supplementary material~\cite{techreport}.)
We explore a design space with banking factors of 1--4 and unrolling factors of 1, 2, 4, 6, and 8.
This design space consists of 32,000 distinct configurations.

We exhaustively evaluated the entire design space using Vivado HLS's estimation mode, which required a total of \dseComputeHours compute hours.
We identify Pareto-optimal configurations according to
their estimated cycle latency and number of lookup tables (LUTs), flip flops (FFs), block RAMs (BRAMs), and arithmetic units (DSPs).

\sys accepts 354 configurations, or about 1.1\% of the unrestricted design space.
But the smaller space is only valuable if it consists of \emph{useful} design points---a broad range of Pareto-optimal configurations.
\Cref{fig:gemm-dse:only-pareto,fig:gemm-dse:only-dahlia} show the Pareto-optimal points and the subset of points that \sys accepts, respectively.
(Pareto optimality is determined using all objectives, but the plot shows only two: LUTs and latency.)
\Cref{fig:gemm-dse:pareto-front} shows a zoomed-in view of the tight cluster of
Pareto points in the bottom-left of the first two graphs. \sys-accepted
points lie primarily on the Pareto frontier and allow area-latency trade-offs.
The optimal points that \sys rejects expend a large number of LUTs to reduce
BRAM consumption which, while Pareto optimal, don't seem to be of practical
use.

\subsection{\sys-Directed DSE \& Programmability}
\label{sec:eval:machsuite}

We port benchmarks from an HLS benchmark suite, MachSuite~\cite{machsuite}, to study \sys's flexibility.
Of the 19 MachSuite benchmarks, one (\bench{backprop}) contains a correctness bug and
two fail to synthesize correctly in Vivado, indicating a bug in the tools.
We successfully ported all 16 of the remaining benchmarks without substantial restructuring.

\begin{figure*}
  \centering
  \begin{subfigure}[b]{0.29\linewidth}
    \centering
    \includegraphics[width=\linewidth]{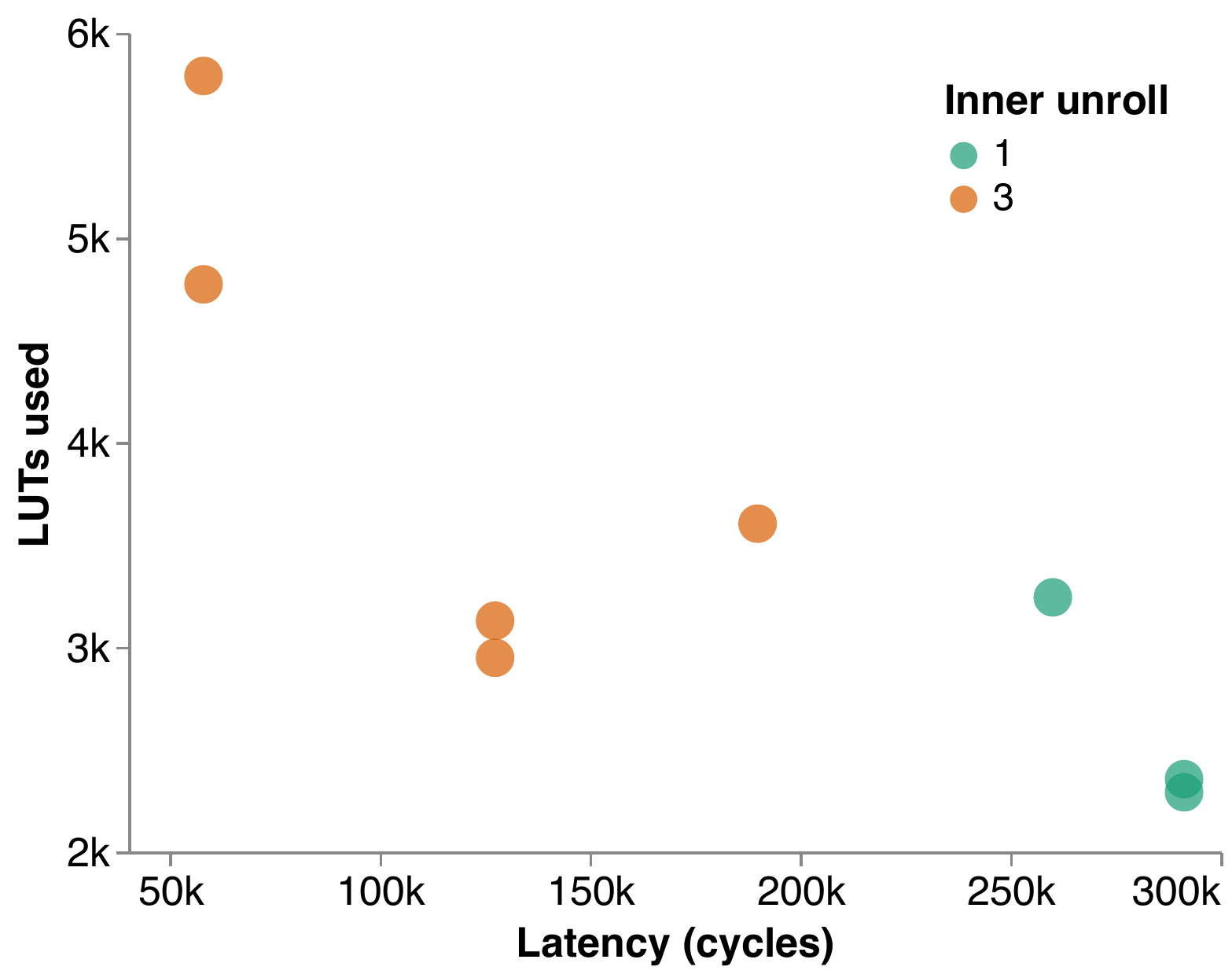}
    \vspace{-15pt}
    \caption{\bench{stencil2d} with inner unroll.}
    \label{fig:dahlia-dse-stencil2d}
  \end{subfigure}
  \hfill
  \begin{subfigure}[b]{0.36\linewidth}
    \centering
    \includegraphics[width=\linewidth]{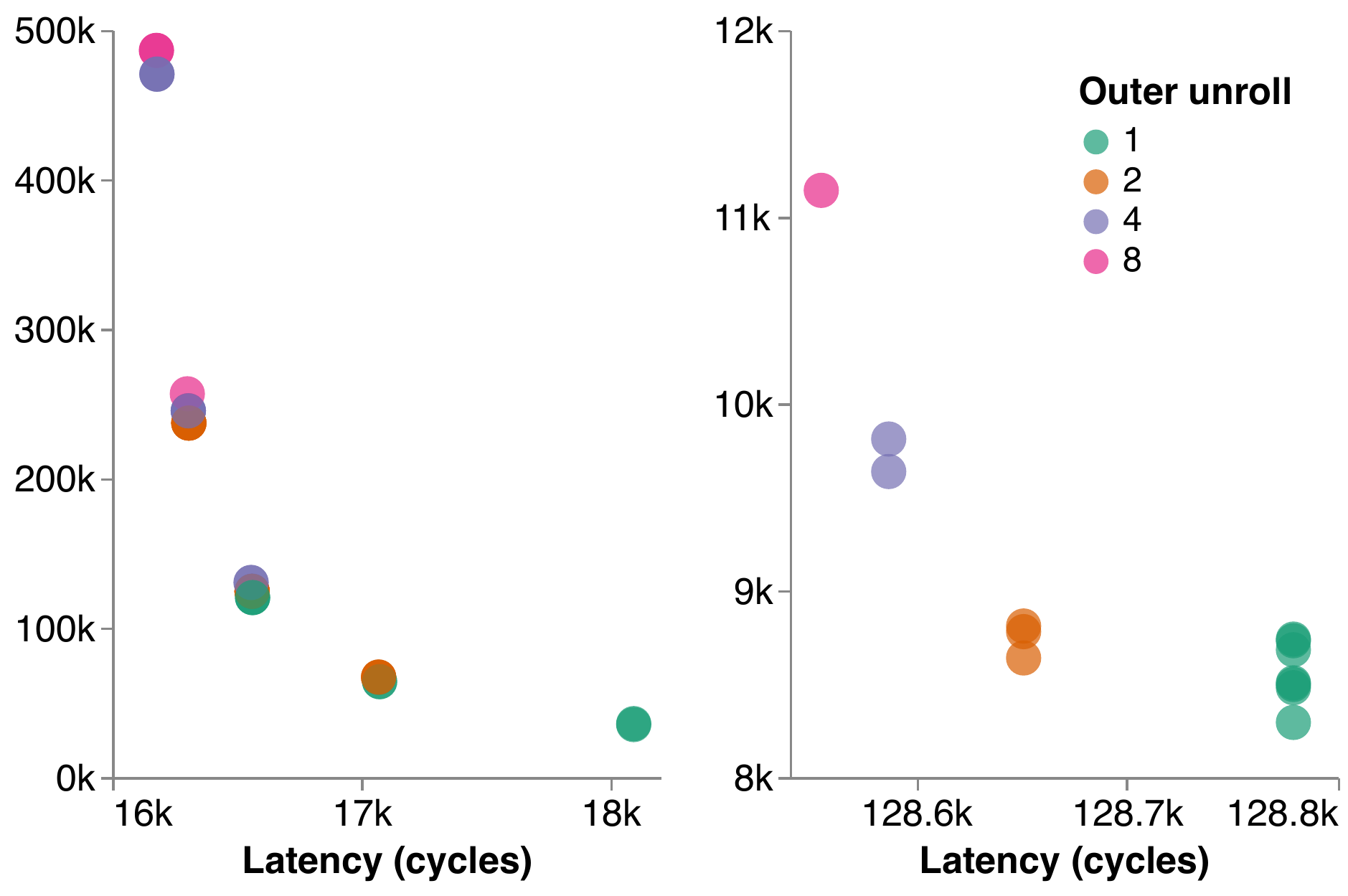}
    \vspace{-15pt}
    \caption{\bench{md-knn} with outer unroll.}
    \label{fig:dahlia-dse-md-knn}
  \end{subfigure}
  \hfill
  \begin{subfigure}[b]{0.29\linewidth}
    \centering
    \includegraphics[width=\linewidth]{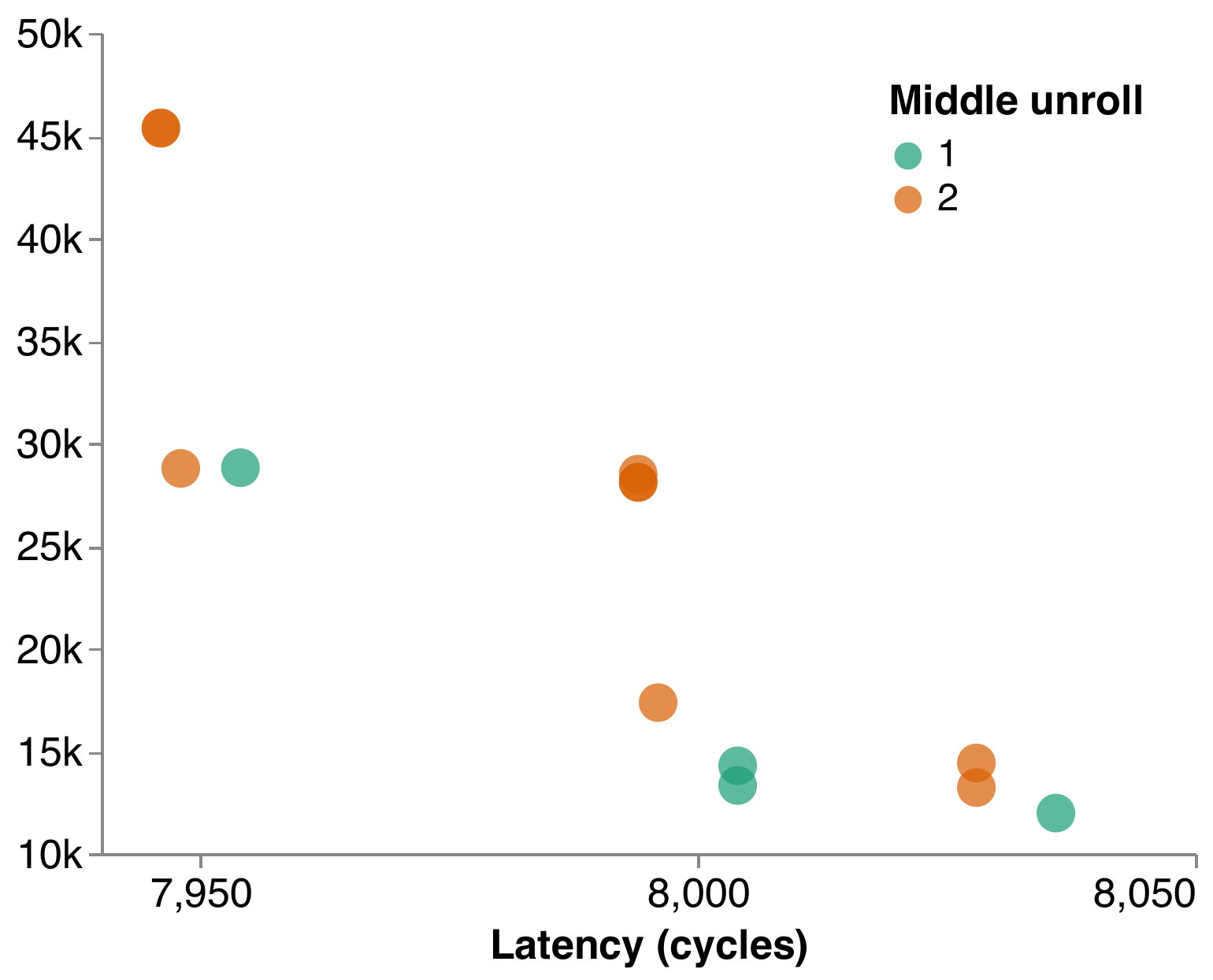}
    \vspace{-15pt}
    \caption{\bench{md-grid} with middle unroll.}
    \label{fig:dahlia-dse-md-grid}
  \end{subfigure}
  \caption{The design spaces for three MachSuite benchmarks.
  Each uses a color to highlight one design parameter.}
  \label{fig:dahlia-dse}
\end{figure*}

From these, we select 3 benchmarks that exhibit the kind of fine-grained, loop-level parallelism that \sys targets as case studies:
\bench{sencil2d}, \bench{md-knn}, and \bench{md-grid}.
As the previous section illustrates, an unrestricted DSE is intractable for even modestly sized benchmarks,
so we instead measure the breadth and performance of the much smaller space of configurations that \sys accepts.
For each benchmark, we find all optimization parameters available in the \sys
port and define a search space.
The type checker rejects some design points, and we measure the remaining space.
We use Vivado HLS's estimation mode to measure the resource counts and estimated latency for each accepted point.
\Cref{fig:dahlia-dse} depicts the Pareto-optimal points in each space.
In each plot, we also highlight the effect a single parameter has on the results.

The rest of this section reports quantitatively on each benchmark's design space
and reports qualitatively on the
programming experience during the port from C to \sys.

\paragraph{stencil2d.}

MachSuite's \bench{stencil2d}
is a filter operation with four nested loops.
The outer loops scan over the input matrix and the inner loops apply a $3 \times 3$ filter. 
Our \sys port unrolls the inner two loops and banks both input memories.
We use unrolling factors from 1 to 3 and bank
each dimension of the
input array by factors 1 to 6.
The resulting design space has 2,916 points.
\sys accepts 18 of these points (0.6\%), of which 8 are Pareto-optimal within the set.

\Cref{fig:dahlia-dse-stencil2d} shows the Pareto frontier among the \sys-accepted points.
The figure uses color to show the unrolling factor for the innermost loop.
This unrolling factor has a large effect on the design's performance, while banking factors and the other loop explain the rest of the variation.

The original C code uses single-dimensional arrays and uses index arithmetic to treat them as matrices:
\begin{lstlisting}[language=C++]
for (r=0; r<row_size-2; r++)
  for (c=0; c<col_size-2; c++)
    for (k1=0; k1<3; k1++)
      for (k2=0; k2<3; k2++)
        mul = filter[k1*3 + k2] *
              orig[(r+k1)*col_size + c+k2];
\end{lstlisting}
In the \sys port, we must use proper two-dimensional arrays because the compiler rejects arbitrary indexing expressions.
Using views, programmers can decouple the storage format from the iteration pattern.
To express the accesses to the input matrix \code{orig}, we create a shifted suffix view (\Cref{sec:views}) for the current window:
\begin{lstlisting}
for (let row = 0..126) {
  for (let col = 0..62) {
    view window = shift orig[by row][by col];
    for (let k1 = 0..3) unroll 3 {
      for (let k2 = 0..3) unroll 3 {
        let mul = filter[k1][k2] * window[k1][k2];
\end{lstlisting}
The view
makes the code's logic more obvious while allowing the \sys type checker to allow unrolling on the inner two loops.
It also clarifies why parallelizing the outer loops would be undesirable: the parallel views would require overlapping regions of the input array, introducing a bank conflict.

\paragraph{md-knn.}

The \bench{md-knn} benchmark implements an $n$-body molecular dynamics simulation with a $k$-nearest neighbors kernel.
The MachSuite implementation uses data-dependent loads in its main loop, which na\"ively seems to prevent parallelization.
In our \sys port, however, we hoist this serial section into a separate loop that runs before the main, parallelizable computation.
\sys's type system helped guide the programmer toward a version of the benchmark where the benefits from parallelization are clear.

For each of the program's four memories, we used banking
factors from 1 to 4.
We unrolled each of the two nested loops with factors from 1 to 8.
The full space has 16,384 points, of which \sys accepts 525 (3\%). 37 of
the \sys-accepted points are Pareto-optimal.

\Cref{fig:dahlia-dse-md-knn} shows two Pareto frontiers that Dahlia accepts at different scales.
The color shows the unrolling factor of the
outer loop. The frontier on the right uses an order of magnitude fewer
resources but is an order of magnitude slower.
In this kernel, the dominant effect is the memory banking (not shown in the
figure), which determines which frontier the designs fall into.
The outer unrolling factor (shown in color) affects the two regimes differently:
on the right, it allows area--latency trade-offs within the frontier;
on the left, it acts as a second-order
effect that expends LUTs to achieve a small increase in performance.

\paragraph{md-grid.}

Another algorithm for the same molecular dynamics problem,
\bench{md-grid},
uses a different strategy
based on a 3D grid implemented with several 4-dimensional arrays.
It calculates forces between neighboring grid cells.
Of its 6 nested loops, the outer three are parallelizable.
We use banking factors of 1 to 4 for each dimension of each array,
and we try unrolling factors from 1 to 8 for both loops.
The full space has 21,952 points, of which \sys accepts 81 (0.4\%). 13 of the
\sys-accepted points are Pareto-optimal.

\Cref{fig:dahlia-dse-md-grid} again shows the Pareto-optimal design points.
The \emph{innermost} loop unrolling factor (not shown in the figure) determines which of three coarse regimes the design falls into.
The color shows the \emph{second} loop unrolling factor, which determines a second-order area--latency trade-off within each regime. Unrolling enables
latency-area trade-offs in both the cases.



\section{Future Work}

\sys{} represents a first step toward high-level semantics for accelerator design languages.
It leaves several avenues for future work on scaling up from kernels to full applications and expressing more hardware implementation techniques.

\paragraph{Modularity.}
\sys{}'s type system relies on a closed-world assumption.
A compositional type system would enable reuse of abstract hardware modules without ``inlining'' them, like functions in a software language.
The primary challenge in modular accelerator design is the balance between abstraction and efficiency:
a more general module is likely to be less efficient.
An abstraction mechanism must also cope with the timing of inter-module interactions: some interfaces are latency-insensitive while others rely on cycle-level timing.

\paragraph{Polymorphism.}
\sys{}'s memory types are monomorphic.
Polymorphism would enable abstraction over memories' banking strategies and sizes.
A polymorphic \sys{}-like language could rule out invalid combinations of abstract implementation parameters before the designer picks concrete values, which would help constrain the search space for design space exploration.

\paragraph{Pipelining.}
Pipelined logic is a critical implementation technique for high-level synthesis.
\sys{} does not reason about the timing of pipeline stages or their resource conflicts.
Extensions to its type system will need to reason about the cycle-level latency of these stages and track the fine-grained sharing of logic resources.

\paragraph{Direct RTL generation.}
The current \sys{} compiler relies on a commercial \cxx-based HLS compiler as its backend.
It generates directives that instruct the HLS tool to generate hardware according to the program's \sys{} types, but the unpredictability of traditional HLS means that results can still vary.
Future compilers for \sys{}-like languages might generate RTL directly and rely on the simpler input language avoid the complexity of unrestricted HLS.

\section{Related Work}

\sys{} builds on a long history of work on safe systems programming.
Substructural type systems are known to be a good fit for controlling system resources ~\cite{linear-haskell, cyclone-region, alms, pony-deny, rust}.
\sys{}'s enforcement of exclusive memory access resembles work on race-free parallel programming using type and effect systems~\cite{dpj}
or concurrent separation logic~\cite{ohearn:csl}.
Safe parallelism on CPUs focuses on data races where concurrent reads and writes
to a memory are unsynchronized.
Conflicts in \sys are different: \emph{any} simultaneous pair of accesses to the same \emph{bank} is illegal.
The distinction influences \sys's capability system and its memory views, which cope with the arrangement of arrays into parallel memory banks.

\sys{} takes inspiration from other approaches to improving the accelerator design process, including HDLs, HLS, DSLs, and
other recent accelerator design languages.

\paragraph{Spatial.}
\label{par:spatial}
\begin{figure}
  \centering
    \centering
    \includegraphics[width=0.65\linewidth]{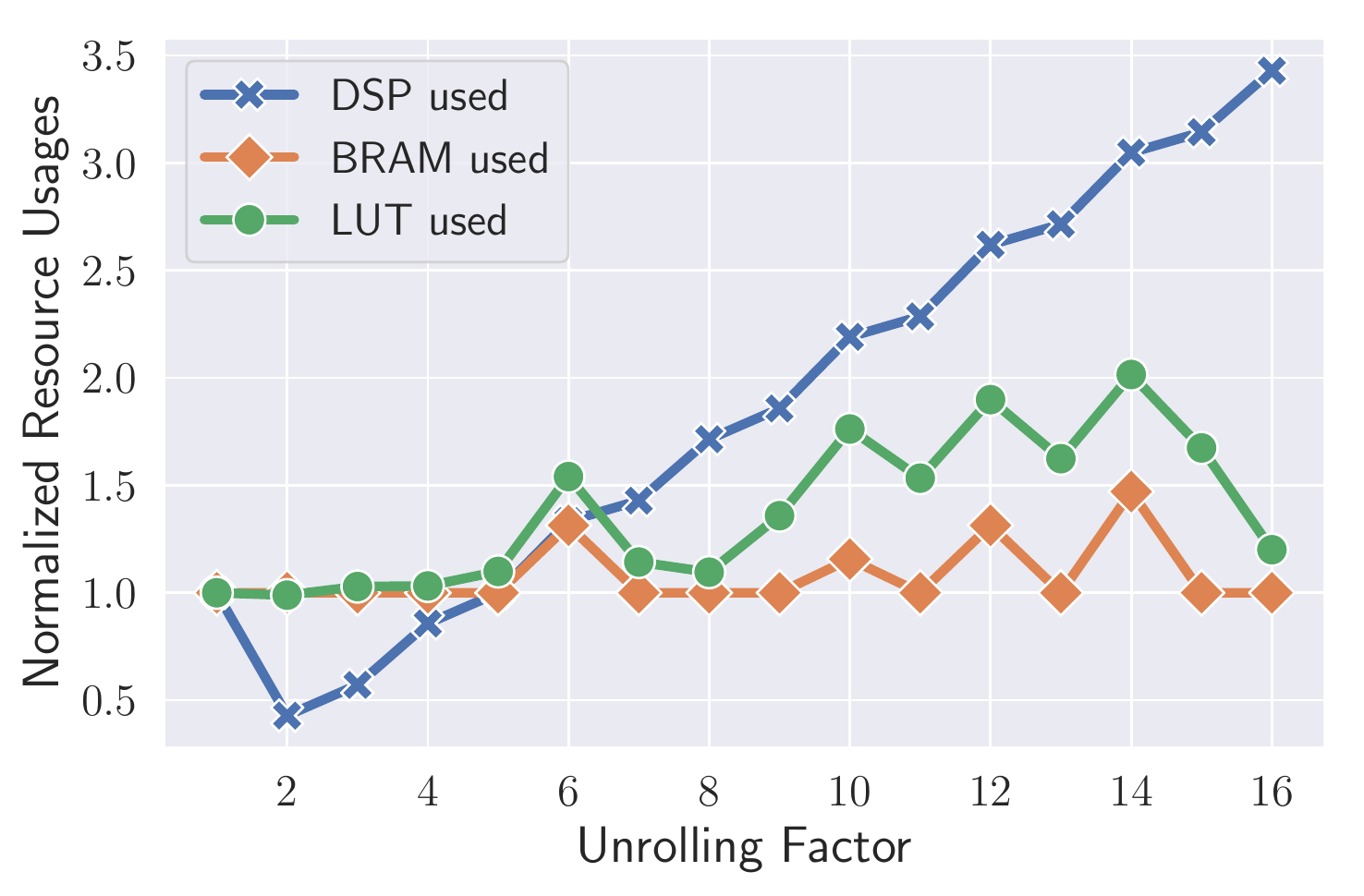}
  \caption{Resource utilization for \bench{gemm-ncubed} in Spatial normalized to the design without unrolling.
  }
  \label{fig:spatial}
\end{figure}

Spatial~\cite{koeplinger:spatial} is a language for designing
accelerators that builds on
\emph{parallel patterns}~\cite{prabhakar:spatial-lang}, which are flexible hardware templates.
Spatial adds some automation beyond traditional HLS: it
infers a banking strategy given some parallel accesses.
Like HLS, Spatial designs can be unpredictable.
\cref{fig:spatial} shows resource usage for the matrix multiplication kernel from \cref{sec:principles} written in Spatial.
(A full experimental setup appears in the supplementary material~\cite{techreport}.)
For unrolling factors that do not evenly divide the memory size, Spatial will sometimes infer a banking factor that is not equal to the unrolling factor.
In these cases, the resource usage abruptly increases.
A type system like \sys could help address these predictability pitfalls in Spatial.

\paragraph{Better HDLs.}

Modern hardware description languages~\cite{bachrach:chisel, pymtl, pyrtl, clash, truong:golden, hardcaml, nikhil:bluespec} aim to address the shortcomings of Verilog and VHDL.
These languages target register transfer level (RTL) design.
\sys{} targets a different level of abstraction and a different use case:
it uses an imperative programming model and focuses exclusively on computational accelerators.
\sys{} is not a good language for implementing a CPU, for example.
Its focus on acceleration requires the language and semantics to more closely resemble software languages.

\paragraph{Traditional HLS.}

Existing commercial~\cite{vivadohls, intelhls, catapulthls, stratushls} and academic~\cite{putnam:chimps, canis:legup, pilato:bambu, zhang:autopilot} high-level synthesis (HLS) tools compile subsets of C, \cxx, OpenCL, or SystemC to RTL.
While their powerful heuristics can be effective, when they fail,
programmers have little insight into what went wrong or how to fix it~\cite{liang:hls}.
\sys{} represents an alternative approach that prioritizes programmer control over black-box optimization.

\paragraph{Targeting hardware from DSLs.}

Compilers to FPGAs and ASICs exist for DSLs for image processing~\cite{hegarty:darkroom, hegarty:rigel, pu:halide-hls, halide-to-hardware}
and machine learning~\cite{george:optimlfpga, dnnweaver}.
\sys{} is not a DSL: it is a general language for implementing accelerators.
While DSLs offer advantages in productivity and compilation for individual application domains,
they do not obviate the need for general languages
to fill in the gaps between popular domains, to offer greater programmer control when appropriate, and to serve as a compilation target for multiple DSLs.

\paragraph{Accelerator design languages.}

Some recent languages also focus on general accelerator design.
HeteroCL~\cite{heterocl} uses a Halide-like~\cite{ragan-kelley:halide} scheduling language to describe how to map algorithms onto HLS-like hardware optimizations, and
T2S~\cite{t2s} similarly lets programs describe how generate a spatial implementation.
Lime~\cite{lime} extends Java to express target-independent streaming accelerators.
CoRAM~\cite{chung:coram} is not a just a language; it extends FPGAs with a programmable memory interface that adapts memory accesses, akin to \sys's memory views.
\sys's focus on predictability and type-driven design makes it unique, as far as we are aware.

\section{Conclusion}

%

\sys exposes predictability as a new design goal for HLS tools.
Predictability comes at a cost---it can rule out design points that perform surprisingly well because of a subtle convergence of heuristics.
We see these outliers as a worthy sacrifice in exchange for an intelligible programming model and robust reasoning tools.
We hope to extend \sys's philosophy to bring predictability to the rest of the reconfigurable hardware system stack, from the language to the LUTs.

\begin{acks}

We thank Drew Zagieboylo and Yi-Hsiang Lai for insightful discussions and Alexa VanHattum, Dietrich
Geisler, and Pedro Henrique Azevedo de Amorim for invaluable comments on early drafts and solidarity in the final hours.
Many thanks to the anonymous PLDI reviewers and our shepherd, Zachary Tatlock, for suggesting important additions.

This work was supported in part by the Center for Applications Driving Architectures (ADA), one of six centers of JUMP, a Semiconductor Research Corporation program co-sponsored by DARPA.
It was also supported by the Intel and NSF joint research center for Computer Assisted Programming for Heterogeneous Architectures (CAPA).
Support included NSF awards \#1723715 and \#1845952.

\end{acks}

\balance
\bibliography{./bib/venues,./bib/papers}

\newpage
\appendix
\section{Semantics}
The following lists the grammar for Filament, the core language of Dahlia.

\begin{small}
\begin{align*}
x &\in \text{variables} \enspace\enspace
a \in \text{memories} \enspace\enspace
n \in \text{numbers} \\
b &::= \ottkw{true} \mid \ottkw{false} \qquad
v ::= n \mid b \mid v_1 \; \ottkw{bop} \; v_2 \\
e &::= v \mid
    \ottkw{bop} \; e_1 \; e_2 \mid
    x \mid
    a[e] \\
c &::= e \mid
    \ottkw{let} \; x = e \mid
    c_1 \seqcomp c_2 \mid
    c_1 \interseq{\rho} c_2 \mid
    c_1 ~\textbf{;}~ c_2 \mid
    \ottkw{if} \; x \; c_1 \; c_2 \mid \\
&\qquad \ottkw{while} \; x \; c \mid
    x := e \mid
    a[e_1] := e_2 \mid
    \ottkw{skip} \\
\tau &::= \ottkw{bit}\langle n \rangle \mid
    \ottkw{float} \mid
    \ottkw{bool} \mid
    \ottkw{mem} \; \tau[n_1]
\end{align*}
\end{small}

\noindent The large-step operational semantics listed below capture the complete evaluation
of an expression or command. The environment $\sigma$ maps variables and memory
names to values, and the context $\rho$ is the set of memories the program has accessed.

\vspace{1em}

\begin{small}

\ottdefnelargereduce

\vspace{1em}

\ottdefnclargereduce

\end{small}

\noindent The small-step operational semantics capture incremental evaluation of an expression
or command and form the basis of the proof of soundness in \Cref{sec:soundness}.

\vspace{1em}

\begin{small}
\ottdefnesmallreduce

\vspace{1em}

\ottdefncsmallreduce

\end{small}

\noindent To enforce Dahlia's safety condition, the typing judgments use the
typing context $\Gamma$ for variables and the affine context $\Delta$ for memories.

\vspace{1em}

\begin{small}
\ottdefnecheck

\vspace{1em}

\ottdefnccheck

\end{small}

\section{Proof of soundness}\label{sec:soundness}
If there exists a typing context $\Gamma$ and affine memory context $\Delta$
under which a command $c$ type-checks, and $\Gamma, \Delta$ is equivalent to an
environment $\sigma$ and context $\rho$, then either
$\sigma, \rho, c \rightarrow^* \sigma', \rho', \gskip$ or $c$ diverges.
\newline To prove this theorem, we will prove the supporting progress and
preservation lemmas (stated below), which together imply soundness.

\vspace{1em}
\paragraph{Supporting definitions}

\begin{itemize}
    \item \underline{Defined:} $a$ is \textit{defined} in $\Delta$ if
    $\exists \: \tau, n$ with $(a \mapsto \textbf{mem} \: \tau[n]) \in \Delta$.
    $x$ is \textit{defined} in $\Gamma$ if $\exists \: \tau$ with
    $(x \mapsto \tau) \in \Gamma$.

    \item \underline{Type-check:} If
    $\Gamma, \Delta \vdash e : \tau \dashv \Delta'$ then
    \textit{$e$ type-checks to $\tau$ under $\Gamma, \Delta$ producing $\Delta'$}.
    If $\Gamma, \Delta \vdash c \dashv \Gamma', \Delta'$ then
    \textit{$c$ type-checks under $\Gamma, \Delta$ producing $\Gamma', \Delta'$}.

    \item \underline{$\sim$ (equivalence):} $\Gamma, \Delta \sim \sigma, \rho$ if
    \begin{enumerate}
        \item $\forall \: x$ with $(x \mapsto \tau) \in \Gamma, \exists \: v$
        with $(x \mapsto v) \in \sigma$ and $v$ type-checks to $\tau$ under $\Gamma, \Delta$
        \item $\forall \: l$ with $(l \mapsto \tau) \in \Delta, l \notin \rho$.
    \end{enumerate}

    \item \underline{$\bar{\rho}$:} $\bar{\rho} = \{a \mapsto \textbf{mem} \: \tau[n] \in \Delta^* \wedge a \notin \rho\}$
    where $\Delta^*$ is the affine context of memories initially available to a program.

    \item \underline{Construction:} $\Gamma, \Delta$ can be \textit{constructed} from $\sigma, \rho$ if
    \begin{enumerate}
        \item $\forall \: (x \mapsto v) \in \sigma, (x \mapsto \tau) \in \Gamma$
        \item $\forall \: l \in \rho, (l \mapsto \textbf{mem} \: \tau[n]) \notin \Delta$
        \item $v$ type-checks to $\tau$ under $\Gamma, \Delta$
    \end{enumerate}
\end{itemize}

\paragraph{Supporting lemmas}
\begin{itemize}
    \item \underline{L1:} If $\Gamma, \Delta \vdash c \dashv \Gamma', \Delta'$,
    then $\Gamma \subseteq \Gamma'$. Proof: The only typing rule that modifies
    $\Gamma$ is \textsc{check\_let}. Under this rule, $\Gamma'$ = $\Gamma$
    extended to add a mapping for a variable $x$. There is no rule that removes
    mapping from $\Gamma$. So $\forall \: m \in \Gamma, m \in \Gamma'$.

    \item \underline{L2}: If $c$ type-checks under $\Gamma, \Delta$, then $c$
    type-checks under $\Gamma', \Delta$ where $\Gamma \subseteq \Gamma'$.
    Proof: The only typing rule that reads $\Gamma$ is \textsc{check\_update},
    which checks in its premises that $x$ is defined in $\Gamma$. By L1, if $x$
    is defined in $\Gamma$ it is defined in $\Gamma'$. There is also no rule
    that changes the type $\tau$ of $x$ in $\Gamma$, so $x$ will have the same
    type $\tau$ in $\Gamma'$.

    \item \underline{L3}: If $\sigma, \rho, e \rightarrow \sigma', \rho', e'$,
    then $\sigma = \sigma'$. Proof: There is no step rule for expressions that
    extends $\sigma$, which by the grammar is the only modification possible to
    memory stores.

    \item \underline{L4}: If $\sigma, \rho, e \rightarrow \sigma', \rho', e'$
    and $\rho' \neq \rho$, then $e$ is a read $a[n]$ and
    $\rho' = \rho \cup \{ a \}$. Proof: the only step rule for expressions that
    adds elements to $\rho$ is \textsc{read2}, by which $\rho' = \rho \cup \{a\}$.
    There is no step rule that removes elements from $\rho$.
\end{itemize}

\subsection{Progress}
If $\exists \: \Gamma, \Delta, \sigma, \rho$ such that
$\Gamma, \Delta \sim \sigma, \rho$ and command $c$ type-checks
under $\Gamma, \Delta$, then either $c$ is a value or
\begin{enumerate}
    \item $\exists \: \sigma', \rho', c'$ with
    $\sigma, \rho, c \rightarrow \sigma', \rho', c'$ or

    \item $c = \gskip \interseq{\rho} c_2$ with
    $c_2 \neq \gskip$ and $\exists \: c_2', \rho''$ with
    $\sigma, \rho'', \gskip \interseq{\rho} c_2 \rightarrow \sigma', \rho'', \gskip \interseq{\rho'} c_2'$.
\end{enumerate}

\subsubsection{Proof}
Inductive hypothesis: Progress holds for sub-forms of any inductive form.
Assumptions: $\Gamma, \Delta \sim \sigma, \rho$ and $c$ type-checks under $\Gamma, \Delta$.

\vspace{1em}
\noindent\case{$c$ is an expression}
    \begin{itemize}
        \item \case{$c$ is a value} Progress holds by assumption.
        \item \case{$c = \gbop{e_1}{e_2}$} For simplicity, we
        ignore the cases in which $\textbf{bop}$ is incompatible with the types
        of $e_1$ and $e_2$. We have three possibilities:
            \begin{enumerate}
                \item Neither $e_1$ nor $e_2$ is a value. By assumption $c$
                type-checks under $\Gamma, \Delta$, so by \textsc{check\_bop}
                $e_1$ type-checks under $\Gamma, \Delta$. By the inductive
                hypothesis, $\sigma, \rho, e_1 \rightarrow \sigma', \rho', e_1'$
                so we have $\sigma, \rho, \gbop{e_1}{e_2} \rightarrow \gbop{e_1'}{e_2}$ as needed.

                \item Only $e_1$ is a value $v_1$. By assumption $c$
                type-checks under $\Gamma, \Delta$ and
                $\Gamma, \Delta \vdash v_1 \dashv \Delta$, so $e_2$ type-checks
                under $\Gamma, \Delta$. By the inductive hypothesis,
                $\sigma, \rho, e_2 \rightarrow \sigma', \rho', e_2$ so
                $\sigma, \rho, \gbop{v_1}{e_2} \rightarrow \sigma', \rho', \gbop{v_1}{e_2'}$.

                \item Both $e_1$ and $e_2$ are values $v_1$ and $v_2$.
                $\sigma, \rho, \gbop{v_1}{v_2} \rightarrow \sigma, \rho, v_1 \: \textbf{bop} \: v_2$ by \textsc{bop3}.
            \end{enumerate}
        \item \case{$c = x$} By assumption $x$ type checks under
        $\Gamma, \Delta$, so $x$ is defined in $\Gamma$. Then $\exists \: v$
        with $(x \mapsto v) \in \sigma$, so
        $\sigma, \rho, x \rightarrow \sigma, \rho, v$ by \textsc{var}.

        \item \case{$c = a[e]$} By assumption $a[e]$ type-checks under
        $\Gamma, \Delta$, so by \textsc{check\_read}, $e$ type-checks under
        $\Gamma, \Delta$ producing $\Delta_2$, and $a$ is defined in
        $\Delta_2$. By the the inductive hypothesis, progress holds for $e$,
        so $\sigma, \rho, e \rightarrow \sigma', \rho', e'$ or $e$ is a value
        $n$. $a$ is defined in $\Delta_2$, so it must be defined in $\Delta$
        since there is no type-checking rule for expressions by which
        $\Gamma, \Delta \vdash e \dashv \Delta_2$ and
        $\exists \: l \notin \Delta, \in \Delta_2$. So $a \notin \rho$. So if
        $e$ is a value, then $\sigma, \rho, a[e] \rightarrow \sigma, \rho \cup \{a\}, \sigma(a)(n)$.
        If $e$ is not a value, then $\sigma, \rho, a[e] \rightarrow \sigma', \rho', a[e']$.

    \end{itemize}

\noindent\case{$c = \glet{x}{e}$}
By assumption this form type-checks under $\Gamma, \Delta$. By
\textsc{check\_let}, $e$ type-checks under $\Gamma, \Delta$. By the inductive
hypothesis, $e$ is either a value $v$ or
$\sigma, \rho, e \rightarrow \sigma', \rho', e'$. In the first case we have
$\sigma, \rho, \glet{x}{v} \rightarrow \sigma[x \mapsto v], \rho, \gskip$.
In the second case we have $\sigma, \rho, \glet{x}{e} \rightarrow \sigma', \rho', \glet{x}{e'}$.

\vspace{1em}
\noindent\case{$c = c_1 \seqcomp c_2$} $\forall \: \sigma, \rho$, $\sigma, \rho, c_1 \seqcomp c_2 \rightarrow \sigma, \rho, c_1 \interseq{\rho} c_2$.

\vspace{1em}
\noindent\case{$c = c_1 \interseq{\rho''} c_2$}
We have three possibilities:
    \begin{itemize}
        \item $c_1 \neq \gskip$. By assumption $c$ type-checks under
        $\Gamma, \Delta$. By \textsc{check\_inter\_seq\_comp}, $c_1$
        type-checks under $\Gamma, \Delta$. By the inductive hypothesis,
        $\sigma, \rho, c_1 \rightarrow \sigma', \rho', c_1'$ and so
        $c_1 \interseq{\rho''} c_2 \rightarrow c_1' \interseq{\rho''} c_2$.

        \item $c = \gskip \interseq{\rho} c_2$. By assumption $c$
        type-checks under $\Gamma, \Delta$. By
        \textsc{check\_inter\_seq\_comp}, $c_2$ type-checks under
        $\Gamma, \bar{\rho}.$ By the inductive hypothesis (and the definition
        of $\sim$, under which we have $\rho$),
        $\sigma, \rho, c_2 \rightarrow \sigma', \rho', c_2'$, so we have
        $\sigma, \rho'', \gskip \interseq{\rho} c_2 \rightarrow \sigma', \rho'', \gskip \interseq{\rho'} c_2'$.

        \item If $c_1 = c_2 = \gskip$, $\forall \: \sigma, \rho$, we
        have $\sigma, \rho, c_1 \interseq{\rho''} c_2 \rightarrow \sigma, \rho \cup \rho'', \gskip$.
    \end{itemize}

\vspace{1em}

\noindent\case{$c = c_1 ; c_2$}
 We have two possibilities:
    \begin{itemize}
        \item $c_1 \neq \gskip$. By assumption, $c$ type-checks under
        $\Gamma, \Delta$, so $c_1$ type-checks under $\Gamma, \Delta$ by
        \textsc{check\_par\_comp}. By the the inductive hypothesis,
        $\sigma, \rho, c_1 \rightarrow \sigma', \rho', c_1'$, so
        $\sigma, \rho, c_1 ; c_2 \rightarrow \sigma', \rho', c_1' ; c_2$.

        \item $c_1 = \gskip$. $\forall \: \sigma, \rho$, we have
        $\sigma, \rho, \gskip ; c_2 \rightarrow \sigma, \rho, c_2$.
    \end{itemize}

\vspace{1em}

\noindent\case{$c = \gif{x}{c_1}{c_2}$}
By assumption, $c$ type-checks under $\Gamma, \Delta$. Then $x$ type-checks to
bool, so $x$ is either \textbf{true} or \textbf{false}.
$\forall \: \sigma, \rho$, we have
$\sigma, \rho, \gif{\gtrue}{c_1}{c_2} \rightarrow \sigma, \rho, c_1$
and $\sigma, \rho, \gif{\gfalse}{c_1}{c_2} \rightarrow \sigma, \rho, c_2$.

\vspace{1em}
\noindent\case{$c = \gwhile{x}{c_1}$}
$\forall \: \sigma, \rho$ we have \\
$\sigma, \rho, \gwhile{x}{c_1} \rightarrow \sigma, \rho, \gif{x}{(c_1 \seqcomp \gwhile{x}{c_1})}{\gskip}$.

\vspace{1em}
\noindent\case{$c = x := e$}
If $e$ is a value $v$, then  $\forall \: \sigma, \rho$ we have
$\sigma, \rho, x := e \rightarrow \sigma[x \mapsto v], \rho, \gskip$.
If $e$ is not a value, then by assumption that $c$ type-checks under
$\Gamma, \Delta$, $e$ type-checks under $\Gamma, \Delta$ by
\textsc{check\_update}. Then by the inductive hypothesis,
$\sigma, \rho, e \rightarrow \sigma', \rho', e'$, so
$\sigma, \rho, x := e \rightarrow \sigma', \rho', x := e'$.

\vspace{1em}
\noindent\case{$c = a[e_1] := e_2$}
We have three possibilities:
    \begin{itemize}
        \item $e_1$ and $e_2$ are values $n$ and $v$. By assumption $c$
        type-checks under $\Gamma, \Delta$, so by \textsc{check\_write},
        $a$ is defined in $\Delta$. By definition of $\sim$, $a \notin \rho$,
        so the premise of \textsc{write3} is satisfied. Then
        $\sigma, \rho, c \rightarrow \sigma[a[n] \mapsto v], \rho \cup \{a\}, \gskip$.

        \item $e_1$ is a value $n$. By assumption $c$ type-checks under
        $\Gamma, \Delta$, so by \textsc{check\_write} $n$ type-checks under
        $\Gamma, \Delta$ producing $\Delta$ and $e_2$ type-checks under
        $\Gamma, \Delta$. By the the inductive hypothesis,
        $\sigma, \rho, e_2 \rightarrow \sigma', \rho', e_2'$, so
        $\sigma, \rho, a[n] := e_2 \rightarrow \sigma', \rho', a[n] := e_2'$.

        \item Neither $e_1$ nor $e_2$ is a value. By assumption $c$ type-checks
        under $\Gamma, \Delta$, so as does $e_1$. By the the inductive
        hypothesis, $\sigma, \rho, e_1 \rightarrow \sigma', \rho', e_1'$, so
        $\sigma, \rho, a[e_1] := e_2 \rightarrow \sigma', \rho', a[e_1'] := e_2$.
    \end{itemize}

\subsection{Preservation}
If:
\begin{enumerate}
    \item $\exists \: \Gamma, \Delta$ such that command $c$ type-checks under $\Gamma, \Delta$
    \item $\exists \: \sigma, \rho$ with $\Gamma, \Delta \sim \sigma, \rho$
    \item $\exists \: \sigma' \rho', c'$ with $\sigma, \rho, c \rightarrow \sigma', \rho', c'$ or $\exists \: \sigma', \rho', \rho'', c_2'$ with $c = \gskip \interseq{\rho} c_2$ and $\sigma, \rho'', \gskip \interseq{\rho} c_2 \rightarrow \sigma', \rho'', \gskip \interseq{\rho'} c_2'$
\end{enumerate}

then $\Gamma', \Delta'$ can be constructed from $\sigma', \rho'$ such that $c'$ type-checks under $\Gamma', \Delta'$.

\subsubsection{Proof}
Inductive hypothesis: Preservation holds for sub-forms of any inductive form. Assumptions: 1., 2., 3.

\vspace{1em}
\noindent\case{$c$ is an expression}
    \begin{itemize}
        \item \case{$c$ is a value} $c$ does not step, so preservation vacuously holds.
        \item \case{$c = \gbop{e_1}{e_2}$} For simplicity, we
        ignore the cases in which $\textbf{bop}$ is incompatible with the types
        of $e_1$ and $e_2$. We have three possibilities:
            \begin{enumerate}
                \item $e_1$ is not a value. By assumption, $c$ type-checks under
                $\Gamma, \Delta$ and
                $\sigma, \rho, c \rightarrow \sigma', \rho', \gbop{e_1'}{e_2}$.
                So $\sigma, \rho, e_1 \rightarrow \sigma', \rho', e_1'$. By
                \textsc{check\_bop} $\Gamma, \Delta \vdash e_1 \dashv \Delta_2$.
                By the inductive hypothesis,
                $\Gamma', \Delta' \vdash e_1' \dashv \Delta_2'$. From L3,
                $\sigma' = \sigma$, so $\Gamma' = \Gamma$. If $\Delta' = \Delta$,
                then $e_2$ type-checks under $\Gamma', \Delta_2'$ and we are
                done. If $\Delta' \neq \Delta$, then $\rho' \neq \rho$. So by
                \textsc{L4} $e_1$ was a read $a[n]$. By assumption and
                \textsc{check\_bop} $\Gamma, \Delta \vdash e_1 \dashv \Delta_2$
                and $e_2$ type-checks under $\Gamma, \Delta_2$. Since
                $e_1 = a[n], a$ could not have been defined in $\Delta_2$. $e_1'$
                must be a value, so $\Gamma', \Delta' \vdash v \dashv \Delta'$,
                so $\Delta' = \Delta_2$. So $e_2$ must type-check under $\Gamma', \Delta'$.

                \item $e_1$ is a value $v_1$. Then by assumption,
                $\sigma, \rho, \gbop{v_1}{e_2} \rightarrow \sigma', \rho', \gbop{v_1}{e_2'}$.
                By assumption, $c$ type-checked under $\Gamma, \Delta$, so
                $e_2$ type-checks under $\Gamma, \Delta$ by
                \textsc{check\_bop}. By the inductive hypothesis, $e_2'$
                type-checks under $\Gamma', \Delta'$, and values always
                type-check, so we are done.

                \item Both $e_1$ and $e_2$ are values $v_1$ and $v_2$. By
                assumption, $\sigma, \rho, c \rightarrow \sigma, \rho, v_1 \: \textbf{bop} \: v_2$.
                Values always type-check, so we are done.
            \end{enumerate}
        \item \case{$c = x$} By assumption $x$ type-checks under
        $\Gamma, \Delta$, so $(x \mapsto \tau) \in \Gamma$ and
        $(x \mapsto v) \in \sigma$, and by assumption $\sigma, \rho, x \rightarrow \sigma, \rho, v$.
        Values always type-check, so we are done.

        \item \case{$c = a[e]$} The first possibility is that $e$ is not a
        value. Then by assumption $\sigma, \rho, a[e] \rightarrow \sigma', \rho', a[e']$,
        and so $\sigma, \rho, e \rightarrow \sigma', \rho', e'$. We need to
        show that $a[e']$ type-checks under $\Gamma', \Delta'$. By the
        inductive hypothesis, $e'$ type-checks under $\Gamma', \Delta'$. To
        satisfy the second premise, it should be that $a$ is defined in
        $\Delta'$. By assumption that $a[e]$ type-checked, we know from
        \textsc{check\_read} that $a$ is defined in $\Delta$ and so
        $a \notin \rho$ (by definition of $\sim$). If $a$ was not defined in
        $\Delta'$, that would mean $a \in \rho'$, but by \textsc{L4} this would
        mean that $e$ was a read $a[n]$, meaning $a$ is not defined in
        $\Delta_2$ where $\Gamma, \Delta \vdash e \dashv Delta_2$ and $c$
        could not type-check under $\Gamma, \Delta$ - this is a contradiction.
        So $a$ must be defined in $\Delta'$ and so $a[e']$ must type-check
        under $\Gamma', \Delta'$. The second possibility is that $e$ is a
        value $n$. Then $\sigma, \rho, a[n] \rightarrow \sigma, \rho, v$ and
        since values always type-check we are done.
    \end{itemize}

\vspace{1em}
\noindent\case{$c = \glet{x}{e}$}
The first possibility is that $e$ is not a value. By assumption $c$
type-checks under $\Gamma, \Delta$. By \textsc{check\_let}, so does $e$. By
assumption $\sigma, \rho, \glet{x}{e} \rightarrow \sigma, \rho, \glet{x}{e'}$,
so $\sigma, \rho, e \rightarrow \sigma', \rho', e'$. By the inductive
hypothesis, $e'$ type-checks under $\Gamma', \Delta'$. Then we have that
$\glet{x}{e'}$ type-checks under $\Gamma', \Delta'$, so we are done.
The second possibility is that $e$ is a value $v$. In this case
$\sigma, \rho, \glet{x}{v} \rightarrow \sigma[x \mapsto v], \rho, \gskip$.
$\gskip$ always type-checks, so we are done.

\vspace{1em}
\noindent\case{$c = c_1 \seqcomp c_2$}
By assumption, $\sigma, \rho, c_1 \seqcomp c_2 \rightarrow \sigma, \rho, c_1 \interseq{\rho} c_2$
and $c$ type-checks under $\Gamma, \Delta$. $\sigma', \rho' = \sigma, \rho$,
so $\Gamma' = \Gamma$ and $\Delta' = \Delta$. By assumption
$\Gamma, \Delta \vdash c_1 \dashv \Gamma_2, \Delta_2$ and $c_2$ type-checks
under $\Gamma_2, \Delta$. We need to show that $c_2$ type-checks under
$\Gamma_2, \bar{\rho}$. Since $c_2$ type-checks under $\Gamma_2, \Delta$, it
does not use any memories in $\rho$ by definition of $\sim$. So it must
type-check under $\Gamma_2, \bar{\rho}$.

\vspace{1em}
\noindent\case{$c = c_1 \interseq{\rho''} c_2$}
We have three possibilities.
\begin{itemize}
    \item Neither $c_1$ nor $c_2$ is \gskip. In this case, we
    have $\sigma, \rho, c_1 \interseq{\rho''} c_2 \rightarrow \sigma', \rho', c_1' \interseq{\rho''} c_2$
    and $\sigma, \rho, c_1 \rightarrow \sigma', \rho', c_1'$. From assumption,
    $c$ type-checks under $\Gamma, \Delta$, so 1) $c_2$ type-checks under
    $\bar{\rho''}$ and 2) by the inductive hypothesis, $c_1'$ type-checks
    under the constructed $\Gamma', \Delta'$. We need to show $c_2$
    type-checks under $\Gamma', \bar{\rho''}$. By L2, if $c_2$ type-checks
    under $\Gamma, \bar{\rho''}$, it will type-check under
    $\Gamma', \bar{\rho''}$, so we are done.

    \item $c_1 = \gskip \neq c_2$. We have that
    $\sigma, \rho'', \gskip \interseq{\rho} c_2 \rightarrow \sigma', \rho'', \gskip \interseq{\rho'} c_2'$,
    so $\sigma, \rho, c_2 \rightarrow \sigma', \rho', c_2'$. By the inductive
    hypothesis, $c_2'$ type-checks under the constructed $\Gamma', \Delta'$
    (so it will type-check under $\Gamma', \bar{\rho'}$) and \gskip
    always type-checks, so we are done.

    \item $c_1 = c_2 = \gskip$. This form steps to \gskip,
    which always type-checks, so we are done.
\end{itemize}

\vspace{1em}
\noindent\case{$c = c_1 ; c_2$}
We have two possibilities:
\begin{itemize}
    \item $c_1 \neq \gskip$. By assumption $\sigma, \rho, c \rightarrow \sigma', \rho', c'$,
    so $\sigma, \rho, c_1 \rightarrow \sigma', \rho', c_1'$. We have by
    assumption that $c_1$ type-checks under $\Gamma, \Delta$ to produce
    $\Gamma_2, \Delta_2$, and $c_2$ type-checks under $\Gamma_2, \Delta_2$.
    By the inductive hypothesis, $c_1'$ type-checks under $\Gamma', \Delta'$
    to produce $\Gamma_2', \Delta_2'$. We need to show that $c_2$ type-checks
    under $\Gamma_2', \Delta_2'$. We have two possibilities: $\rho' = \rho$
    and $\rho' \neq \rho$. Consider the first possibility. We would have
    $\Delta = \Delta'$, so $\Delta_2 = \Delta_2'$. By L2, $c_2$ type-checks
    under $\Gamma', \Delta_2'$ as needed. With the second possibility, it can
    only be that $\rho \subset \rho'$. There are then only two cases to consider:
    \begin{enumerate}
        \item $c_1$ contained a read or write involving $a[n]$ and $c_1'$ is a
        value $v$. Then $\rho' = \rho \cup \{a\}$ and $a$ is not defined in
        $\Delta'$. By \textsc{check\_write} and \textsc{check\_read}, $a$
        could not have been defined in $\Delta_2$.
        Since $\Gamma', \Delta' \vdash v \dashv \Delta'$, $\Delta_2 = \Delta'$.
        So $c_2$ must type-check under $\Gamma', \Delta'$.

        \item $c_1 = \gskip \interseq{\rho''} \gskip$. By
        \textsc{inter\_seq3} $c_1' = \gskip$ and $\sigma' = \sigma$,
        so $\Gamma' = \Gamma$. By assumption $c_2$ type-checks under
        $\Gamma_2, \Delta_2$ where $\Gamma, \Delta \vdash c_1 \dashv \Gamma_2, \Delta_2$.
        By \textsc{check\_inter\_ seq\_comp} $\Gamma_2 = \Gamma$.
        \newline We need to show $c_2$ type-checks under $\Gamma_2', \Delta_2' = \Gamma', \Delta' = \Gamma, \Delta'$
        (since $\Gamma, \Delta \vdash \gskip \dashv \Gamma, \Delta$).
        For this to be the case, $c_2$ cannot use any memories in $\rho$ or
        $\rho''$ (by definition of construction).
        \newline 1) Because $\Gamma, \Delta \sim \sigma, \rho$ and $c_1$
        type-checks under $\Gamma, \Delta$, $c_1$ does not use any memories
        in $\rho$. Then by assumption that $c_2$ type-checks under
        $\Gamma_2, \Delta_2$ and by \textsc{check\_par\_comp}, $c_2$ also
        cannot use any memories in $\rho$.
        \newline 2) By assumption $c_1$ type-checks under $\Gamma, \Delta$ to
        produce $\Gamma_2, \Delta_2$. By \textsc{check\_inter\_seq\_comp} and
        \textsc{small\_seq} $\Delta_2 \subseteq \bar{\rho''}$, and by
        assumption $c_2$ type-checks under $\Gamma_2, \Delta_2$, so $c_2$ does
        not use any memories in $\rho''$. So $c_2$ type-checks under
        $\Gamma', \Delta'$ as needed.
    \end{enumerate}

    \item $c_1 = \gskip$. By assumption $\gskip ; c_2$
    type-checks under $\Gamma, \Delta$, so $c_2$ type-checks under
    $\Gamma, \Delta$. $\sigma, \rho, \gskip ; c_2 \rightarrow \sigma, \rho, c_2$
    so $\Gamma' = \Gamma$ and $\Delta' = \Delta$. Then we need to show
    $c_2$ type-checks under $\Gamma', \Delta' = \Gamma, \Delta$, which we
    have from assumption, so we are done.
\end{itemize}

\vspace{1em}
\noindent\case{$c = \gif{x}{c_1}{c_2}$}
By assumption $c$ type-checks under $\Gamma, \Delta$, so $c_1$ and $c_2$ both
type-check under $\Gamma, \Delta$, and $x$ is either \textbf{true} or
\textbf{false} by \textsc{check\_if}. If true, $\sigma, \rho, c \rightarrow \sigma, \rho, c_1$.
If false, $\sigma, \rho, c \rightarrow \sigma, \rho, c_2$. We need to show
that $c_1$ and $c_2$ type-check under $\Gamma, \Delta$ ($\sigma', \rho' = \sigma, \rho$,
so $\Gamma', \Delta' = \Gamma, \Delta$). This is given by assumption so we are done.

\vspace{1em}
\noindent\case{$c = \gwhile{x}{c_1}$}
By assumption $c$ type-checks under $\Gamma, \Delta$,
so by \textsc{check\_while} $x$ type-checks to bool and $c_1$ type-checks
under $\Gamma, \Delta$ to produce $\Gamma_2, \Delta_2$. We need to show that
$\gif{x}{(c_1 \seqcomp \gwhile{x}{c_1})}{\gskip}$
type-checks under $\Gamma, \Delta$ ($\sigma', \rho' = \sigma, \rho$, so
$\Gamma', \Delta' = \Gamma, \Delta$). For this, it should be the case that $x$
type-checks to bool. This is already given. It should also be the case that
$(c_1 \seqcomp \gwhile{x}{c_1})$ type-checks under $\Gamma, \Delta$.
This requires that $c_1$ type-checks under $\Gamma, \Delta$, which is given by
assumption, and that $\gwhile{x}{c_1}$ type-checks under
$\Gamma_2, \Delta$. By L2 if $\gwhile{x}{c_1}$ type-checks under
$\Gamma, \Delta$ (which it does by assumption), it type-checks under
$\Gamma_2, \Delta$, so we are done.

\vspace{1em}
\noindent\case{$c = x := e$}
By assumption $c$ type-checks under $\Gamma, \Delta$, so $(x \mapsto \tau) \in \Gamma$
and $e$ type-checks under $\Gamma, \Delta$ producing $\Delta_2$. We have two
possibilities:
\begin{itemize}
    \item $e$ is not a value. By assumption and \textsc{check\_update}
    $\sigma, \rho, x := e \rightarrow \sigma', \rho', x := e'$ and
    $\sigma, \rho, e \rightarrow \sigma', \rho', e'$. By the inductive
    hypothesis, $e'$ type-checks under the constructed $\Gamma', \Delta'$.
    Then by L1 and L2, if $\Gamma, \Delta \vdash e : \tau \dashv \Delta_2$
    then $\Gamma', \Delta' \vdash e' : \tau ; \dashv \Delta_2'$. So
    $(x \mapsto \tau) \in \Gamma'$ and $c'$ type-checks under
    $\Gamma', \Delta'$ as needed.

    \item $e$ is a value $v$. By assumption $x := v$ type-checks under
    $\Gamma, \Delta$ so $\Gamma, \Delta \vdash v : \tau \dashv \Delta$ and
    ($x \mapsto \tau) \in \Gamma$. $\sigma, \rho, x := v \rightarrow \sigma[x \mapsto v], \rho, \gskip$,
    and $\gskip$ always type-checks, so we are done.
\end{itemize}

\vspace{1em}
\noindent\case{$c = a[e_1] := e_2$}
By assumption $c$ type-checks under $\Gamma, \Delta$, so $e_1$ type-checks
under $\Gamma, \Delta$ producing $\Delta_2$ and $e_2$ type-checks under
$\Gamma, \Delta_2$ by \textsc{check\_write}. Additionally, $a$ is defined in
$\Delta$ and $\Delta_2$ so neither $e_1$ nor $e_2$ use memory $a$. We then
have three possibilities:
\begin{itemize}
    \item Neither $e_1$ nor $e_2$ is a value. Then
    $\sigma, \rho, a[e_1] := e_2 \rightarrow \sigma', \rho', a[e_1'] := e_2$
    and $\sigma, \rho, e_1 \rightarrow \sigma', \rho', e_1'$. By the inductive
    hypothesis, $e_1'$ type-checks under the constructed $\Gamma', \Delta'$ to
    produce $\Delta_2'$. Either $\rho' = \rho$ or not. If $\rho' = \rho$, then
    by \textsc{L2}, $e_2$ will type-check under $\Gamma', \Delta_2'$ since
    $\Gamma \subseteq \Gamma'$ and $\Delta' = \Delta$. If not, then by L4
    $e_1$ is a read $a_1[n]$. By assumption and \textsc{check\_write}
    $\Gamma, \Delta \vdash e_1 \dashv \Delta_2$ and $e_2$ type-checks under
    $\Gamma, \Delta_2$. Since $e_1 = a_1[n], a_1$ could not have been defined
    in $\Delta_2$. $e_1'$ then must be a value, so
    $\Gamma', \Delta' \vdash v \dashv \Delta'$, so $\Delta_2 = \Delta'$. So
    $e_2$ must type-check under $\Gamma', \Delta'$.

    \item $e_1$ is a value $n$ and $e_2$ is not a value. Then
    $\sigma, \rho, a[e_1] := e_2 \rightarrow \sigma', \rho', a[e_1] := e_2'$
    and $\sigma, \rho, e_2 \rightarrow \sigma', \rho', e_2'$. By the inductive
    hypothesis, $e_2$ type-checks under $\Gamma', \Delta'$. $e_1$ is a value
    and always type-checks, so we are done.

    \item $e_1$ is a value $n$ and $e_2$ is a value $v$. Assuming this
    type-checks, we step to \gskip, which always type-checks, so we are
    done.
\end{itemize}

\section{\dseKernel Design Space Exploration}
\begin{figure*}
\begin{lstlisting}[language=C,escapeinside={<@}{@>},numbers=none]
ap_int<32> m1[128][128];
#pragma HLS resource variable=m1 core=RAM_1P_BRAM
#pragma HLS ARRAY_PARTITION variable=m1 cyclic factor=<@::\dseParam{BANK11}::@> dim=1
#pragma HLS ARRAY_PARTITION variable=m1 cyclic factor=<@::\dseParam{BANK12}::@> dim=2
ap_int<32> m2[128][128];
#pragma HLS resource variable=m2 core=RAM_1P_BRAM
#pragma HLS ARRAY_PARTITION variable=m2 cyclic factor=<@::\dseParam{BANK11}::@> dim=1
#pragma HLS ARRAY_PARTITION variable=m2 cyclic factor=<@::\dseParam{BANK12}::@> dim=2
ap_int<32> prod[128][128];
#pragma HLS resource variable=prod core=RAM_1P_BRAM
#pragma HLS ARRAY_PARTITION variable=prod cyclic factor=<@::\dseParam{BANK21}::@> dim=1
#pragma HLS ARRAY_PARTITION variable=prod cyclic factor=<@::\dseParam{BANK22}::@> dim=2

for (int jj = 0; jj < 16; jj++) {
  for (int kk = 0; kk < 16; kk++) {
    for (int i = 0; i < 128; i++) {
      #pragma HLS UNROLL factor=<@::\dseParam{UNROLL1}::@> skip_exit_check
      for (int j = 0; j < 8; j++) {
        #pragma HLS UNROLL factor=<@::\dseParam{UNROLL2}::@> skip_exit_check
        for (int k = 0; k < 8; k++) {
          #pragma HLS UNROLL factor=<@::\dseParam{UNROLL3}::@> skip_exit_check
          ap_int<32> mul = m1[i][8 * kk + k] * m2[8 * kk + k][8 * jj + j];
          prod[i][8 * jj + j] += mul;
        }
      }
    }
  }
}
\end{lstlisting}
\caption{\dseKernel kernel used for exhaustive DSE. The highlighted tokens indicate parameters for exploration.}
\label{fig:lst:dse-kernel}
\end{figure*}

\Cref{fig:lst:dse-kernel} lists the parameterized code for the \dseKernel
exhaustive design space exploration discussed in the evaluation. The code is
adapted from MachSuite~\cite{machsuite}.

\section{MachSuite Ports}
\begin{figure*}
  \centering
  \begin{subfigure}{0.32\linewidth}
    \centering
    \includegraphics[width=\linewidth]{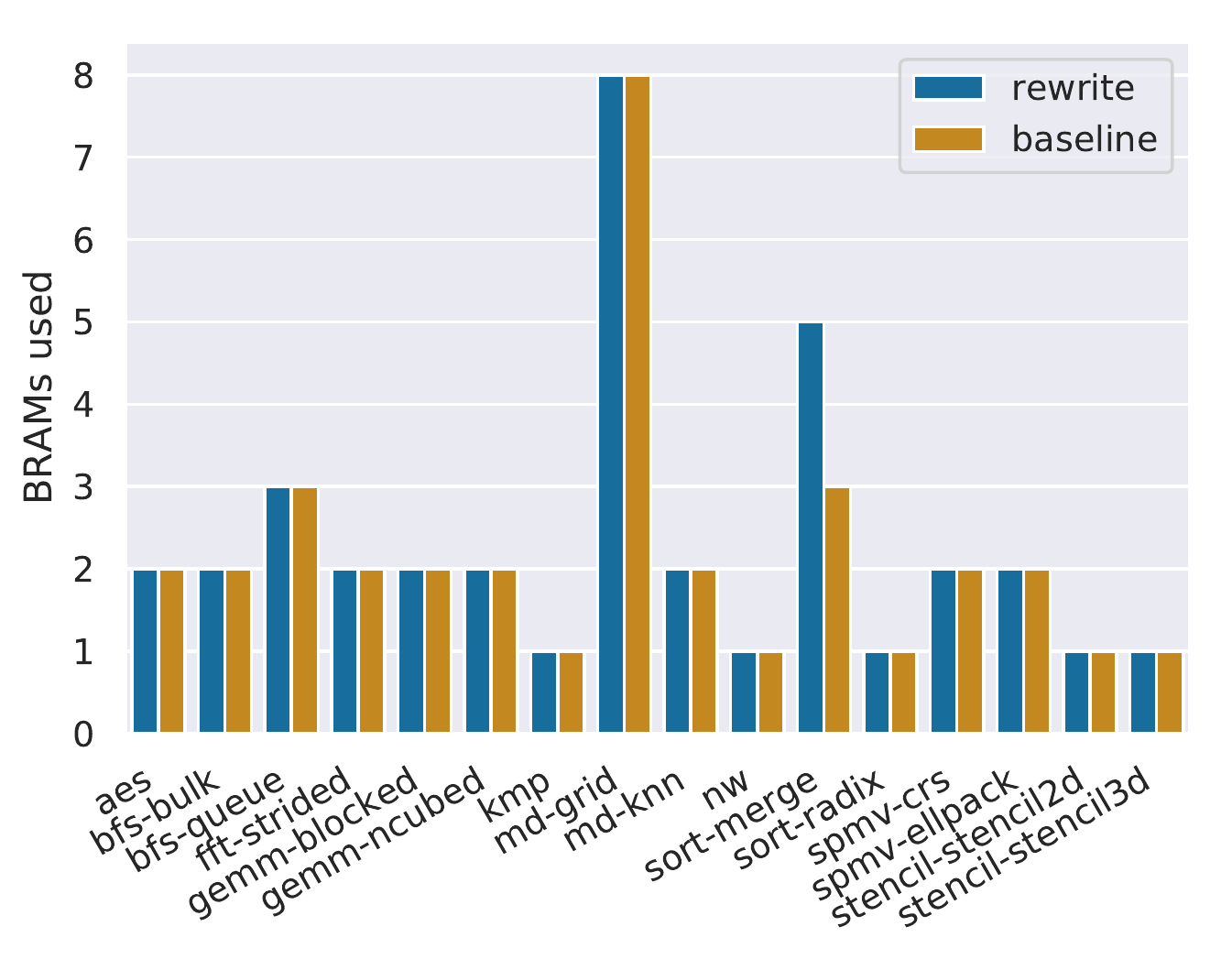}
    \caption{Comparison of BRAMs used.}
    \label{fig:machsuite-rewrites:bram}
  \end{subfigure}
  \begin{subfigure}{0.32\linewidth}
    \centering
    \includegraphics[width=\linewidth]{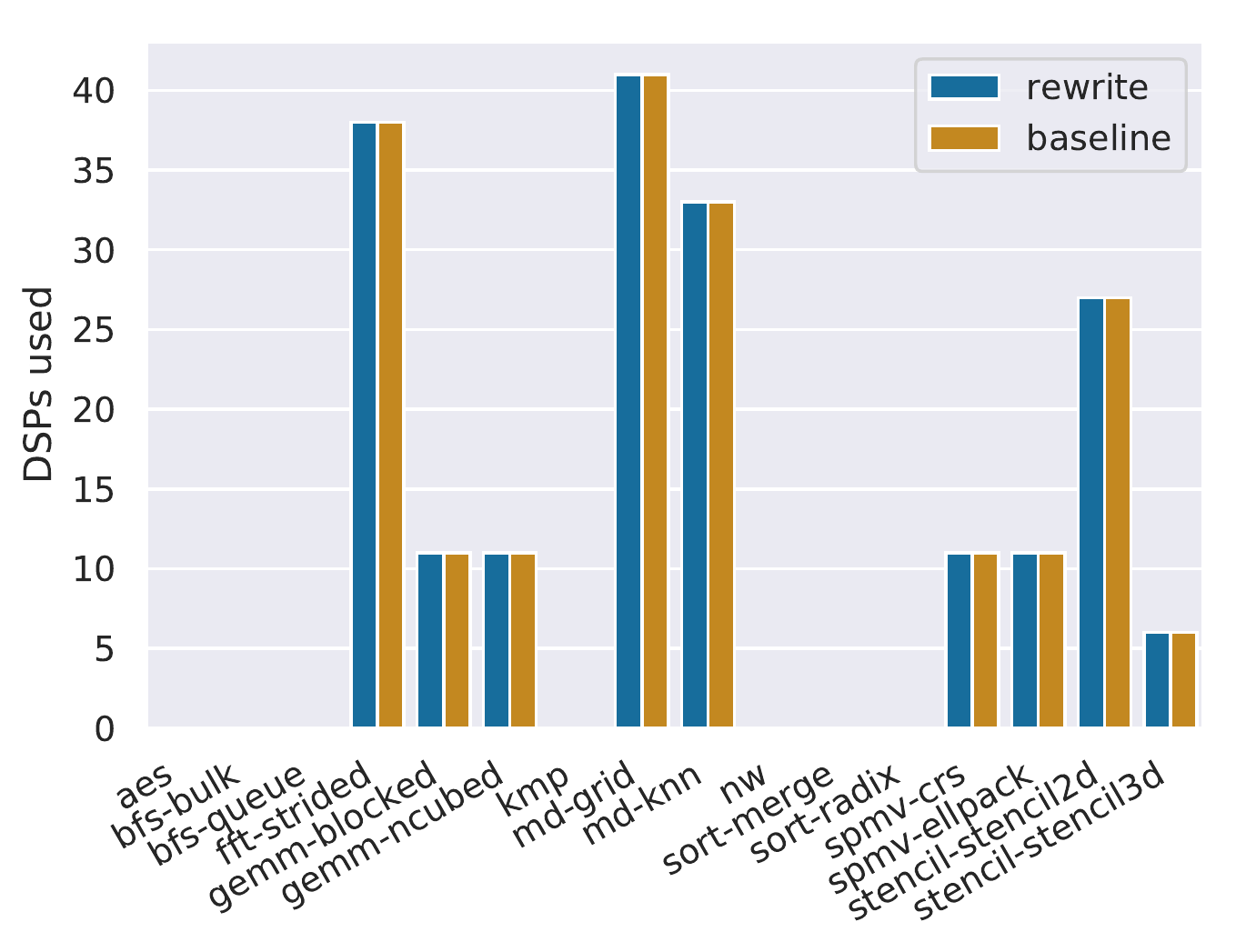}
    \caption{Comparison of DSPs used.}
    \label{fig:machsuite-rewrites:dsp}
  \end{subfigure}
  \begin{subfigure}{0.32\linewidth}
    \centering
    \includegraphics[width=\linewidth]{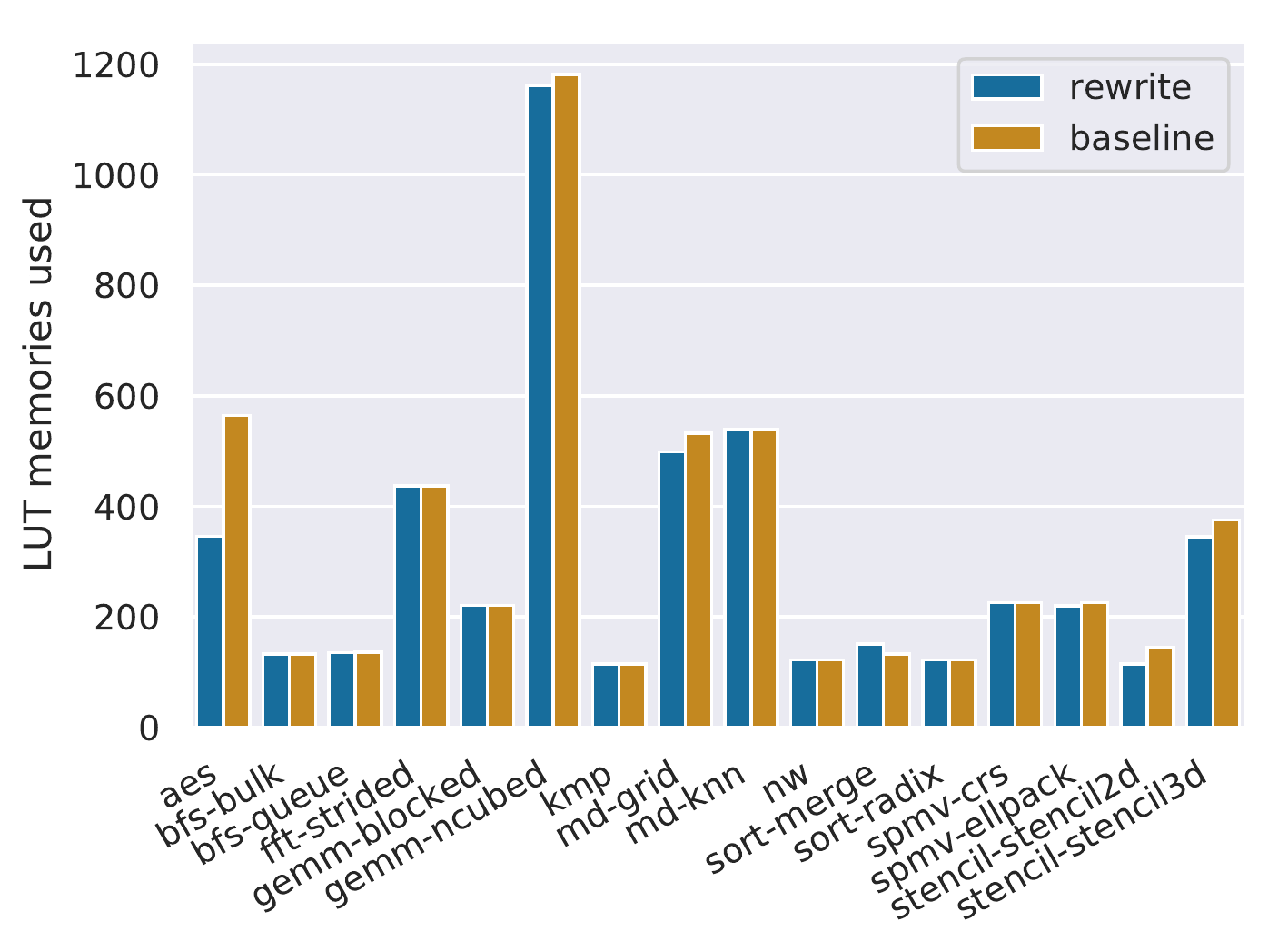}
    \caption{Comparison of LUT Mems used.}
    \label{fig:machsuite-rewrites:lutmem}
  \end{subfigure}
  \begin{subfigure}{0.32\linewidth}
    \centering
    \includegraphics[width=\linewidth]{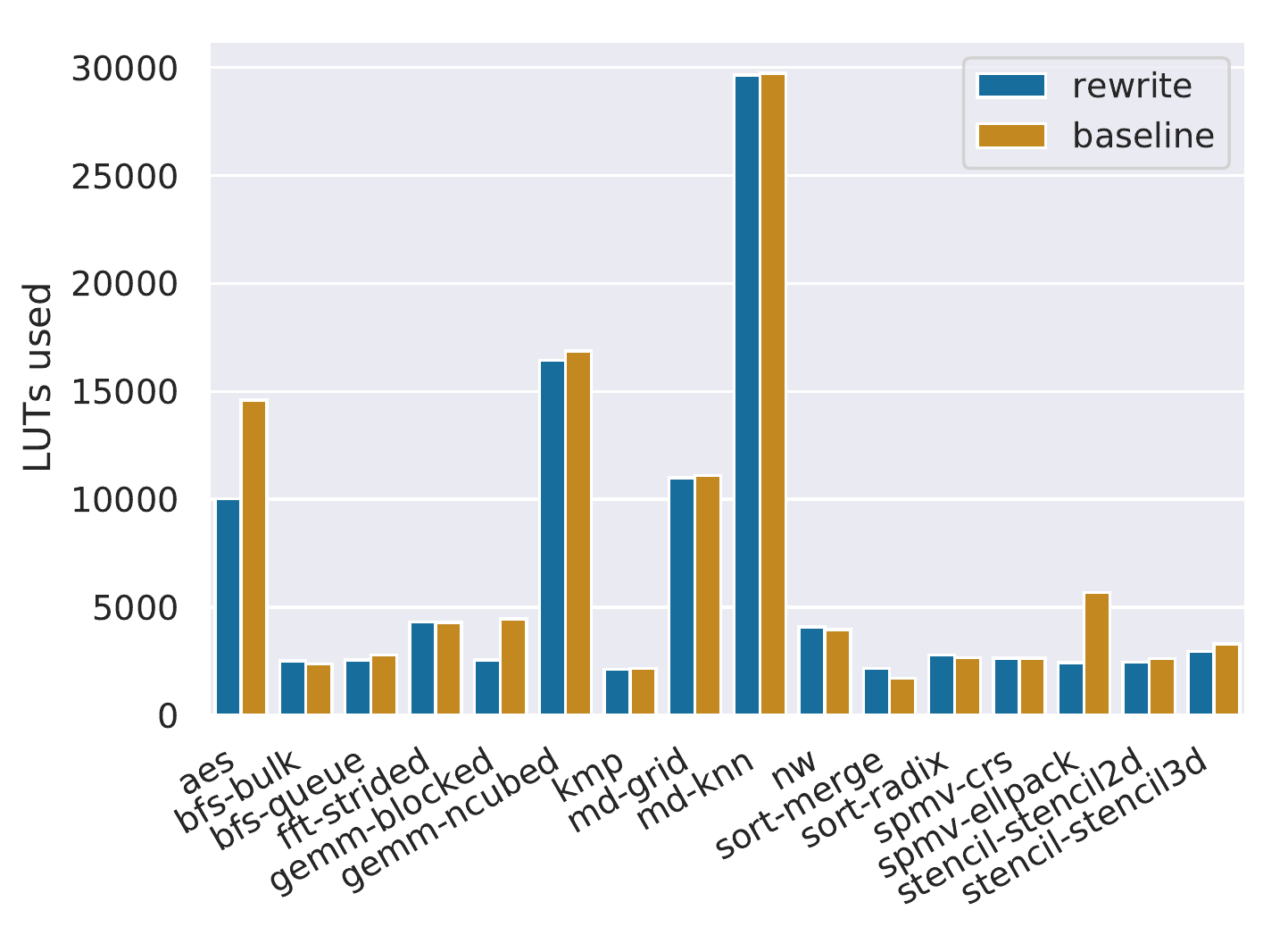}
    \caption{Comparison of LUTs used.}
    \label{fig:machsuite-rewrites:lut}
  \end{subfigure}
  \begin{subfigure}{0.32\linewidth}
    \centering
    \includegraphics[width=\linewidth]{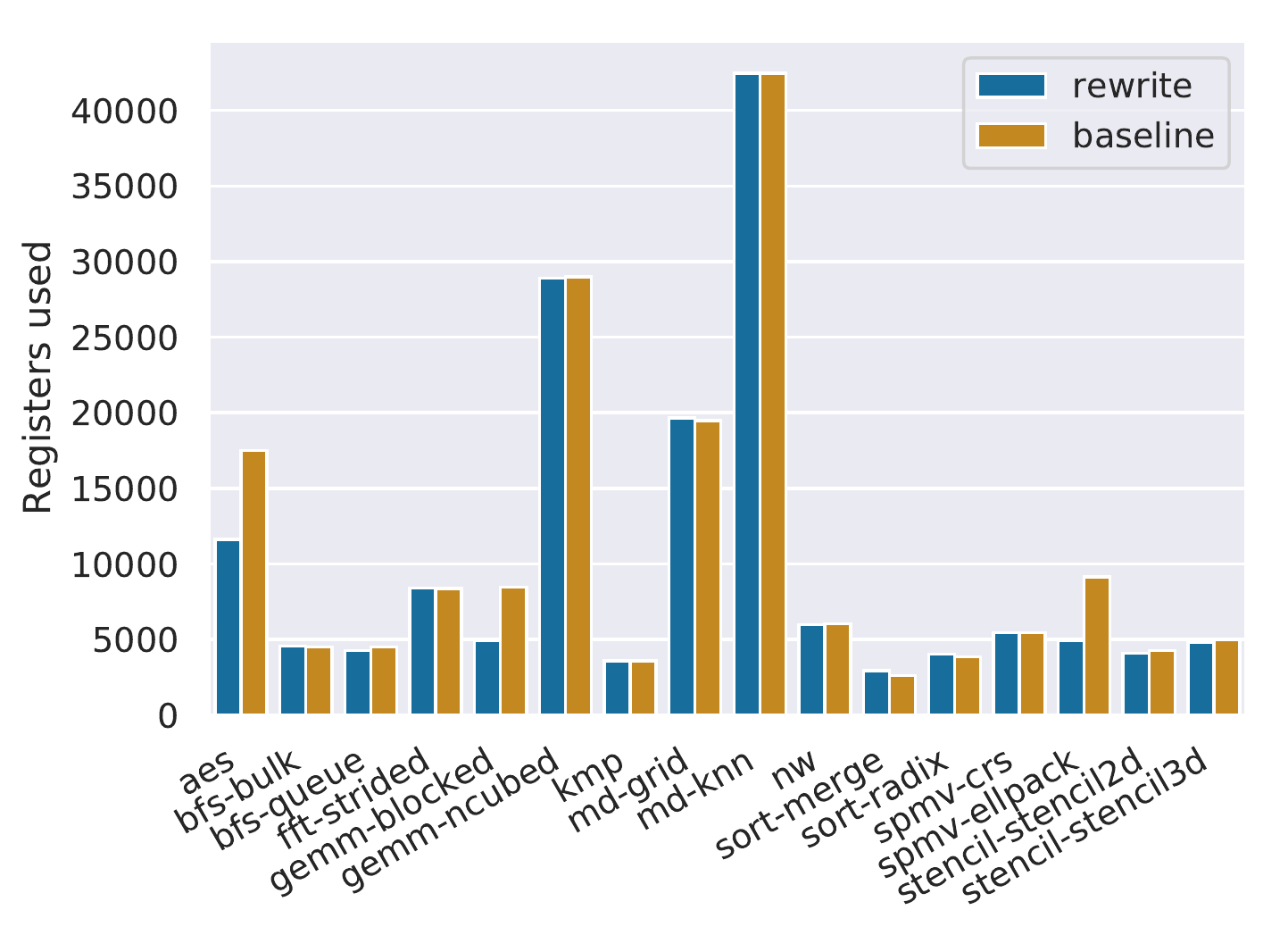}
    \caption{Comparison of Registers used.}
    \label{fig:machsuite-rewrites:reg}
  \end{subfigure}
  \begin{subfigure}{0.32\linewidth}
    \centering
    \includegraphics[width=\linewidth]{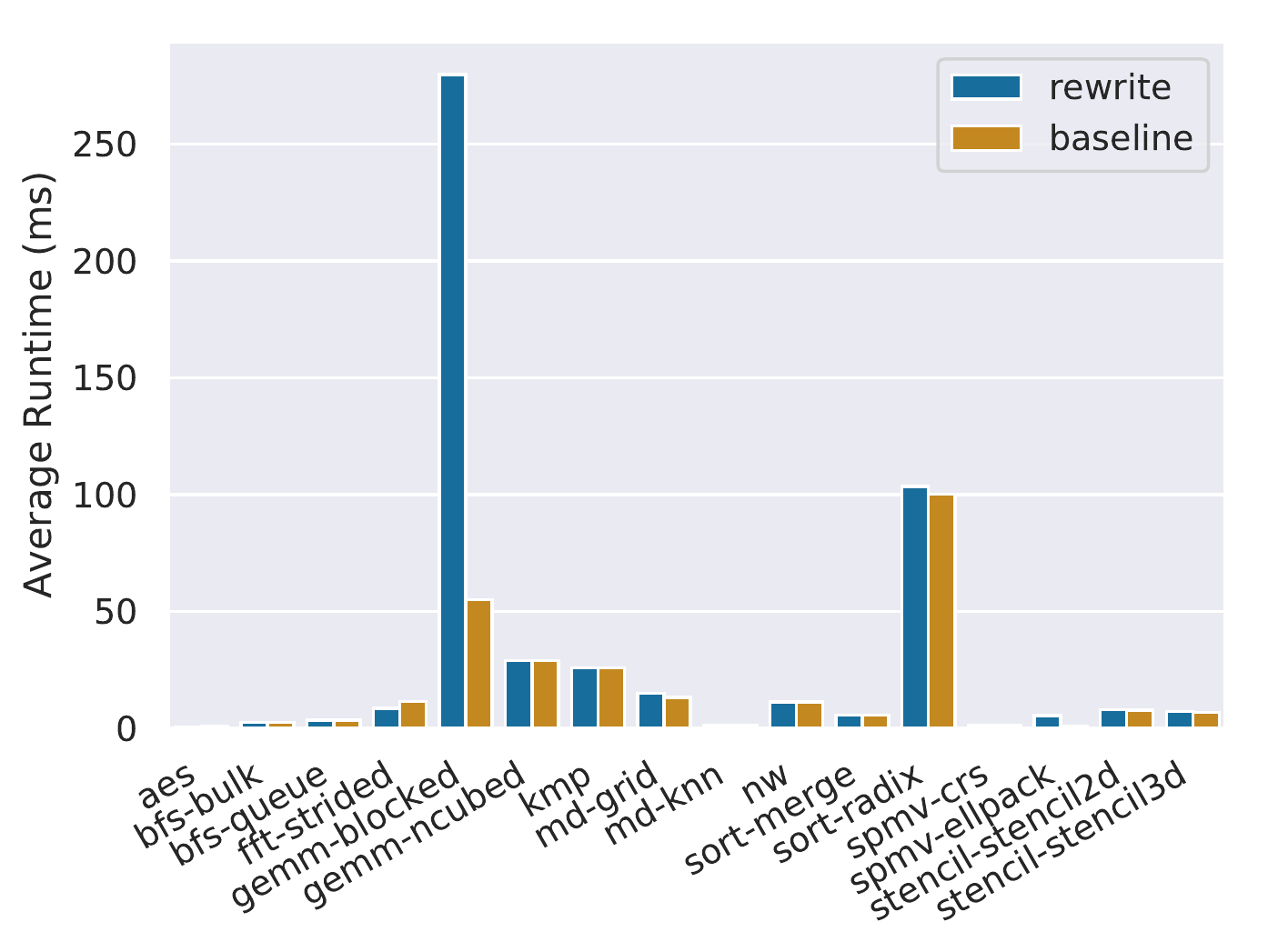}
    \caption{Comparison of runtime averages.}
    \label{fig:machsuite-rewrites:runtime}
  \end{subfigure}
  \caption{Resource usage comparison between baseline MachSuite implementations
    in Vivado HLS and rewrite implementations in Dahlia. The kernel names are on
    the X-axis and the resource usage is on the Y-axis. Excludes \bench{backprop} (functionally incorrect), \bench{fft-transpose}, and \bench{viterbi} (mis-synthesized by Vivado HLS)}
  \label{fig:spatial}
\end{figure*}

We ported 16 Machsuite benchmarks to Dahlia and compared their resource usage
against baseline implementations after a full synthesis flow targeting Xilinx's
UltraScale+ VU9P for both rewritten and baseline implementations. We present the
comparison in
\cref{fig:machsuite-rewrites:bram,fig:machsuite-rewrites:dsp,fig:machsuite-rewrites:lutmem,fig:machsuite-rewrites:lut,fig:machsuite-rewrites:reg,fig:machsuite-rewrites:runtime}.
The benchmarks highlighted in red represent benchmarks the completed synthesis
but failed their correctness checks because of a miscompilation from the Vivado
toolchain.

The graphs show that most of the benchmarks perform identically when rewritten
in Dahlia. This is because Dahlia generates C++ which goes through the same
synthesis flow as the baseline implementations.

\section{Spatial Experiments}

\begin{figure*}
  \begin{lstlisting}[language=scala,escapeinside={<@}{@>},numbers=none]
  @spatial object GEMM_NCubed_16 extends SpatialApp {
    type T = FixPt[TRUE,_16,_16]
    def main(args: Array[String]): Unit = {
      val (a_dram, b_data, c_data) = (DRAM[T](128,128), DRAM[T](128,128), DRAM[T](128,128))
      // Generate random data for the input matrices.
      val (a_data, b_data) = ((0::dim,0::dim){(i,j) => random[T](5)}, (0::dim,0::dim){(i,j) => random[T](5)})
      // Set data in the input matrices.
      setMem(a_dram, a_data)
      setMem(b_dram, b_data)
      // Accelerator kernel.
      Accel {
        val (a_sram, b_sram, c_sram) = (SRAM[T](dim,dim), SRAM[T](dim,dim), SRAM[T](dim,dim))
        // Load data into memories.
        a_sram.load(a_dram)
        b_sram.load(b_dram)
        // Computation loop.
        Foreach(dim by 1) { i => Foreach(dim by 1) { j =>
            // DSE parameter: Innermost loop parallelism. Try values from [1, 16].
            val sum = Reduce(Reg[T](0))(dim by 1 par <@::\dseParam{UNROLL}::@>) { k => a_sram(i,k) * b_sram(k,j) }{_+_}
            c_sram(i,j) = sum
        }}
        c_dram store c_sram
      }
      // Compute the expected value of the computation.
      val c_gold = (0::dim,0::dim){(i,j) => Array.tabulate(dim){k => a_data(i,k) * b_data(k,j)}.reduce{_+_}}
      // Check that the computed values are within some range of the expected value.
      val c_result = getMatrix(c_dram)
      val cksum = c_gold.zip(c_result){(a,b) => abs(a-b) < 0.5.to[T]}.reduce{_&&_}
      // Fail if the computed value is different.
      assert(cksum)
    }
  }
  \end{lstlisting}
\label{fig:list:spatial-dse}
\caption{Kernel used to collect resource utilization number for the Spatial
evaluation.}
\end{figure*}

\begin{figure*}
  \centering
  \begin{subfigure}{0.32\linewidth}
    \centering
    \includegraphics[width=\linewidth]{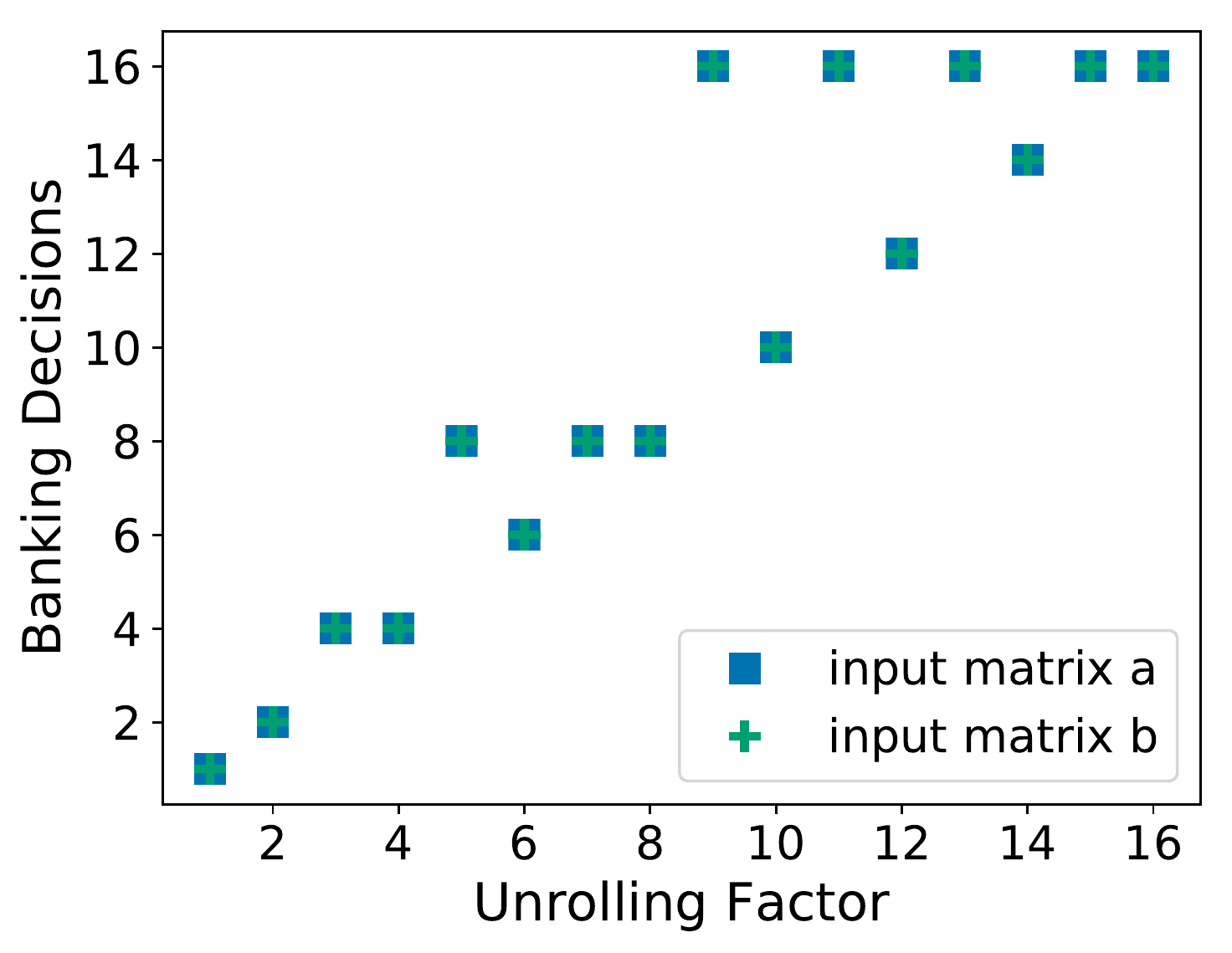}
    \caption{Banking decision Spatial made given the unrolling factors.}
    \label{fig:spatial:banking}
  \end{subfigure}
  \begin{subfigure}{0.32\linewidth}
    \centering
    \includegraphics[width=\linewidth]{fig/paper-normalized.pdf}
    \caption{Resource usage normalized against no unrolling.}
    \label{fig:spatial:normalized}
  \end{subfigure}
  \begin{subfigure}{0.32\linewidth}
    \centering
    \includegraphics[width=\linewidth]{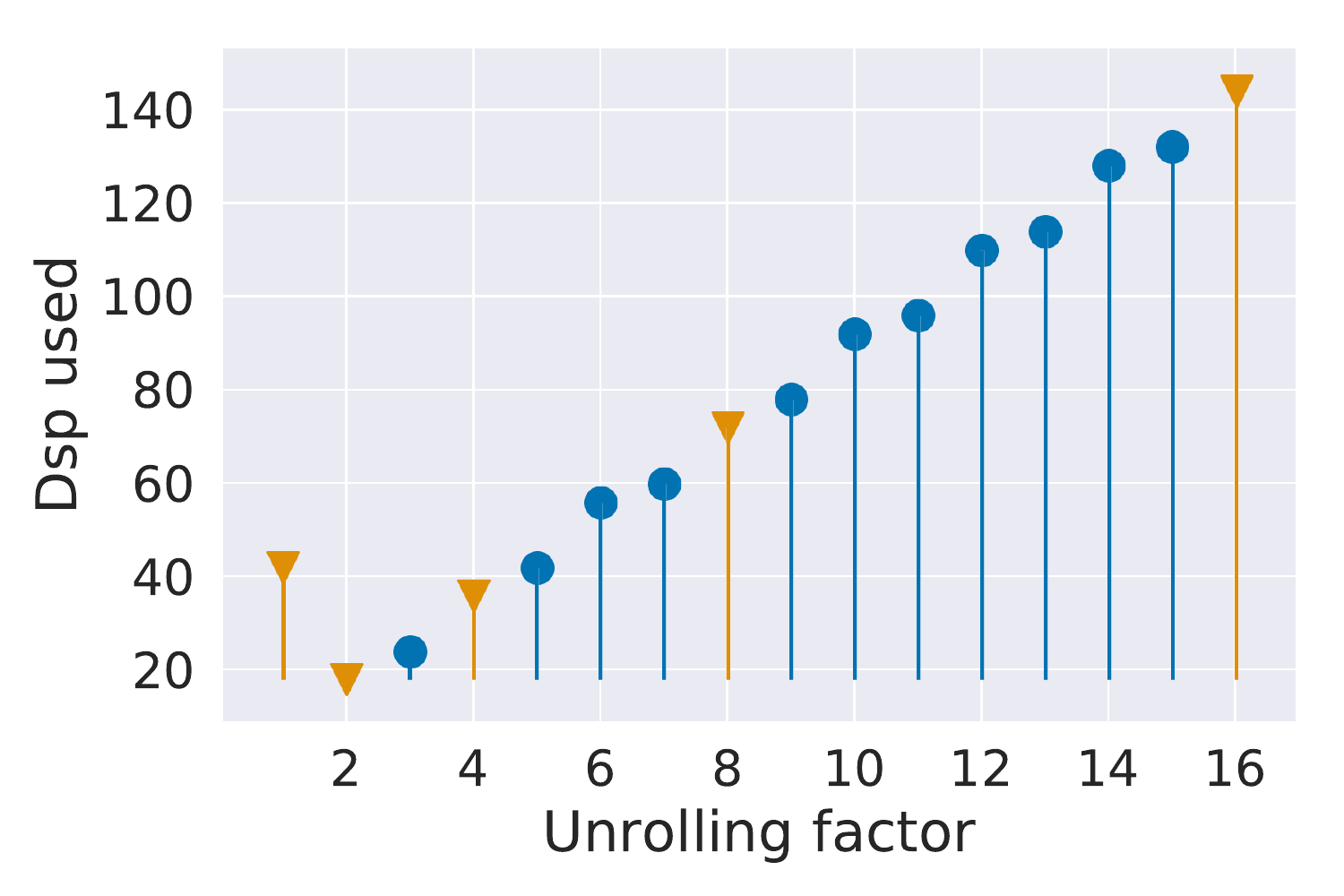}
    \caption{Absolute DSP usage. Predictable points highlighted.}
    \label{fig:spatial:absolute-dsp}
  \end{subfigure}
  \begin{subfigure}{0.32\linewidth}
    \centering
    \includegraphics[width=\linewidth]{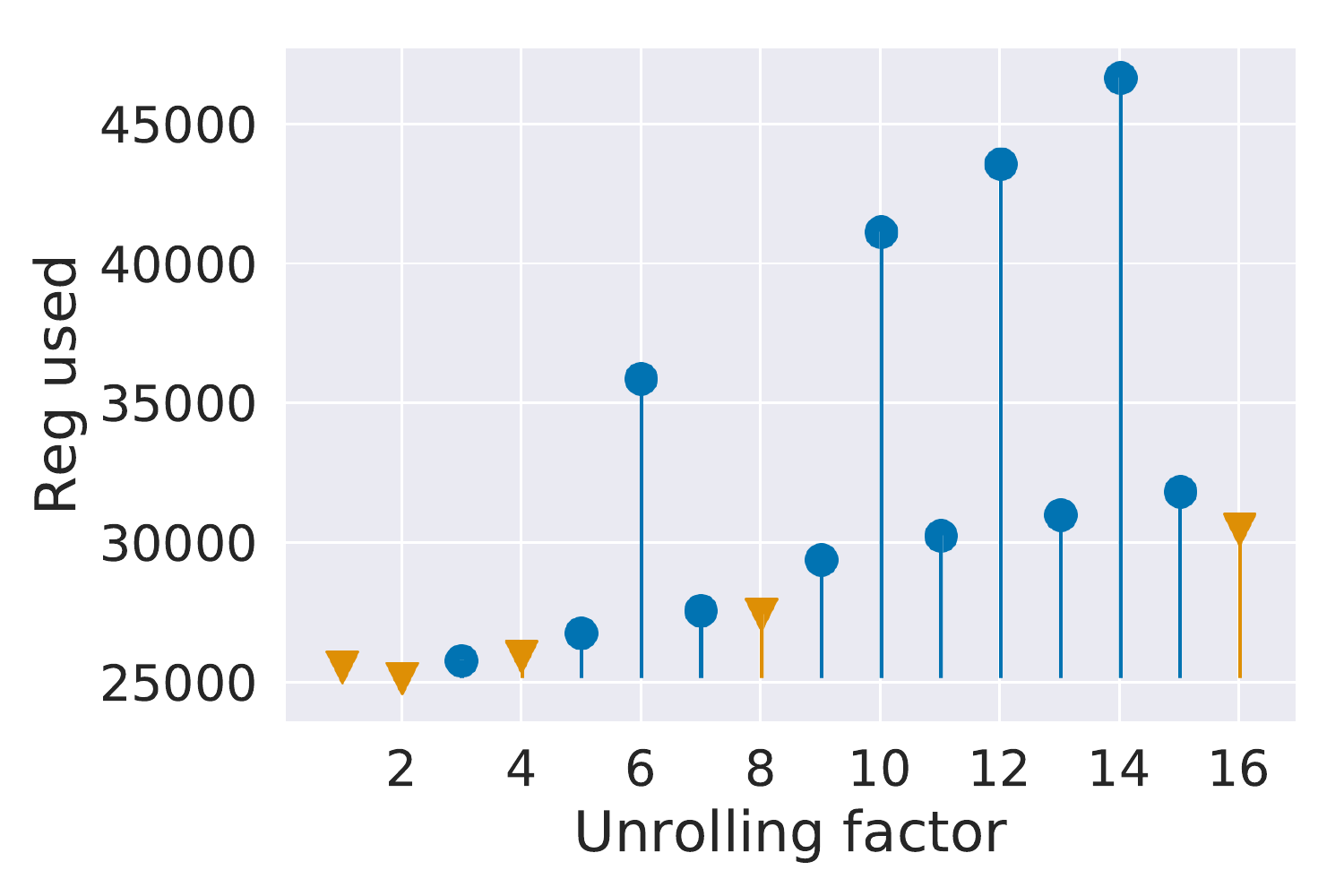}
    \caption{Absolute REG usage. Predictable points highlighted.}
    \label{fig:spatial:absolute-reg}
  \end{subfigure}
  \begin{subfigure}{0.32\linewidth}
    \centering
    \includegraphics[width=\linewidth]{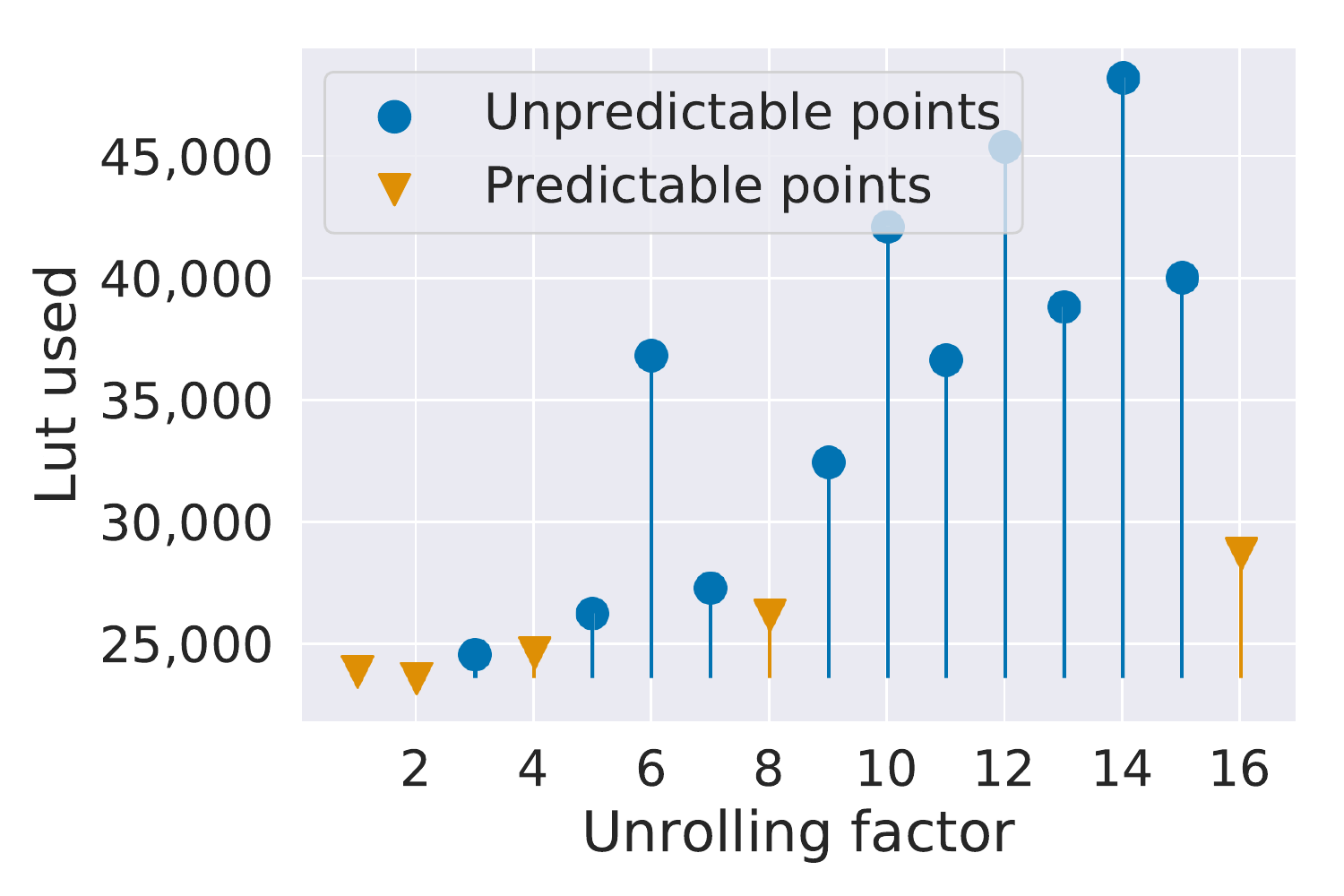}
    \caption{Absolute LUT usage. Predictable points highlighted.}
    \label{fig:spatial:absolute-lut}
  \end{subfigure}
  \begin{subfigure}{0.32\linewidth}
    \centering
    \includegraphics[width=\linewidth]{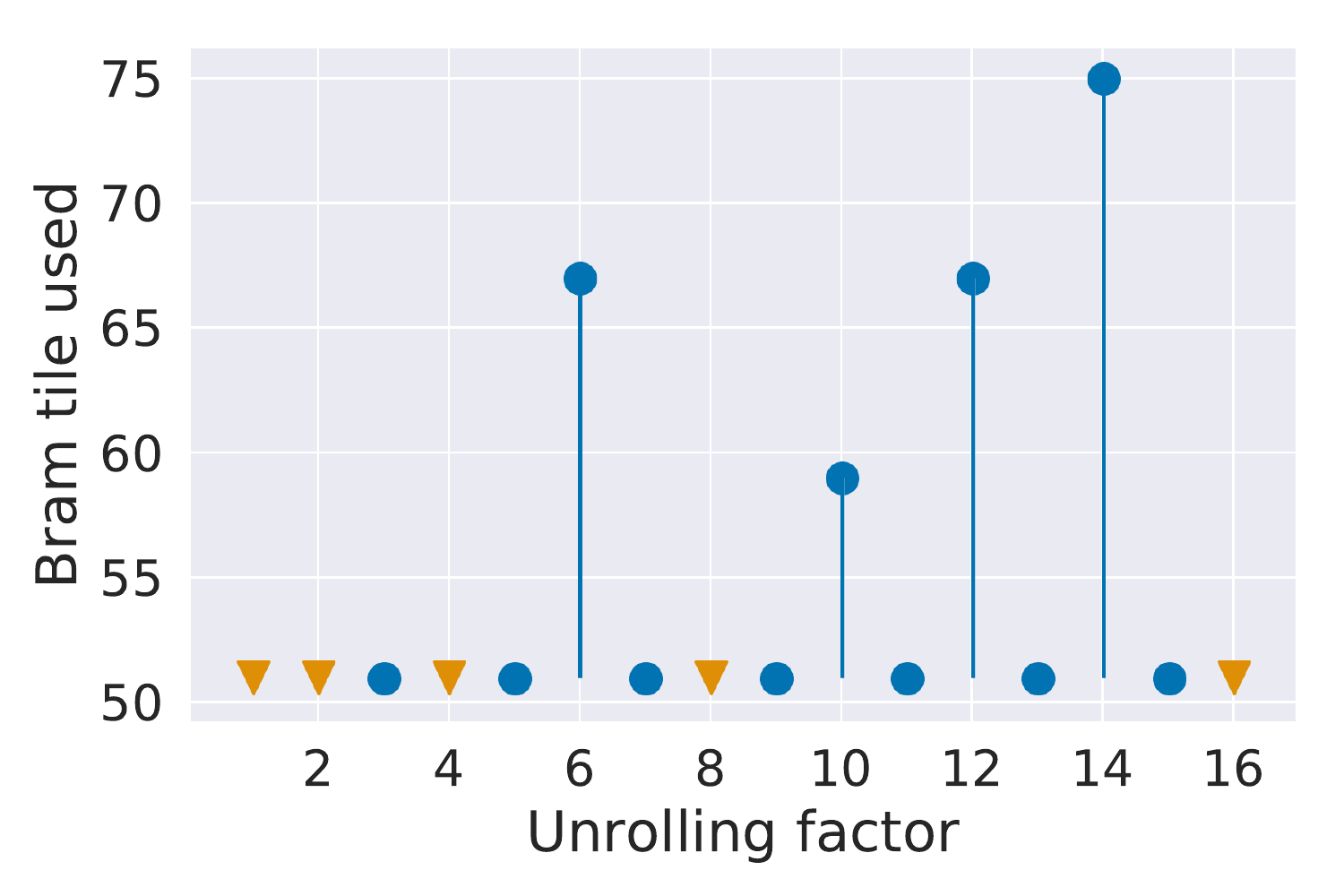}
    \caption{Absolute BRAM usage. Predictable points highlighted.}
    \label{fig:spatial:absolute-bram}
  \end{subfigure}
  \caption{Resource utilization for \bench{gemm-ncubed} design in Spatial on a Zynq-7000. Absolute resource utilization shows extreme variation between adjacent design points.}
  \label{fig:spatial}
\end{figure*}

We perform a simple design sweep over a general matrix multiply kernel written
in Spatial and vary the unrolling factor from $1$ to $16$.

We run the generated designs through a full synthesis flow by targeting
Xilinx's Zynq-7000 SoC~\cite{zynq}. We extended the Spatial quick-start template
to generate designs for the DSE comparison~\cite{dahlia-spatial}. We were
unable to get Spatial designs to pass through the AWS F1 based synthesis
flow due to numerous issues~\cite{spatial-error-1,spatial-error-2,spatial-error-3}.

\Cref{fig:spatial:banking} shows the banking factors inferred by Spatial
for a given unrolling factor. \cref{fig:spatial:normalized} plots the
resources usages normalized against the spatial design with no unrolling.
When Spatial's inferred banking decisions are not aligned with
the unrolling factor, the resource usages abruptly increase.

We also plot the absolute LUT, DSP, REG and BRAM usage against unrolling factor
in
\Cref{fig:spatial:absolute-lut,fig:spatial:absolute-dsp,fig:spatial:absolute-reg,fig:spatial:absolute-bram}.
The designs use significantly fewer LUTs when
the unrolling factor is a factor of the size of the memory.
Furthermore, Spatial designs use up to $10\times$ more LUTs and $2\times$ more
DSPs than the equivalent designs generated by \sys's Vivado HLS backend.

Spatial uses automated design space exploration tools to find optimal
parameters for the design. \sys's type system can be used to eliminate
unpredictable design points and drastically reduce the search spaces with
such automated tools.

\end{document}

%% file: format-hacks.tex
\newcommand{\case}[1]{\underline{Case: #1}.} 

\newcommand{\interseq}[1]{\overset{#1}{\sim}} 

\newcommand{\seqcomp}{~\rule[0.5ex]{1.5em}{0.55pt}~}

\newcommand{\glet}[2]{\textbf{let}~#1 = #2}
\newcommand{\gskip}{\textbf{skip}}
\newcommand{\gif}[3]{\textbf{if}~#1~#2~#3}
\newcommand{\gtrue}{\textbf{true}}
\newcommand{\gfalse}{\textbf{false}}
\newcommand{\gwhile}[2]{\textbf{while}~#1~#2}
\newcommand{\gbop}[2]{\textbf{bop}~#1~#2}

%% file: semantics.tex
\newcommand{\ottdrule}[4][]{{\displaystyle\frac{\begin{array}{l}#2\end{array}}{#3}\quad\ottdrulename{#4}}}
\newcommand{\ottusedrule}[1]{\[#1\]}
\newcommand{\ottpremise}[1]{ #1 \\}
\newenvironment{ottdefnblock}[3][]{ \framebox{\mbox{#2}} \quad #3 \\[0pt]}{}
\newenvironment{ottfundefnblock}[3][]{ \framebox{\mbox{#2}} \quad #3 \\[0pt]\begin{displaymath}\begin{array}{l}}{\end{array}\end{displaymath}}
\newcommand{\ottfunclause}[2]{ #1 \equiv #2 \\}
\newcommand{\ottnt}[1]{\mathit{#1}}
\newcommand{\ottmv}[1]{\mathit{#1}}
\newcommand{\ottkw}[1]{\mathbf{#1}}
\newcommand{\ottsym}[1]{#1}
\newcommand{\ottcom}[1]{\text{#1}}
\newcommand{\ottdrulename}[1]{\textsc{#1}}
\newcommand{\ottcomplu}[5]{\overline{#1}^{\,#2\in #3 #4 #5}}
\newcommand{\ottcompu}[3]{\overline{#1}^{\,#2<#3}}
\newcommand{\ottcomp}[2]{\overline{#1}^{\,#2}}
\newcommand{\ottgrammartabular}[1]{\begin{supertabular}{llcllllll}#1\end{supertabular}}
\newcommand{\ottmetavartabular}[1]{\begin{supertabular}{ll}#1\end{supertabular}}
\newcommand{\ottrulehead}[3]{$#1$ & & $#2$ & & & \multicolumn{2}{l}{#3}}
\newcommand{\ottprodline}[6]{& & $#1$ & $#2$ & $#3 #4$ & $#5$ & $#6$}
\newcommand{\ottfirstprodline}[6]{\ottprodline{#1}{#2}{#3}{#4}{#5}{#6}}
\newcommand{\ottlongprodline}[2]{& & $#1$ & \multicolumn{4}{l}{$#2$}}
\newcommand{\ottfirstlongprodline}[2]{\ottlongprodline{#1}{#2}}
\newcommand{\ottbindspecprodline}[6]{\ottprodline{#1}{#2}{#3}{#4}{#5}{#6}}
\newcommand{\ottprodnewline}{\\}
\newcommand{\ottinterrule}{\\[5.0mm]}
\newcommand{\ottafterlastrule}{\\}
\newcommand{\ottmetavars}{
\ottmetavartabular{
 $ \ottmv{x} $ & \ottcom{variables} \\
 $ \ottmv{a} $ & \ottcom{memories} \\
 $ \ottmv{n} $ & \ottcom{numbers} \\
}}

\newcommand{\ottb}{
\ottrulehead{\ottnt{b}}{::=}{}\ottprodnewline
\ottfirstprodline{|}{\ottkw{true}}{}{}{}{}\ottprodnewline
\ottprodline{|}{\ottkw{false}}{}{}{}{}}

\newcommand{\ottv}{
\ottrulehead{\ottnt{v}}{::=}{}\ottprodnewline
\ottfirstprodline{|}{\ottmv{n}}{}{}{}{}\ottprodnewline
\ottprodline{|}{\ottnt{b}}{}{}{}{}\ottprodnewline
\ottprodline{|}{\ottnt{v_{{\mathrm{1}}}} \, \ottkw{bop} \, \ottnt{v_{{\mathrm{2}}}}}{}{}{}{}}

\newcommand{\ottc}{
\ottrulehead{\ottnt{c}}{::=}{}\ottprodnewline
\ottfirstprodline{|}{\ottnt{e}}{}{}{}{}\ottprodnewline
\ottprodline{|}{ \textbf{let}~ \ottmv{x}  =  \ottnt{e} }{}{}{}{}\ottprodnewline
\ottprodline{|}{ \ottnt{c_{{\mathrm{1}}}} ~\rule[0.5ex]{1.5em}{0.55pt}~ \ottnt{c_{{\mathrm{2}}}} }{}{}{}{}\ottprodnewline
\ottprodline{|}{ \ottnt{c_{{\mathrm{1}}}}  \overset{ \rho }{\sim}  \ottnt{c_{{\mathrm{2}}}} }{}{}{}{}\ottprodnewline
\ottprodline{|}{\ottnt{c_{{\mathrm{1}}}}  ~\textbf{;}~  \ottnt{c_{{\mathrm{2}}}}}{}{}{}{}\ottprodnewline
\ottprodline{|}{\ottkw{if} \, \ottmv{x} \, \ottnt{c_{{\mathrm{1}}}} \, \ottnt{c_{{\mathrm{2}}}}}{}{}{}{}\ottprodnewline
\ottprodline{|}{\ottkw{while} \, \ottmv{x} \, \ottnt{c}}{}{}{}{}\ottprodnewline
\ottprodline{|}{\ottmv{x}  \ottsym{:=}  \ottnt{e}}{}{}{}{}\ottprodnewline
\ottprodline{|}{\ottmv{a}  \ottsym{[}  \ottnt{e_{{\mathrm{1}}}}  \ottsym{]}  \ottsym{:=}  \ottnt{e_{{\mathrm{2}}}}}{}{}{}{}\ottprodnewline
\ottprodline{|}{ \textbf{skip} }{}{}{}{}\ottprodnewline
\ottprodline{|}{\ottsym{(}  \ottnt{c}  \ottsym{)}}{}{}{}{}}

\newcommand{\otte}{
\ottrulehead{\ottnt{e}}{::=}{}\ottprodnewline
\ottfirstprodline{|}{\ottnt{v}}{}{}{}{}\ottprodnewline
\ottprodline{|}{\ottkw{bop} \, \ottnt{e_{{\mathrm{1}}}} \, \ottnt{e_{{\mathrm{2}}}}}{}{}{}{}\ottprodnewline
\ottprodline{|}{\ottmv{x}}{}{}{}{}\ottprodnewline
\ottprodline{|}{\ottmv{a}  \ottsym{[}  \ottnt{e}  \ottsym{]}}{}{}{}{}}

\newcommand{\ottt}{
\ottrulehead{\tau}{::=}{}\ottprodnewline
\ottfirstprodline{|}{ \ottkw{bit} \langle  \ottmv{n}  \rangle }{}{}{}{}\ottprodnewline
\ottprodline{|}{\ottkw{float}}{}{}{}{}\ottprodnewline
\ottprodline{|}{\ottkw{bool}}{}{}{}{}\ottprodnewline
\ottprodline{|}{\ottkw{mem} \, \tau  \ottsym{[}  \ottmv{n_{{\mathrm{1}}}}  \ottsym{]}}{}{}{}{}}

\newcommand{\ottsto}{
\ottrulehead{\sigma}{::=}{\ottcom{Memory Store}}\ottprodnewline
\ottfirstprodline{|}{ \sigma [  \ottmv{x}  \mapsto  \ottnt{v}  ] }{}{}{}{}\ottprodnewline
\ottprodline{|}{ \sigma [  \ottmv{a}  [  \ottmv{n}  ] \mapsto  \ottnt{v}  ] }{}{}{}{}}

\newcommand{\otttau}{
\ottrulehead{\Gamma}{::=}{\ottcom{Type Environment}}\ottprodnewline
\ottfirstprodline{|}{ \Gamma_{{\mathrm{1}}}  \cup  \Gamma_{{\mathrm{2}}} }{}{}{}{}\ottprodnewline
\ottprodline{|}{ \Gamma_{{\mathrm{1}}} [x \mapsto  \tau ] }{}{}{}{}}

\newcommand{\ottloc}{
\ottrulehead{\rho}{::=}{\ottcom{Consumed locations}}\ottprodnewline
\ottfirstprodline{|}{ \rho  \cup \{  \ottmv{a}  \} }{}{}{}{}\ottprodnewline
\ottprodline{|}{ \rho_{{\mathrm{1}}}  \cup  \rho_{{\mathrm{2}}} }{}{}{}{}}

\newcommand{\otttloc}{
\ottrulehead{\Delta}{::=}{\ottcom{Affine Typing Contexts}}\ottprodnewline
\ottfirstprodline{|}{ \Delta^* }{}{}{}{}\ottprodnewline
\ottprodline{|}{ \Delta_{{\mathrm{1}}}  \cap  \Delta_{{\mathrm{2}}} }{}{}{}{}\ottprodnewline
\ottprodline{|}{ \bar{ \rho } }{}{}{}{}\ottprodnewline
\ottprodline{|}{ \Delta }{}{}{}{}}

\newcommand{\ottgrammar}{\ottgrammartabular{
\ottb\ottinterrule
\ottv\ottinterrule
\ottc\ottinterrule
\otte\ottinterrule
\ottt\ottinterrule
\ottsto\ottinterrule
\otttau\ottinterrule
\ottloc\ottinterrule
\otttloc\ottafterlastrule
}}

\newcommand{\ottdrulelargeXXbop}[1]{\ottdrule[#1]{%
\ottpremise{\sigma_{{\mathrm{1}}}  \ottsym{,}  \rho_{{\mathrm{1}}}  \ottsym{,}  \ottnt{e_{{\mathrm{1}}}}  \Downarrow  \sigma_{{\mathrm{2}}}  \ottsym{,}  \rho_{{\mathrm{2}}}  \ottsym{,}  \ottnt{v_{{\mathrm{1}}}}}%
\ottpremise{\sigma_{{\mathrm{2}}}  \ottsym{,}  \rho_{{\mathrm{2}}}  \ottsym{,}  \ottnt{e_{{\mathrm{2}}}}  \Downarrow  \sigma_{{\mathrm{3}}}  \ottsym{,}  \rho_{{\mathrm{3}}}  \ottsym{,}  \ottnt{v_{{\mathrm{2}}}}}%
\ottpremise{\ottnt{v_{{\mathrm{3}}}}  \ottsym{=}  \ottnt{v_{{\mathrm{1}}}} \, \ottkw{bop} \, \ottnt{v_{{\mathrm{2}}}}}%
}{
\sigma_{{\mathrm{1}}}  \ottsym{,}  \rho_{{\mathrm{1}}}  \ottsym{,}  \ottkw{bop} \, \ottnt{e_{{\mathrm{1}}}} \, \ottnt{e_{{\mathrm{2}}}}  \Downarrow  \sigma_{{\mathrm{3}}}  \ottsym{,}  \rho_{{\mathrm{3}}}  \ottsym{,}  \ottnt{v_{{\mathrm{3}}}}}{%
{\ottdrulename{large\_bop}}{}%
}}

\newcommand{\ottdrulelargeXXvar}[1]{\ottdrule[#1]{%
\ottpremise{\sigma  \ottsym{(x)}  \ottsym{=}  \ottnt{v}}%
}{
\sigma  \ottsym{,}  \rho  \ottsym{,}  \ottmv{x}  \Downarrow  \sigma  \ottsym{,}  \rho  \ottsym{,}  \ottnt{v}}{%
{\ottdrulename{large\_var}}{}%
}}

\newcommand{\ottdrulelargeXXread}[1]{\ottdrule[#1]{%
\ottpremise{\ottmv{a}  \not\in  \rho_{{\mathrm{1}}}}%
\ottpremise{\sigma_{{\mathrm{1}}}  \ottsym{,}  \rho_{{\mathrm{1}}}  \ottsym{,}  \ottnt{e}  \Downarrow  \sigma_{{\mathrm{2}}}  \ottsym{,}  \rho_{{\mathrm{2}}}  \ottsym{,}  \ottmv{n}}%
\ottpremise{\sigma_{{\mathrm{2}}}  \ottsym{(a)(n)}  \ottsym{=}  \ottnt{v}}%
}{
\sigma_{{\mathrm{1}}}  \ottsym{,}  \rho_{{\mathrm{1}}}  \ottsym{,}  \ottmv{a}  \ottsym{[}  \ottnt{e}  \ottsym{]}  \Downarrow  \sigma_{{\mathrm{2}}}  \ottsym{,}   \rho_{{\mathrm{2}}}  \cup \{  \ottmv{a}  \}   \ottsym{,}  \ottnt{v}}{%
{\ottdrulename{large\_read}}{}%
}}

\newcommand{\ottdefnelargereduce}[1]{\begin{ottdefnblock}[#1]{$\sigma_{{\mathrm{1}}}  \ottsym{,}  \rho_{{\mathrm{1}}}  \ottsym{,}  \ottnt{e}  \Downarrow  \sigma_{{\mathrm{2}}}  \ottsym{,}  \rho_{{\mathrm{2}}}  \ottsym{,}  \ottnt{v}$}{}
\ottusedrule{\ottdrulelargeXXbop{}}
\ottusedrule{\ottdrulelargeXXvar{}}
\ottusedrule{\ottdrulelargeXXread{}}
\end{ottdefnblock}}

\newcommand{\ottdrulelargeXXlet}[1]{\ottdrule[#1]{%
\ottpremise{\sigma_{{\mathrm{1}}}  \ottsym{,}  \rho_{{\mathrm{1}}}  \ottsym{,}  \ottnt{e}  \Downarrow  \sigma_{{\mathrm{2}}}  \ottsym{,}  \rho_{{\mathrm{2}}}  \ottsym{,}  \ottnt{v}}%
}{
\sigma_{{\mathrm{1}}}  \ottsym{,}  \rho_{{\mathrm{1}}}  \ottsym{,}   \textbf{let}~ \ottmv{x}  =  \ottnt{e}   \Downarrow   \sigma_{{\mathrm{2}}} [  \ottmv{x}  \mapsto  \ottnt{v}  ]   \ottsym{,}  \rho_{{\mathrm{2}}}}{%
{\ottdrulename{large\_let}}{}%
}}

\newcommand{\ottdrulelargeXXseq}[1]{\ottdrule[#1]{%
\ottpremise{\sigma_{{\mathrm{1}}}  \ottsym{,}  \rho_{{\mathrm{1}}}  \ottsym{,}  \ottnt{c_{{\mathrm{1}}}}  \Downarrow  \sigma_{{\mathrm{2}}}  \ottsym{,}  \rho_{{\mathrm{2}}}}%
\ottpremise{\sigma_{{\mathrm{2}}}  \ottsym{,}  \rho_{{\mathrm{1}}}  \ottsym{,}  \ottnt{c_{{\mathrm{2}}}}  \Downarrow  \sigma_{{\mathrm{3}}}  \ottsym{,}  \rho_{{\mathrm{3}}}}%
}{
\sigma_{{\mathrm{1}}}  \ottsym{,}  \rho_{{\mathrm{1}}}  \ottsym{,}   \ottnt{c_{{\mathrm{1}}}} ~\rule[0.5ex]{1.5em}{0.55pt}~ \ottnt{c_{{\mathrm{2}}}}   \Downarrow  \sigma_{{\mathrm{3}}}  \ottsym{,}   \rho_{{\mathrm{2}}}  \cup  \rho_{{\mathrm{3}}} }{%
{\ottdrulename{large\_seq}}{}%
}}

\newcommand{\ottdrulelargeXXinterXXseq}[1]{\ottdrule[#1]{%
\ottpremise{\sigma_{{\mathrm{1}}}  \ottsym{,}  \rho_{{\mathrm{1}}}  \ottsym{,}  \ottnt{c_{{\mathrm{1}}}}  \Downarrow  \sigma_{{\mathrm{2}}}  \ottsym{,}  \rho_{{\mathrm{2}}}}%
\ottpremise{\sigma_{{\mathrm{2}}}  \ottsym{,}  \rho  \ottsym{,}  \ottnt{c_{{\mathrm{2}}}}  \Downarrow  \sigma_{{\mathrm{3}}}  \ottsym{,}  \rho_{{\mathrm{3}}}}%
}{
\sigma_{{\mathrm{1}}}  \ottsym{,}  \rho_{{\mathrm{1}}}  \ottsym{,}   \ottnt{c_{{\mathrm{1}}}}  \overset{ \rho }{\sim}  \ottnt{c_{{\mathrm{2}}}}   \Downarrow  \sigma_{{\mathrm{3}}}  \ottsym{,}   \rho_{{\mathrm{2}}}  \cup  \rho_{{\mathrm{3}}} }{%
{\ottdrulename{large\_inter\_seq}}{}%
}}

\newcommand{\ottdrulelargeXXpar}[1]{\ottdrule[#1]{%
\ottpremise{\sigma_{{\mathrm{1}}}  \ottsym{,}  \rho_{{\mathrm{1}}}  \ottsym{,}  \ottnt{c_{{\mathrm{1}}}}  \Downarrow  \sigma_{{\mathrm{2}}}  \ottsym{,}  \rho_{{\mathrm{2}}}}%
\ottpremise{\sigma_{{\mathrm{2}}}  \ottsym{,}  \rho_{{\mathrm{2}}}  \ottsym{,}  \ottnt{c_{{\mathrm{2}}}}  \Downarrow  \sigma_{{\mathrm{3}}}  \ottsym{,}  \rho_{{\mathrm{3}}}}%
}{
\sigma_{{\mathrm{1}}}  \ottsym{,}  \rho_{{\mathrm{1}}}  \ottsym{,}  \ottnt{c_{{\mathrm{1}}}}  ~\textbf{;}~  \ottnt{c_{{\mathrm{2}}}}  \Downarrow  \sigma_{{\mathrm{3}}}  \ottsym{,}  \rho_{{\mathrm{3}}}}{%
{\ottdrulename{large\_par}}{}%
}}

\newcommand{\ottdrulelargeXXifXXtrue}[1]{\ottdrule[#1]{%
\ottpremise{\sigma_{{\mathrm{1}}}  \ottsym{,}  \rho_{{\mathrm{1}}}  \ottsym{,}  \ottnt{e_{{\mathrm{1}}}}  \Downarrow  \sigma_{{\mathrm{2}}}  \ottsym{,}  \rho_{{\mathrm{2}}}  \ottsym{,}  \ottkw{true}}%
\ottpremise{\sigma_{{\mathrm{2}}}  \ottsym{,}  \rho_{{\mathrm{2}}}  \ottsym{,}  \ottnt{c_{{\mathrm{1}}}}  \Downarrow  \sigma_{{\mathrm{3}}}  \ottsym{,}  \rho_{{\mathrm{3}}}}%
}{
\sigma_{{\mathrm{1}}}  \ottsym{,}  \rho_{{\mathrm{1}}}  \ottsym{,}  \ottkw{if} \, \ottmv{x} \, \ottnt{c_{{\mathrm{1}}}} \, \ottnt{c_{{\mathrm{2}}}}  \Downarrow  \sigma_{{\mathrm{3}}}  \ottsym{,}  \rho_{{\mathrm{3}}}}{%
{\ottdrulename{large\_if\_true}}{}%
}}

\newcommand{\ottdrulelargeXXifXXfalse}[1]{\ottdrule[#1]{%
\ottpremise{\sigma_{{\mathrm{1}}}  \ottsym{,}  \rho_{{\mathrm{1}}}  \ottsym{,}  \ottnt{e_{{\mathrm{1}}}}  \Downarrow  \sigma_{{\mathrm{2}}}  \ottsym{,}  \rho_{{\mathrm{2}}}  \ottsym{,}  \ottkw{false}}%
\ottpremise{\sigma_{{\mathrm{2}}}  \ottsym{,}  \rho_{{\mathrm{2}}}  \ottsym{,}  \ottnt{c_{{\mathrm{2}}}}  \Downarrow  \sigma_{{\mathrm{3}}}  \ottsym{,}  \rho_{{\mathrm{3}}}}%
}{
\sigma_{{\mathrm{1}}}  \ottsym{,}  \rho_{{\mathrm{1}}}  \ottsym{,}  \ottkw{if} \, \ottmv{x} \, \ottnt{c_{{\mathrm{1}}}} \, \ottnt{c_{{\mathrm{2}}}}  \Downarrow  \sigma_{{\mathrm{3}}}  \ottsym{,}  \rho_{{\mathrm{3}}}}{%
{\ottdrulename{large\_if\_false}}{}%
}}

\newcommand{\ottdrulelargeXXwhileXXtrue}[1]{\ottdrule[#1]{%
\ottpremise{ \sigma_{{\mathrm{1}}}  \ottsym{,}  \rho_{{\mathrm{1}}}  \ottsym{,}  \ottnt{e_{{\mathrm{1}}}}  \Downarrow  \sigma_{{\mathrm{2}}}  \ottsym{,}  \rho_{{\mathrm{2}}}  \ottsym{,}  \ottkw{true}  \ottlinebreakhack }%
\ottpremise{\sigma_{{\mathrm{2}}}  \ottsym{,}  \rho_{{\mathrm{2}}}  \ottsym{,}   \ottnt{c} ~\rule[0.5ex]{1.5em}{0.55pt}~ \ottkw{while} \, \ottmv{x} \, \ottnt{c}   \Downarrow  \sigma_{{\mathrm{3}}}  \ottsym{,}  \rho_{{\mathrm{3}}}}%
}{
\sigma_{{\mathrm{1}}}  \ottsym{,}  \rho_{{\mathrm{1}}}  \ottsym{,}  \ottkw{while} \, \ottmv{x} \, \ottnt{c}  \Downarrow  \sigma_{{\mathrm{3}}}  \ottsym{,}  \rho_{{\mathrm{3}}}}{%
{\ottdrulename{large\_while\_true}}{}%
}}

\newcommand{\ottdrulelargeXXwhileXXfalse}[1]{\ottdrule[#1]{%
\ottpremise{\sigma_{{\mathrm{1}}}  \ottsym{,}  \rho_{{\mathrm{1}}}  \ottsym{,}  \ottnt{e_{{\mathrm{1}}}}  \Downarrow  \sigma_{{\mathrm{2}}}  \ottsym{,}  \rho_{{\mathrm{2}}}  \ottsym{,}  \ottkw{false}}%
}{
\sigma_{{\mathrm{1}}}  \ottsym{,}  \rho_{{\mathrm{1}}}  \ottsym{,}  \ottkw{while} \, \ottmv{x} \, \ottnt{c}  \Downarrow  \sigma_{{\mathrm{2}}}  \ottsym{,}  \rho_{{\mathrm{2}}}}{%
{\ottdrulename{large\_while\_false}}{}%
}}

\newcommand{\ottdrulelargeXXupdate}[1]{\ottdrule[#1]{%
\ottpremise{\sigma_{{\mathrm{1}}}  \ottsym{,}  \rho_{{\mathrm{1}}}  \ottsym{,}  \ottnt{e}  \Downarrow  \sigma_{{\mathrm{2}}}  \ottsym{,}  \rho_{{\mathrm{2}}}  \ottsym{,}  \ottnt{v}}%
}{
\sigma_{{\mathrm{1}}}  \ottsym{,}  \rho_{{\mathrm{1}}}  \ottsym{,}  \ottmv{x}  \ottsym{:=}  \ottnt{e}  \Downarrow   \sigma_{{\mathrm{2}}} [  \ottmv{x}  \mapsto  \ottnt{v}  ]   \ottsym{,}  \rho_{{\mathrm{2}}}}{%
{\ottdrulename{large\_update}}{}%
}}

\newcommand{\ottdrulelargeXXwrite}[1]{\ottdrule[#1]{%
\ottpremise{\sigma_{{\mathrm{1}}}  \ottsym{,}  \rho_{{\mathrm{1}}}  \ottsym{,}  \ottnt{e_{{\mathrm{1}}}}  \Downarrow  \sigma_{{\mathrm{2}}}  \ottsym{,}  \rho_{{\mathrm{2}}}  \ottsym{,}  \ottmv{n}}%
\ottpremise{\sigma_{{\mathrm{2}}}  \ottsym{,}  \rho_{{\mathrm{2}}}  \ottsym{,}  \ottnt{e_{{\mathrm{2}}}}  \Downarrow  \sigma_{{\mathrm{3}}}  \ottsym{,}  \rho_{{\mathrm{3}}}  \ottsym{,}  \ottnt{v}}%
\ottpremise{\ottmv{a}  \not\in  \rho_{{\mathrm{3}}}}%
}{
\sigma_{{\mathrm{1}}}  \ottsym{,}  \rho_{{\mathrm{1}}}  \ottsym{,}  \ottmv{a}  \ottsym{[}  \ottnt{e_{{\mathrm{1}}}}  \ottsym{]}  \ottsym{:=}  \ottnt{e_{{\mathrm{2}}}}  \Downarrow   \sigma_{{\mathrm{3}}} [  \ottmv{a}  [  \ottmv{n}  ] \mapsto  \ottnt{v}  ]   \ottsym{,}   \rho_{{\mathrm{3}}}  \cup \{  \ottmv{a}  \} }{%
{\ottdrulename{large\_write}}{}%
}}

\newcommand{\ottdefnclargereduce}[1]{\begin{ottdefnblock}[#1]{$\sigma_{{\mathrm{1}}}  \ottsym{,}  \rho_{{\mathrm{1}}}  \ottsym{,}  \ottnt{c}  \Downarrow  \sigma_{{\mathrm{2}}}  \ottsym{,}  \rho_{{\mathrm{2}}}$}{}
\ottusedrule{\ottdrulelargeXXlet{}}
\ottusedrule{\ottdrulelargeXXseq{}}
\ottusedrule{\ottdrulelargeXXinterXXseq{}}
\ottusedrule{\ottdrulelargeXXpar{}}
\ottusedrule{\ottdrulelargeXXifXXtrue{}}
\ottusedrule{\ottdrulelargeXXifXXfalse{}}
\ottusedrule{\ottdrulelargeXXwhileXXtrue{}}
\ottusedrule{\ottdrulelargeXXwhileXXfalse{}}
\ottusedrule{\ottdrulelargeXXupdate{}}
\ottusedrule{\ottdrulelargeXXwrite{}}
\end{ottdefnblock}}

\newcommand{\ottdrulesmallXXreadOne}[1]{\ottdrule[#1]{%
\ottpremise{\sigma  \ottsym{,}  \rho  \ottsym{,}  \ottnt{e}  \rightarrow  \sigma'  \ottsym{,}  \rho'  \ottsym{,}  \ottnt{e'}}%
}{
\sigma  \ottsym{,}  \rho  \ottsym{,}  \ottmv{a}  \ottsym{[}  \ottnt{e}  \ottsym{]}  \rightarrow  \sigma'  \ottsym{,}  \rho'  \ottsym{,}  \ottmv{a}  \ottsym{[}  \ottnt{e'}  \ottsym{]}}{%
{\ottdrulename{small\_read1}}{}%
}}

\newcommand{\ottdrulesmallXXreadTwo}[1]{\ottdrule[#1]{%
\ottpremise{\ottmv{a}  \not\in  \rho}%
}{
\sigma  \ottsym{,}  \rho  \ottsym{,}  \ottmv{a}  \ottsym{[}  \ottmv{n}  \ottsym{]}  \rightarrow  \sigma  \ottsym{,}   \rho  \cup \{  \ottmv{a}  \}   \ottsym{,}  \ottnt{v}}{%
{\ottdrulename{small\_read2}}{}%
}}

\newcommand{\ottdrulesmallXXbopOne}[1]{\ottdrule[#1]{%
\ottpremise{\sigma  \ottsym{,}  \rho  \ottsym{,}  \ottnt{e_{{\mathrm{1}}}}  \rightarrow  \sigma'  \ottsym{,}  \rho'  \ottsym{,}  \ottnt{e'_{{\mathrm{1}}}}}%
}{
\sigma  \ottsym{,}  \rho  \ottsym{,}  \ottkw{bop} \, \ottnt{e_{{\mathrm{1}}}} \, \ottnt{e_{{\mathrm{2}}}}  \rightarrow  \sigma'  \ottsym{,}  \rho'  \ottsym{,}  \ottkw{bop} \, \ottnt{e'_{{\mathrm{1}}}} \, \ottnt{e_{{\mathrm{2}}}}}{%
{\ottdrulename{small\_bop1}}{}%
}}

\newcommand{\ottdrulesmallXXbopTwo}[1]{\ottdrule[#1]{%
\ottpremise{\sigma  \ottsym{,}  \rho  \ottsym{,}  \ottnt{e_{{\mathrm{2}}}}  \rightarrow  \sigma'  \ottsym{,}  \rho'  \ottsym{,}  \ottnt{e'_{{\mathrm{2}}}}}%
}{
\sigma  \ottsym{,}  \rho  \ottsym{,}  \ottkw{bop} \, \ottnt{v_{{\mathrm{1}}}} \, \ottnt{e_{{\mathrm{2}}}}  \rightarrow  \sigma'  \ottsym{,}  \rho'  \ottsym{,}  \ottkw{bop} \, \ottnt{v_{{\mathrm{1}}}} \, \ottnt{e'_{{\mathrm{2}}}}}{%
{\ottdrulename{small\_bop2}}{}%
}}

\newcommand{\ottdrulesmallXXbopThree}[1]{\ottdrule[#1]{%
\ottpremise{\ottnt{v_{{\mathrm{3}}}}  \ottsym{=}  \ottnt{v_{{\mathrm{1}}}} \, \ottkw{bop} \, \ottnt{v_{{\mathrm{2}}}}}%
}{
\sigma  \ottsym{,}  \rho  \ottsym{,}  \ottkw{bop} \, \ottnt{v_{{\mathrm{1}}}} \, \ottnt{v_{{\mathrm{2}}}}  \rightarrow  \sigma  \ottsym{,}  \rho  \ottsym{,}  \ottnt{v_{{\mathrm{3}}}}}{%
{\ottdrulename{small\_bop3}}{}%
}}

\newcommand{\ottdrulevar}[1]{\ottdrule[#1]{%
\ottpremise{\sigma  \ottsym{(x)}  \ottsym{=}  \ottnt{v}}%
}{
\sigma  \ottsym{,}  \rho  \ottsym{,}  \ottmv{x}  \rightarrow  \sigma  \ottsym{,}  \rho  \ottsym{,}  \ottnt{v}}{%
{\ottdrulename{var}}{}%
}}

\newcommand{\ottdefnesmallreduce}[1]{\begin{ottdefnblock}[#1]{$\sigma  \ottsym{,}  \rho  \ottsym{,}  \ottnt{e}  \rightarrow  \sigma'  \ottsym{,}  \rho'  \ottsym{,}  \ottnt{e'}$}{}
\ottusedrule{\ottdrulesmallXXreadOne{}}
\ottusedrule{\ottdrulesmallXXreadTwo{}}
\ottusedrule{\ottdrulesmallXXbopOne{}}
\ottusedrule{\ottdrulesmallXXbopTwo{}}
\ottusedrule{\ottdrulesmallXXbopThree{}}
\ottusedrule{\ottdrulevar{}}
\end{ottdefnblock}}

\newcommand{\ottdrulesmallXXwriteOne}[1]{\ottdrule[#1]{%
\ottpremise{\sigma  \ottsym{,}  \rho  \ottsym{,}  \ottnt{e_{{\mathrm{1}}}}  \rightarrow  \sigma'  \ottsym{,}  \rho'  \ottsym{,}  \ottnt{e'_{{\mathrm{1}}}}}%
}{
\sigma  \ottsym{,}  \rho  \ottsym{,}  \ottmv{a}  \ottsym{[}  \ottnt{e_{{\mathrm{1}}}}  \ottsym{]}  \ottsym{:=}  \ottnt{e_{{\mathrm{2}}}}  \rightarrow  \sigma'  \ottsym{,}  \rho'  \ottsym{,}  \ottmv{a}  \ottsym{[}  \ottnt{e'_{{\mathrm{1}}}}  \ottsym{]}  \ottsym{:=}  \ottnt{e_{{\mathrm{2}}}}}{%
{\ottdrulename{small\_write1}}{}%
}}

\newcommand{\ottdrulesmallXXwriteTwo}[1]{\ottdrule[#1]{%
\ottpremise{\sigma  \ottsym{,}  \rho  \ottsym{,}  \ottnt{e}  \rightarrow  \sigma'  \ottsym{,}  \rho'  \ottsym{,}  \ottnt{e'}}%
}{
\sigma  \ottsym{,}  \rho  \ottsym{,}  \ottmv{a}  \ottsym{[}  \ottmv{n}  \ottsym{]}  \ottsym{:=}  \ottnt{e}  \rightarrow  \sigma'  \ottsym{,}  \rho'  \ottsym{,}  \ottmv{a}  \ottsym{[}  \ottmv{n}  \ottsym{]}  \ottsym{:=}  \ottnt{e'}}{%
{\ottdrulename{small\_write2}}{}%
}}

\newcommand{\ottdrulesmallXXwriteThree}[1]{\ottdrule[#1]{%
\ottpremise{\ottmv{a}  \not\in  \rho}%
}{
\sigma  \ottsym{,}  \rho  \ottsym{,}  \ottmv{a}  \ottsym{[}  \ottmv{n}  \ottsym{]}  \ottsym{:=}  \ottnt{v}  \rightarrow   \sigma [  \ottmv{a}  [  \ottmv{n}  ] \mapsto  \ottnt{v}  ]   \ottsym{,}   \rho  \cup \{  \ottmv{a}  \}   \ottsym{,}   \textbf{skip} }{%
{\ottdrulename{small\_write3}}{}%
}}

\newcommand{\ottdrulesmallXXletOne}[1]{\ottdrule[#1]{%
\ottpremise{\sigma  \ottsym{,}  \rho  \ottsym{,}  \ottnt{e}  \rightarrow  \sigma'  \ottsym{,}  \rho'  \ottsym{,}  \ottnt{e'}}%
}{
\sigma  \ottsym{,}  \rho  \ottsym{,}   \textbf{let}~ \ottmv{x}  =  \ottnt{e}   \rightarrow  \sigma'  \ottsym{,}  \rho'  \ottsym{,}   \textbf{let}~ \ottmv{x}  =  \ottnt{e'} }{%
{\ottdrulename{small\_let1}}{}%
}}

\newcommand{\ottdrulesmallXXletTwo}[1]{\ottdrule[#1]{%
}{
\sigma  \ottsym{,}  \rho  \ottsym{,}   \textbf{let}~ \ottmv{x}  =  \ottnt{v}   \rightarrow   \sigma [  \ottmv{x}  \mapsto  \ottnt{v}  ]   \ottsym{,}  \rho  \ottsym{,}   \textbf{skip} }{%
{\ottdrulename{small\_let2}}{}%
}}

\newcommand{\ottdrulesmallXXparOne}[1]{\ottdrule[#1]{%
\ottpremise{\sigma  \ottsym{,}  \rho  \ottsym{,}  \ottnt{c_{{\mathrm{1}}}}  \rightarrow  \sigma'  \ottsym{,}  \rho'  \ottsym{,}  \ottnt{c'_{{\mathrm{1}}}}}%
}{
\sigma  \ottsym{,}  \rho  \ottsym{,}  \ottnt{c_{{\mathrm{1}}}}  ~\textbf{;}~  \ottnt{c_{{\mathrm{2}}}}  \rightarrow  \sigma'  \ottsym{,}  \rho'  \ottsym{,}  \ottnt{c'_{{\mathrm{1}}}}  ~\textbf{;}~  \ottnt{c_{{\mathrm{2}}}}}{%
{\ottdrulename{small\_par1}}{}%
}}

\newcommand{\ottdrulesmallXXparTwo}[1]{\ottdrule[#1]{%
}{
\sigma  \ottsym{,}  \rho  \ottsym{,}   \textbf{skip}   ~\textbf{;}~  \ottnt{c_{{\mathrm{2}}}}  \rightarrow  \sigma  \ottsym{,}  \rho  \ottsym{,}  \ottnt{c_{{\mathrm{2}}}}}{%
{\ottdrulename{small\_par2}}{}%
}}

\newcommand{\ottdrulesmallXXseq}[1]{\ottdrule[#1]{%
}{
\sigma  \ottsym{,}  \rho  \ottsym{,}   \ottnt{c_{{\mathrm{1}}}} ~\rule[0.5ex]{1.5em}{0.55pt}~ \ottnt{c_{{\mathrm{2}}}}   \rightarrow  \sigma  \ottsym{,}  \rho  \ottsym{,}   \ottnt{c_{{\mathrm{1}}}}  \overset{ \rho }{\sim}  \ottnt{c_{{\mathrm{2}}}} }{%
{\ottdrulename{small\_seq}}{}%
}}

\newcommand{\ottdrulesmallXXinterXXseqOne}[1]{\ottdrule[#1]{%
\ottpremise{\sigma  \ottsym{,}  \rho  \ottsym{,}  \ottnt{c_{{\mathrm{1}}}}  \rightarrow  \sigma'  \ottsym{,}  \rho'  \ottsym{,}  \ottnt{c'_{{\mathrm{1}}}}}%
}{
\sigma  \ottsym{,}  \rho  \ottsym{,}   \ottnt{c_{{\mathrm{1}}}}  \overset{ \rho'' }{\sim}  \ottnt{c_{{\mathrm{2}}}}   \rightarrow  \sigma'  \ottsym{,}  \rho'  \ottsym{,}   \ottnt{c'_{{\mathrm{1}}}}  \overset{ \rho'' }{\sim}  \ottnt{c_{{\mathrm{2}}}} }{%
{\ottdrulename{small\_inter\_seq1}}{}%
}}

\newcommand{\ottdrulesmallXXinterXXseqTwo}[1]{\ottdrule[#1]{%
\ottpremise{\sigma  \ottsym{,}  \rho''  \ottsym{,}  \ottnt{c_{{\mathrm{2}}}}  \rightarrow  \sigma'  \ottsym{,}  \rho'''  \ottsym{,}  \ottnt{c'_{{\mathrm{2}}}}}%
}{
\sigma  \ottsym{,}  \rho  \ottsym{,}    \textbf{skip}   \overset{ \rho'' }{\sim}  \ottnt{c_{{\mathrm{2}}}}   \rightarrow  \sigma'  \ottsym{,}  \rho  \ottsym{,}    \textbf{skip}   \overset{ \rho''' }{\sim}  \ottnt{c'_{{\mathrm{2}}}} }{%
{\ottdrulename{small\_inter\_seq2}}{}%
}}

\newcommand{\ottdrulesmallXXinterXXseqThree}[1]{\ottdrule[#1]{%
}{
\sigma  \ottsym{,}  \rho  \ottsym{,}    \textbf{skip}   \overset{ \rho'' }{\sim}   \textbf{skip}    \rightarrow  \sigma  \ottsym{,}   \rho  \cup  \rho''   \ottsym{,}   \textbf{skip} }{%
{\ottdrulename{small\_inter\_seq3}}{}%
}}

\newcommand{\ottdrulesmallXXifOne}[1]{\ottdrule[#1]{%
\ottpremise{\sigma  \ottsym{(x)}  \ottsym{=}  \ottkw{true}}%
}{
\sigma  \ottsym{,}  \rho  \ottsym{,}  \ottkw{if} \, \ottmv{x} \, \ottnt{c_{{\mathrm{1}}}} \, \ottnt{c_{{\mathrm{2}}}}  \rightarrow  \sigma  \ottsym{,}  \rho  \ottsym{,}  \ottnt{c_{{\mathrm{1}}}}}{%
{\ottdrulename{small\_if1}}{}%
}}

\newcommand{\ottdrulesmallXXifTwo}[1]{\ottdrule[#1]{%
\ottpremise{\sigma_{{\mathrm{1}}}  \ottsym{(x)}  \ottsym{=}  \ottkw{false}}%
}{
\sigma  \ottsym{,}  \rho  \ottsym{,}  \ottkw{if} \, \ottmv{x} \, \ottnt{c_{{\mathrm{1}}}} \, \ottnt{c_{{\mathrm{2}}}}  \rightarrow  \sigma  \ottsym{,}  \rho  \ottsym{,}  \ottnt{c_{{\mathrm{2}}}}}{%
{\ottdrulename{small\_if2}}{}%
}}

\newcommand{\ottdrulesmallXXwhile}[1]{\ottdrule[#1]{%
}{
\sigma  \ottsym{,}  \rho  \ottsym{,}  \ottkw{while} \, \ottmv{x} \, \ottnt{c}  \rightarrow  \sigma  \ottsym{,}  \rho  \ottsym{,}  \ottkw{if} \, \ottmv{x} \, \ottsym{(}   \ottnt{c} ~\rule[0.5ex]{1.5em}{0.55pt}~ \ottkw{while} \, \ottmv{x} \, \ottnt{c}   \ottsym{)} \,  \textbf{skip} }{%
{\ottdrulename{small\_while}}{}%
}}

\newcommand{\ottdefncsmallreduce}[1]{\begin{ottdefnblock}[#1]{$\sigma_{{\mathrm{1}}}  \ottsym{,}  \rho_{{\mathrm{1}}}  \ottsym{,}  \ottnt{c}  \rightarrow  \sigma'  \ottsym{,}  \rho'  \ottsym{,}  \ottnt{c'}$}{}
\ottusedrule{\ottdrulesmallXXwriteOne{}}
\ottusedrule{\ottdrulesmallXXwriteTwo{}}
\ottusedrule{\ottdrulesmallXXwriteThree{}}
\ottusedrule{\ottdrulesmallXXletOne{}}
\ottusedrule{\ottdrulesmallXXletTwo{}}
\ottusedrule{\ottdrulesmallXXparOne{}}
\ottusedrule{\ottdrulesmallXXparTwo{}}
\ottusedrule{\ottdrulesmallXXseq{}}
\ottusedrule{\ottdrulesmallXXinterXXseqOne{}}
\ottusedrule{\ottdrulesmallXXinterXXseqTwo{}}
\ottusedrule{\ottdrulesmallXXinterXXseqThree{}}
\ottusedrule{\ottdrulesmallXXifOne{}}
\ottusedrule{\ottdrulesmallXXifTwo{}}
\ottusedrule{\ottdrulesmallXXwhile{}}
\end{ottdefnblock}}

\newcommand{\ottdrulecheckXXval}[1]{\ottdrule[#1]{%
}{
\Gamma  \ottsym{,}  \Delta  \vdash  \ottnt{v}  \ottsym{:}  \tau  \dashv  \Delta}{%
{\ottdrulename{check\_val}}{}%
}}

\newcommand{\ottdrulecheckXXbop}[1]{\ottdrule[#1]{%
\ottpremise{\Gamma  \ottsym{,}  \Delta_{{\mathrm{1}}}  \vdash  \ottnt{e_{{\mathrm{1}}}}  \ottsym{:}  \tau  \dashv  \Delta_{{\mathrm{2}}}}%
\ottpremise{\Gamma  \ottsym{,}  \Delta_{{\mathrm{2}}}  \vdash  \ottnt{e_{{\mathrm{2}}}}  \ottsym{:}  \tau  \dashv  \Delta_{{\mathrm{3}}}}%
\ottpremise{ \textbf{bop} :  \tau  \rightarrow  \tau  \rightarrow  \tau }%
}{
\Gamma  \ottsym{,}  \Delta_{{\mathrm{1}}}  \vdash  \ottkw{bop} \, \ottnt{e_{{\mathrm{1}}}} \, \ottnt{e_{{\mathrm{2}}}}  \ottsym{:}  \tau  \dashv  \Delta_{{\mathrm{3}}}}{%
{\ottdrulename{check\_bop}}{}%
}}

\newcommand{\ottdrulecheckXXvar}[1]{\ottdrule[#1]{%
\ottpremise{\Gamma  \ottsym{(x)}  \ottsym{=}  \tau}%
}{
\Gamma  \ottsym{,}  \Delta_{{\mathrm{1}}}  \vdash  \ottmv{x}  \ottsym{:}  \tau  \dashv  \Delta_{{\mathrm{1}}}}{%
{\ottdrulename{check\_var}}{}%
}}

\newcommand{\ottdrulecheckXXread}[1]{\ottdrule[#1]{%
\ottpremise{\Gamma  \ottsym{,}  \Delta_{{\mathrm{1}}}  \vdash  \ottnt{e_{{\mathrm{1}}}}  \ottsym{:}   \ottkw{bit} \langle  \ottmv{n}  \rangle   \dashv  \Delta_{{\mathrm{2}}}}%
\ottpremise{ \Delta_{{\mathrm{2}}}  =  \Delta_{{\mathrm{3}}}  \cup \{  \ottmv{a}  \mapsto  \ottkw{mem} \, \tau  \ottsym{[}  \ottmv{n_{{\mathrm{1}}}}  \ottsym{]}  \} }%
}{
\Gamma  \ottsym{,}  \Delta_{{\mathrm{1}}}  \vdash  \ottmv{a}  \ottsym{[}  \ottnt{e}  \ottsym{]}  \ottsym{:}  \tau  \dashv  \Delta_{{\mathrm{3}}}}{%
{\ottdrulename{check\_read}}{}%
}}

\newcommand{\ottdefnecheck}[1]{\begin{ottdefnblock}[#1]{$\Gamma  \ottsym{,}  \Delta_{{\mathrm{1}}}  \vdash  \ottnt{e}  \ottsym{:}  \tau  \dashv  \Delta_{{\mathrm{2}}}$}{}
\ottusedrule{\ottdrulecheckXXval{}}
\ottusedrule{\ottdrulecheckXXbop{}}
\ottusedrule{\ottdrulecheckXXvar{}}
\ottusedrule{\ottdrulecheckXXread{}}
\end{ottdefnblock}}

\newcommand{\ottdrulecheckXXskip}[1]{\ottdrule[#1]{%
}{
\Gamma  \ottsym{,}  \Delta  \vdash   \textbf{skip}   \dashv  \Gamma  \ottsym{,}  \Delta}{%
{\ottdrulename{check\_skip}}{}%
}}

\newcommand{\ottdrulecheckXXwrite}[1]{\ottdrule[#1]{%
\ottpremise{\Gamma  \ottsym{,}  \Delta_{{\mathrm{1}}}  \vdash  \ottnt{e_{{\mathrm{1}}}}  \ottsym{:}   \ottkw{bit} \langle  \ottmv{n}  \rangle   \dashv  \Delta_{{\mathrm{2}}}}%
\ottpremise{ \Gamma  \ottsym{,}  \Delta_{{\mathrm{2}}}  \vdash  \ottnt{e_{{\mathrm{2}}}}  \ottsym{:}  \tau  \dashv  \Delta_{{\mathrm{3}}}  \ottlinebreakhack }%
\ottpremise{ \Delta_{{\mathrm{3}}}  =  \Delta_{{\mathrm{4}}}  \cup \{  \ottmv{a}  \mapsto  \ottkw{mem} \, \tau  \ottsym{[}  \ottmv{n_{{\mathrm{1}}}}  \ottsym{]}  \} }%
}{
\Gamma  \ottsym{,}  \Delta_{{\mathrm{1}}}  \vdash  \ottmv{a}  \ottsym{[}  \ottnt{e_{{\mathrm{1}}}}  \ottsym{]}  \ottsym{:=}  \ottnt{e_{{\mathrm{2}}}}  \dashv  \Gamma  \ottsym{,}  \Delta_{{\mathrm{4}}}}{%
{\ottdrulename{check\_write}}{}%
}}

\newcommand{\ottdrulecheckXXlet}[1]{\ottdrule[#1]{%
\ottpremise{\Gamma  \ottsym{,}  \Delta_{{\mathrm{1}}}  \vdash  \ottnt{e}  \ottsym{:}  \tau  \dashv  \Delta_{{\mathrm{2}}}}%
\ottpremise{ ( \ottmv{x}  \rightarrow  \tau ) \notin  \Gamma }%
}{
\Gamma  \ottsym{,}  \Delta_{{\mathrm{1}}}  \vdash   \textbf{let}~ \ottmv{x}  =  \ottnt{e}   \dashv   \Gamma [x \mapsto  \tau ]   \ottsym{,}  \Delta_{{\mathrm{2}}}}{%
{\ottdrulename{check\_let}}{}%
}}

\newcommand{\ottdrulecheckXXparXXcomp}[1]{\ottdrule[#1]{%
\ottpremise{\Gamma_{{\mathrm{1}}}  \ottsym{,}  \Delta_{{\mathrm{1}}}  \vdash  \ottnt{c_{{\mathrm{1}}}}  \dashv  \Gamma_{{\mathrm{2}}}  \ottsym{,}  \Delta_{{\mathrm{2}}}}%
\ottpremise{\Gamma_{{\mathrm{2}}}  \ottsym{,}  \Delta_{{\mathrm{2}}}  \vdash  \ottnt{c_{{\mathrm{2}}}}  \dashv  \Gamma_{{\mathrm{3}}}  \ottsym{,}  \Delta_{{\mathrm{3}}}}%
}{
\Gamma_{{\mathrm{1}}}  \ottsym{,}  \Delta_{{\mathrm{1}}}  \vdash  \ottnt{c_{{\mathrm{1}}}}  ~\textbf{;}~  \ottnt{c_{{\mathrm{2}}}}  \dashv  \Gamma_{{\mathrm{3}}}  \ottsym{,}  \Delta_{{\mathrm{3}}}}{%
{\ottdrulename{check\_par\_comp}}{}%
}}

\newcommand{\ottdrulecheckXXseqXXcomp}[1]{\ottdrule[#1]{%
\ottpremise{\Gamma_{{\mathrm{1}}}  \ottsym{,}  \Delta_{{\mathrm{1}}}  \vdash  \ottnt{c_{{\mathrm{1}}}}  \dashv  \Gamma_{{\mathrm{2}}}  \ottsym{,}  \Delta_{{\mathrm{2}}}}%
\ottpremise{\Gamma_{{\mathrm{2}}}  \ottsym{,}  \Delta_{{\mathrm{1}}}  \vdash  \ottnt{c_{{\mathrm{2}}}}  \dashv  \Gamma_{{\mathrm{3}}}  \ottsym{,}  \Delta_{{\mathrm{3}}}}%
}{
\Gamma_{{\mathrm{1}}}  \ottsym{,}  \Delta_{{\mathrm{1}}}  \vdash   \ottnt{c_{{\mathrm{1}}}} ~\rule[0.5ex]{1.5em}{0.55pt}~ \ottnt{c_{{\mathrm{2}}}}   \dashv  \Gamma_{{\mathrm{3}}}  \ottsym{,}   \Delta_{{\mathrm{2}}}  \cap  \Delta_{{\mathrm{3}}} }{%
{\ottdrulename{check\_seq\_comp}}{}%
}}

\newcommand{\ottdrulecheckXXinterXXseqXXcomp}[1]{\ottdrule[#1]{%
\ottpremise{\Gamma_{{\mathrm{1}}}  \ottsym{,}  \Delta_{{\mathrm{1}}}  \vdash  \ottnt{c_{{\mathrm{1}}}}  \dashv  \Gamma_{{\mathrm{2}}}  \ottsym{,}  \Delta_{{\mathrm{2}}}}%
\ottpremise{\Gamma_{{\mathrm{2}}}  \ottsym{,}   \bar{ \rho }   \vdash  \ottnt{c_{{\mathrm{2}}}}  \dashv  \Gamma_{{\mathrm{3}}}  \ottsym{,}  \Delta_{{\mathrm{3}}}}%
}{
\Gamma_{{\mathrm{1}}}  \ottsym{,}  \Delta_{{\mathrm{1}}}  \vdash   \ottnt{c_{{\mathrm{1}}}}  \overset{ \rho }{\sim}  \ottnt{c_{{\mathrm{2}}}}   \dashv  \Gamma_{{\mathrm{3}}}  \ottsym{,}   \Delta_{{\mathrm{2}}}  \cap  \Delta_{{\mathrm{3}}} }{%
{\ottdrulename{check\_inter\_seq\_comp}}{}%
}}

\newcommand{\ottdrulecheckXXif}[1]{\ottdrule[#1]{%
\ottpremise{\Gamma  \ottsym{,}  \Delta_{{\mathrm{1}}}  \vdash  \ottmv{x}  \ottsym{:}  \ottkw{bool}  \dashv  \Delta_{{\mathrm{2}}}}%
\ottpremise{ \Gamma  \ottsym{,}  \Delta_{{\mathrm{2}}}  \vdash  \ottnt{c_{{\mathrm{1}}}}  \dashv  \Gamma_{{\mathrm{2}}}  \ottsym{,}  \Delta_{{\mathrm{3}}}  \ottlinebreakhack }%
\ottpremise{\Gamma  \ottsym{,}  \Delta_{{\mathrm{2}}}  \vdash  \ottnt{c_{{\mathrm{2}}}}  \dashv  \Gamma_{{\mathrm{3}}}  \ottsym{,}  \Delta_{{\mathrm{4}}}}%
}{
\Gamma  \ottsym{,}  \Delta_{{\mathrm{1}}}  \vdash  \ottkw{if} \, \ottmv{x} \, \ottnt{c_{{\mathrm{1}}}} \, \ottnt{c_{{\mathrm{2}}}}  \dashv  \Gamma  \ottsym{,}     \Delta_{{\mathrm{2}}}  \cap  \Delta_{{\mathrm{3}}}    \cap  \Delta_{{\mathrm{4}}} }{%
{\ottdrulename{check\_if}}{}%
}}

\newcommand{\ottdrulecheckXXifXXalt}[1]{\ottdrule[#1]{%
\ottpremise{\Gamma  \ottsym{,}  \Delta_{{\mathrm{1}}}  \vdash  \ottmv{x}  \ottsym{:}  \ottkw{bool}  \dashv  \Delta_{{\mathrm{2}}}}%
\ottpremise{ \Gamma  \ottsym{,}  \Delta_{{\mathrm{2}}}  \vdash  \ottnt{c_{{\mathrm{1}}}}  \dashv  \Gamma_{{\mathrm{2}}}  \ottsym{,}  \Delta_{{\mathrm{3}}}  \ottlinebreakhack }%
\ottpremise{\Gamma  \ottsym{,}  \Delta_{{\mathrm{3}}}  \vdash  \ottnt{c_{{\mathrm{2}}}}  \dashv  \Gamma_{{\mathrm{3}}}  \ottsym{,}  \Delta_{{\mathrm{4}}}}%
}{
\Gamma  \ottsym{,}  \Delta_{{\mathrm{1}}}  \vdash  \ottkw{if} \, \ottmv{x} \, \ottnt{c_{{\mathrm{1}}}} \, \ottnt{c_{{\mathrm{2}}}}  \dashv  \Gamma  \ottsym{,}  \Delta_{{\mathrm{4}}}}{%
{\ottdrulename{check\_if\_alt}}{}%
}}

\newcommand{\ottdrulecheckXXupdate}[1]{\ottdrule[#1]{%
\ottpremise{\Gamma  \ottsym{,}  \Delta_{{\mathrm{1}}}  \vdash  \ottnt{e}  \ottsym{:}  \tau  \dashv  \Delta_{{\mathrm{2}}}}%
\ottpremise{\Gamma  \ottsym{(x)}  \ottsym{=}  \tau}%
}{
\Gamma  \ottsym{,}  \Delta_{{\mathrm{1}}}  \vdash  \ottmv{x}  \ottsym{:=}  \ottnt{e}  \dashv  \Gamma  \ottsym{,}  \Delta_{{\mathrm{2}}}}{%
{\ottdrulename{check\_update}}{}%
}}

\newcommand{\ottdrulecheckXXwhile}[1]{\ottdrule[#1]{%
\ottpremise{\Gamma  \ottsym{,}  \Delta_{{\mathrm{1}}}  \vdash  \ottmv{x}  \ottsym{:}  \ottkw{bool}  \dashv  \Delta_{{\mathrm{2}}}}%
\ottpremise{\Gamma  \ottsym{,}  \Delta_{{\mathrm{2}}}  \vdash  \ottnt{c}  \dashv  \Gamma_{{\mathrm{3}}}  \ottsym{,}  \Delta_{{\mathrm{3}}}}%
}{
\Gamma  \ottsym{,}  \Delta_{{\mathrm{1}}}  \vdash  \ottkw{while} \, \ottmv{x} \, \ottnt{c}  \dashv  \Gamma  \ottsym{,}   \Delta_{{\mathrm{3}}}  \cap  \Delta_{{\mathrm{2}}} }{%
{\ottdrulename{check\_while}}{}%
}}

\newcommand{\ottdefnccheck}[1]{\begin{ottdefnblock}[#1]{$\Gamma_{{\mathrm{1}}}  \ottsym{,}  \Delta_{{\mathrm{1}}}  \vdash  \ottnt{c}  \dashv  \Gamma_{{\mathrm{2}}}  \ottsym{,}  \Delta_{{\mathrm{2}}}$}{}
\ottusedrule{\ottdrulecheckXXskip{}}
\ottusedrule{\ottdrulecheckXXwrite{}}
\ottusedrule{\ottdrulecheckXXlet{}}
\ottusedrule{\ottdrulecheckXXparXXcomp{}}
\ottusedrule{\ottdrulecheckXXseqXXcomp{}}
\ottusedrule{\ottdrulecheckXXinterXXseqXXcomp{}}
\ottusedrule{\ottdrulecheckXXif{}}
\ottusedrule{\ottdrulecheckXXifXXalt{}}
\ottusedrule{\ottdrulecheckXXupdate{}}
\ottusedrule{\ottdrulecheckXXwhile{}}
\end{ottdefnblock}}

\newcommand{\ottdefnsJop}{
\ottdefnelargereduce{}\ottdefnclargereduce{}\ottdefnesmallreduce{}\ottdefncsmallreduce{}\ottdefnecheck{}\ottdefnccheck{}}

\newcommand{\ottdefnss}{
\ottdefnsJop
}

\newcommand{\ottall}{\ottmetavars\\[0pt]
\ottgrammar\\[5.0mm]
\ottdefnss}